\documentclass[usenatbib]{mn2e}
\usepackage{graphicx}	
\usepackage{amsmath}	
\usepackage{amssymb}	
\usepackage{multicol}        
\usepackage{bm}		
\usepackage{pdflscape}	
\usepackage{ulem} 

\usepackage[T1]{fontenc}
\usepackage{ae,aecompl}

\usepackage{newtxtext,newtxmath}

\title[Clocking the formation of today's largest galaxies]{Clocking the formation of today's largest galaxies: Wide field integral spectroscopy of Brightest Cluster Galaxies and their surroundings}

\author [Edwards et al.]{Louise O.V. Edwards$^1$\thanks{E-mail:ledwar04@calpoly.edu},
Matthew Salinas$^1$, Steffanie Stanley$^1$, Priscilla E. Holguin West$^1$,
\newauthor
Isabella Trierweiler$^{2,3}$, Hannah Alpert$^{2,4}$, Paula Coelho$^5$, Saisneha Koppaka$^2$,
\newauthor
Grant R. Tremblay$^6$, Hugo Martel$^{7,8}$, Yuan Li$^9$
\\
$^1$Department of Physics, Cal Poly State University, San Luis Obispo, CA\\
$^2$Department of Physics and Astronomy, Yale University, New Haven, CT\\
$^3$Division of Astronomy and Astrophysics, University of California Los Angeles, Los Angeles, CA\\
$^4$Aeronautics and Astronautics Department, Stanford University, Stanford, CA\\
$^5$Universidade de S\~{a}o Paulo, IAG, S\~{a}o Paulo, Brazil\\
$^6$Harvard-Smithsonian Center for Astrophysics, Cambridge, MA\\
$^7$D\'{e}partement de physique, de g\'{e}nie physique et d'optique, Universit\'{e} Laval, Qu\'{e}bec, QC, Canada\\ 
$^8$ Centre de Recherche en Astrophysique du Qu\'{e}bec, Qu\'{e}bec, QC, Canada\\ 
$^9$ Center for Computational Astronomy (CCA), Flatiron Institute, New York, NY, USA\\ 
}

\date{\today}

\begin{document}
\maketitle

\begin{abstract}

The formation and evolution of local brightest cluster galaxies (BCGs) is investigated by determining the stellar populations and dynamics from the galaxy core, though the outskirts and into the intracluster light (ICL).  Integral spectroscopy of 23 BCGs observed out to $4\, r_{e}$ is collected and high signal-to-noise regions are identified. Stellar population synthesis codes are used to determine the age, metallicity, velocity, and velocity dispersion of stars within each region. The intracluster light (ICL) spectra are best modeled with populations that are younger and less metal-rich than those of the BCG cores. The average BCG core age of the sample is $\rm 13.3\pm 2.8\,Gyr$ and the average metallicity is $\rm [Fe/H] = 0.30\pm0.09$, whereas for the ICL the average age is $\rm 9.2\pm3.5\,Gyr$ and the average metallicity is $\rm [Fe/H] =  0.18\pm0.16$. The velocity dispersion profile is seen to be rising or flat in most of the sample (17/23), and those with rising values reach the value of the host cluster's velocity dispersion in several cases. The most extended BCGs are closest to the peak of the cluster's X-ray luminosity. The results are consistent with the idea that the BCG cores and inner regions formed quickly and long ago, with the outer regions and ICL forming more recently, and continuing to assemble through minor merging. Any recent star formation in the BCGs is a minor component, and is associated with the cluster cool core status.

\end{abstract}


\section{Introduction}

The deep gravitational potential wells at the centre of rich galaxy clusters are home to a dynamic environment, with components unobserved elsewhere in the universe. The brightest, largest, and most massive galaxy in the cluster (the BCG) is usually found at the cluster core. An extended component of intracluster light (ICL) is also found at the bottom of the gravitational potential of the cluster, often coincident with the BCG. Many other massive red galaxies are found within the virial radius of the cluster, in addition to some dwarf and spiral galaxies. These galaxies interact with each other and with the hot intracluster medium. 

BCGs form a distinct class of galaxies \citep{hau78,pos95}. They are red and have smooth brightness distributions, however, both observations \citep{von06,edw09,oli15,bar16} and models \citep{mur07,ton12,con14} show that BCGs have complex formation histories best described with a variety of stellar populations, which include components from major mergers, in situ star formation, and contributions from low-mass and L$_{*}$ galaxies. Very old stellar populations, high metallicities and $\alpha$-enhancements are observed in longslit studies of BCGs \citep{bro07b,lou08,lou12} which reach out to the effective radius of the galaxy. Less is known about the stellar populations and dynamics of the outer parts of the BCG, and whether there is a smooth transition between the BCG outskirts and the intracluster light (ICL).

The ICL is very faint, so it is difficult to get accurate surface brightness profiles and herculean efforts are required to obtain a spectrum. It is only recently that \citet{klu19} has a presented a large photometric study of 170 BCG+ICL systems observed in g'. Tests for separating the BCG from the ICL have been performed, and indeed breaks in the luminosity profile from the BCG core have been seen \citep{kri06,kri07}. Recent observations of 10 clusters observed with the Hubble Space Telescope as part of CLASH \citep[]{pos12} and Frontier Fields \citep{lot16} programs have provided exquisite photometric measurements of negative colour and metallicity gradients of the ICL \citep{mon14,dem15,mon18}. The ICL metallicities found in the Frontier Fields clusters echo those of Milky-Way mass outskirts, and younger ages imply accretion within $z<1$. The negative metallicity gradients support an origin from dwarf or $\rm L_{*}$ galaxies as predicted in models of \citet{hir15} and are consistent with the colours measured by \citet{dem15}, but is difficult to explain if the source is the BCG.  None the less, this analysis has only been performed using photometric measurements, and a true test would be to compare the ages and metallicities of the BCG and nearby galaxies with the ICL derived from spectroscopic observations. 

Currently, only a few spectra of the ICL exist. \citet{mel12} study a $z=0.3$ cluster, finding stellar populations to be old and metal-rich, very similar to the nearby BCG. Observations of the BCGs in Coma  by \citet{coc10a} and \citet{coc10b} paint a different picture, of old stars with metal-poor or solar metallicities beyond one effective radius ($1\, r_{e}$). For the cD galaxy in Hydra~I, negative age gradients are observed beyond $1\, r_{e}$, suggesting a component from dwarf galaxies \citep{bar16}. 


The formation of BCGs and the origin of the ICL has been studied numerically using various techniques, including gravity-only simulations and gas dynamical simulations, often combined with semi-analytical models (SAMs) for the evolution of galaxies \citep{del04,wil04,som05,cro06,rud06,tut07,puc10,dol10,rud11,hir16,tan18,rag18}. Most of the work on BCGs focuses on reconstructing their merger and star formation histories. By combining a SAM with merger trees extracted from cosmological simulations, \citet{del07} studied the hierarchical growth of BCGs in a $\Lambda$CDM universe. They found that most stars in BCGs
formed early inside progenitors, that later merge together to form the BCGs. In these simulations, 80\% of the stars destined to end up in BCGs were already formed at redshifts $z=3$,
but at $z=0.5$, only half the mass of the final BCGs had been assembled. However, these results depend on the assumptions behind the SAM, and other studies have found that BCGs assemble their mass at earlier redshifts (e.g. \citealt{ton12}). \citet{rag18} found that at $z>2$, most stars that are destined to end up in the BCGs form in progenitors that have not merged yet, while at lower redshifts 
 most stars form inside the BCGs themselves. More recently, \citet{con18} estimated that low-mass BCGs build-up 35\% of their present-day stellar mass by mergers, while the fraction reaches 70\% for massive BCGs. The cosmological hydrodynamical simulations of \citet{cui14} find that a double Maxwellian distribution fits the stellar distributions in the core of the clusters, and identify the ICL with the high velocity dispersion component, and the BCG with the low velocity dispersion component. The resulting ICL is found to be composed of a slightly younger and less metal-rich stellar population than the BCG.

As for the origin of the ICL, several scenarios ranging from in situ star formation to cluster evaporation and 3-body ejection are currently debated (see \citealt{tut11} for a review). While all these scenarios might contribute to the ICL to some extent, there is a general consensus that the bulk of the ICL originates from dynamical processes, either stellar mass loss during mergers or tidal disruption during close encounters. The simulations of \citet{mur04} show the presence of a diffuse stellar component around the BCG that can account for the ICL, and that probably originates from tidal interactions. With their later simulations \citep{mur07}, they reached the different conclusion that the ICL forms in parallel with the BCG and other massive galaxies, and that the bulk of the ICL is formed during mergers, with a small contribution from tidal disruption. Other studies \citep{bar09,rud09,mar12,con18} found that tidal disruption can account for 40\% to 100\% of the observed ICL in some clusters, depending on the assumptions of the model. Most of the ICL produced by this process comes from intermediate-mass galaxies, since lower-mass galaxies, even though much more numerous, tend to be resistant to tidal disruption because of their compactness \citep{pur07,mar12,con14,con18}. 

Putting together ideas from both theory and observations, one interesting paradigm for the formation of today's most massive galaxies is the following: a high $z$~phase where gas accretion is high, a moderate $z$~phase where star formation peaks, and a low $z$~phase where buildup is mostly via mergers and accretion \citep{van16,coo19}. Very early starbursts associated with dusty star forming galaxies at $z > 3$ \citep{tof14}  which would already have become compact quiescent galaxies by $z\sim2$ have been observed \citep{kri06,van08}. Further time allows these dense red nuggets to merge together with smaller \citep{bez09} and newly quenched galaxies \citep{kro14}, building the larger and less dense systems observed today. \citet{vul14} have compared central galaxies in massive haloes at $z=0.6$ with those at $z\sim0$, finding that the central parts of the galaxy form first and quickly, but that the central velocity dispersion evolves quite slowly, consistent with these findings.  Negative metallicity gradients have been measured in several massive galaxies \citep{lab12,gre15}, with a flattening beyond $2\, r_{e}$ \citep{oya19}, which further corroborates this picture if the outskirts form in shallower potentials \citep{tre04}. It is possible that local BCGs as a group will show negative metallicity gradients in their outskirts as well, as massive BCGs are expected to have undergone more mergers \citep{edw12, jim13} than large ellipticals, and \citet{lou12} find negative metallicity gradients in local BCGs within $0.5\, r_{e}$.  

Thus, mapping the stellar populations and kinematics of BCGs from the core to beyond one effective radius ($1\, r_{e}$) has great potential to constrain the history of mass assembly.  Stellar population synthesis is still hampered by issues such as the age-metallicity degeneracy, and assumptions of dust morphologies and a universal initial mass function (IMF); however, uniformly analysed samples can still provide useful overall trends of stellar populations in galaxies. Previous spectroimaging studies of BCGs include those of Hydra I by \citet{bar16}, 9 BCGs by \citet{oli15}, 9 BCGs by \citet{edw09} and 6 by \citet{hat07}. All authors find a mix of old and intermediate stellar populations in the core, no significant age gradient, and a mix of metallicity gradients (shallow to flat) out to $1\, r_{e}$. The BCG core populations are beginning to be well understood, but the connection between the BCG and ICL has not yet been resolved, which motivates observations over a larger area, into the galaxy outskirts and ICL. 

 Broadening our perspective and including an analysis of the BCG's environment may provide important insights. Assuming the galaxy is in equilibrium and symmetric, the Jeans equation with an NFW \citep{nav96} density profile for the dark matter gives velocity dispersion gradients that are falling. This might explain why dwarf galaxies, the Milky-Way and many isolated ellipticals show decreasing velocity dispersion gradients.  The velocity dispersion profile of galaxies is usually interpreted as a signature of the underlying mass profile (of the stars and dark matter of the galaxy itself, and of the cluster mass in the case of centrally located BCGs). The work of \citet{ben15} on the cD of Abell~2199, and \citet{rit11} on the Hydra~I cD find that the velocity dispersion rises to match that of the cluster - a kinematic signature of a different component at a distance of $\sim20\,^{\prime\prime}-70\,^{\prime\prime}$ from the BCG core. While rising velocity dispersions in BCGs are often found \citep{dre79,bro07b,new13,bar16,lou18,hil18}, this is not always the case \citep{fis95a,bro07b,lou08,vea18}. Velocity dispersion can also be influenced by dynamical activity, such as recent cluster-scale or galaxy-galaxy mergers and \citet{hil18} find substructure in a 2D map of the velocity dispersion in the halo of the Hydra~I cD galaxy, NGC~3311. 

Furthermore, in a relaxed cluster in equilibrium, the most massive galaxy should be at rest in the centre of the cluster. This is indeed the case for most observed BCGs, and this constitutes the basis for the ``central galaxy paradigm,'' upon which most semi-analytical galaxy evolution models rely. However, several cases of BCGs located off-centre, or at the centre of clusters but with a large velocity, have been reported \citep{coz09,ski11}. These clusters are probably remnants of major mergers between clusters of comparable masses, which have not yet relaxed to equilibrium
\citep{mar14}. The effect of such mergers on the BCG outskirts and ICL is unknown at present.

\begin{table*}
  \centering
  \caption{Observational Data. Twenty three BCGs were observed using SparsePak on the dates shown. The RA and Dec positions are given for the centre of the BCG. The integration time listed is the total number of seconds per position and the rotator offset angle is given in degrees clockwise from North. The redshift and E(B$-$V) are listed next. The FOV listed is the $x$ and $y$ dimensions for the main body of SparsePak fibres, converted to kpc.}
  \begin{tabular}{l l l l c c c c c}
    \hline \hline
    Cluster & Dates & RA & Dec& Int Time & Rot. & $z$ & E(B$-$V) & FOV\\ 
      & (dd/mm/yr) & (h:m:s) & (h:m:s) & (s/pos) & (deg) & & &($\rm kpc \times kpc$)\\ 
      \hline
    A75 & 14/10/14  & 00:39:42.35 & +21:14:06.68 & 2700 & 0 & 0.0626&0.031&83.4$\times$82.5\\
    A85 & 25/11/13 & 00:41:50.54 & $-$09:18:13.07 & 3600 & $-$15 & 0.0551&0.034&73.9$\times$73.2\\
    A160A &  04/12/16 & 01:12:59.56  & +15:29:28.64  & 3600 & 0 &0.0447&0.076&61.6$\times$61.0\\
    A193 &  28/11/13 & 01:25:07.22 & +08:41:57.80 & 3600 & $-$75 & 0.0486&0.043&66.6$\times$65.9\\
    A376 &  27/11/13 & 02:46:03.59 & +36:54:12.06 & 3600 & 150 & 0.0484&0.063&65.3$\times$64.7\\
    A407 &  25+28/11/13 & 03:01:52.10 & +35:50:22.96 & 7200 & 90 & 0.0462&0.172&63.9$\times$62.2\\
    A602  &   04/12/16 &  07:53:26.59 & +29:21:33.59   & 3600 & 0 & 0.0619&0.050&82.2$\times$81.4\\
    A671   &  04/12/16 & 08:28:31.84 &+30:25:48.61  &3600 & 0 &0.0502&0.042&67.8$\times$67.1\\
    A757 &   05/12/16 & 09:13:07.74 & +47:42:30.32  & 3600 & 0 &0.0517&0.015&70.2$\times$69.6\\
    A1668 &  21+22/05/14 & 13:03:46.56 & +19:16:16.75 & 2524 & 0 & 0.0634&0.081&83.4$\times$82.5\\
    A1795 & 01+02/05/13 & 13:48:52.45 & +26:35:34.71 & 3600 & 107.6 &0.0625&0.012&82.2$\times$81.4\\
    A2199 &  21/05/14 & 16:28:38.40 & +39:33:03.97 & 2700 & 0 &0.0301&0.003&42.3$\times$41.9\\
    A2457 &  13/10/14 & 22:35:40.74 & +01:29:05.76 & 3600 & 0 &0.0594&0.074&78.7$\times$77.9\\
    A2589 &   05/12/16 &  23:23:57.43 & +16:46:29.24  & 3600 & 0 &0.0414&0.026&56.6$\times$56.0\\
    A2622 & 16/10/14 & 23:35:01.71 & +27:22:15.58 & 3600 & $-$55 &0.0620&0.050&82.2$\times$81.4\\
    A2626 & 13/10/14 & 23:36:30.45 & +21:08:46.97 & 2700 & 0 &0.0553&0.057&73.9$\times$73.2\\
    A2634 & 06/12/16 & 23:38:29.23 &+27:01:50.15 & 3600 &0 &0.0314&0.062&43.6$\times$43.2\\    
    A2665 & 16/10/14 & 23:50:50.79 & +06:09:03.35 & 2700 & $-$75 &0.0556&0.071&75.1$\times$74.3\\
    IIZw108 &  27/11/13+15/10/14 & 21:13:54.99 & +02:33:54.79 & 5400 & 103 & 0.0494&0.063&66.6$\times$65.9\\
    MKW3 &  21+22/05/14 & 15:21:51.53 & +07:42:34.78 & 3600 & 0 &0.0450&0.072&60.3$\times$59.8\\
    UGC03957 &  05/12/16 &07:40:58.37  & +55:25:38.27  & 3600 & 0 & 0.0341&0.043&47.5$\times$47.1\\
    Z2844 & 02/05/13 & 10:02:36.51 & +32:42:24.82 & 1800 & 65.6 & 0.0500&0.013&67.8$\times$67.1\\
    Z8338 & 22/05/14 & 18:11:05.20 & +49:54:33.55 & 4100 & 0 & 0.0470&0.192&64.1$\times$63.5\\
\hline
\label{tab:obdat}  
\end{tabular}
 \end{table*}

This paper presents a sample of local BCGs in X-ray bright clusters observed using integral field spectroscopy. The observations span a $\sim70\,$kpc field around each source, so that nearby neighbours and the BCG outskirts are captured in addition to the radius out to where the ICL is expected to become dominant. These results constitute the largest sample of spectroscopically observed ICL currently available. The analysis of the first three clusters is published in \citet[hereafter Paper~I]{edw16}. Section~\ref{obs} describes the dataset, observations,  data reduction, use of population synthesis models, and a discussion of errors and uncertainties. Section~\ref{res} presents the age and metallicity profiles of the BCGs, which generally fall from the core to the outskirts.   
The average velocity dispersion profiles are generally rising, but negative velocity gradients are seen in smaller, luminous BCGs hosted by clusters with low X-ray luminosity ($\rm L_{X}$). Most of the BCGs are found close to the X-ray peak: cool cores (CCs) and the presence of optical emission lines are observed in this dataset for 11 of the 23 systems (as listed in Table~\ref{tab:ObjectData}). The few BCGs with large separations from the X-ray luminosity peak have the same average central velocity dispersion as the sample at large, but a smaller average effective radius. Section~\ref{disc} uses the key results to argue for a two-phase formation model for the BCG and its outskirts and ICL where the core forms first and the outskirts assemble later. Stellar population and kinematic arguments are presented for the ICL as a distinct component. Section~\ref{con} provides our conclusions. Values of $\rm H_{0}=70\,$km/s/Mpc, $\Omega_{m}=0.3, \Omega_{\Lambda}=0.7$ are used throughout.

\section{Observations and Data Reduction}\label{obs}
The uniqueness of this dataset lies in its large field of view (FOV), which reaches the galaxy outskirts, far enough that ICL begins to dominate the light. Furthermore, it extends to $\sim20-50\,$kpc beyond the BCG core, the region occupied by bound companions that are likely to merge within $0.5\,$Gyr \citep{edw12}. Integral-field observations of twenty three BCGs are performed using the SparsePak instrument on the $3.5\,$m WIYN telescope. The reduced spectra are matched to synthetic spectra in order to deduce stellar populations and galaxy kinematics. This section discusses the sample selection, data reduction, population synthesis modeling, and known errors and uncertainties within the data and analysis techniques.
 
 \subsection{Sample Selection}
 \begin{figure}
\includegraphics[width=3.5in]{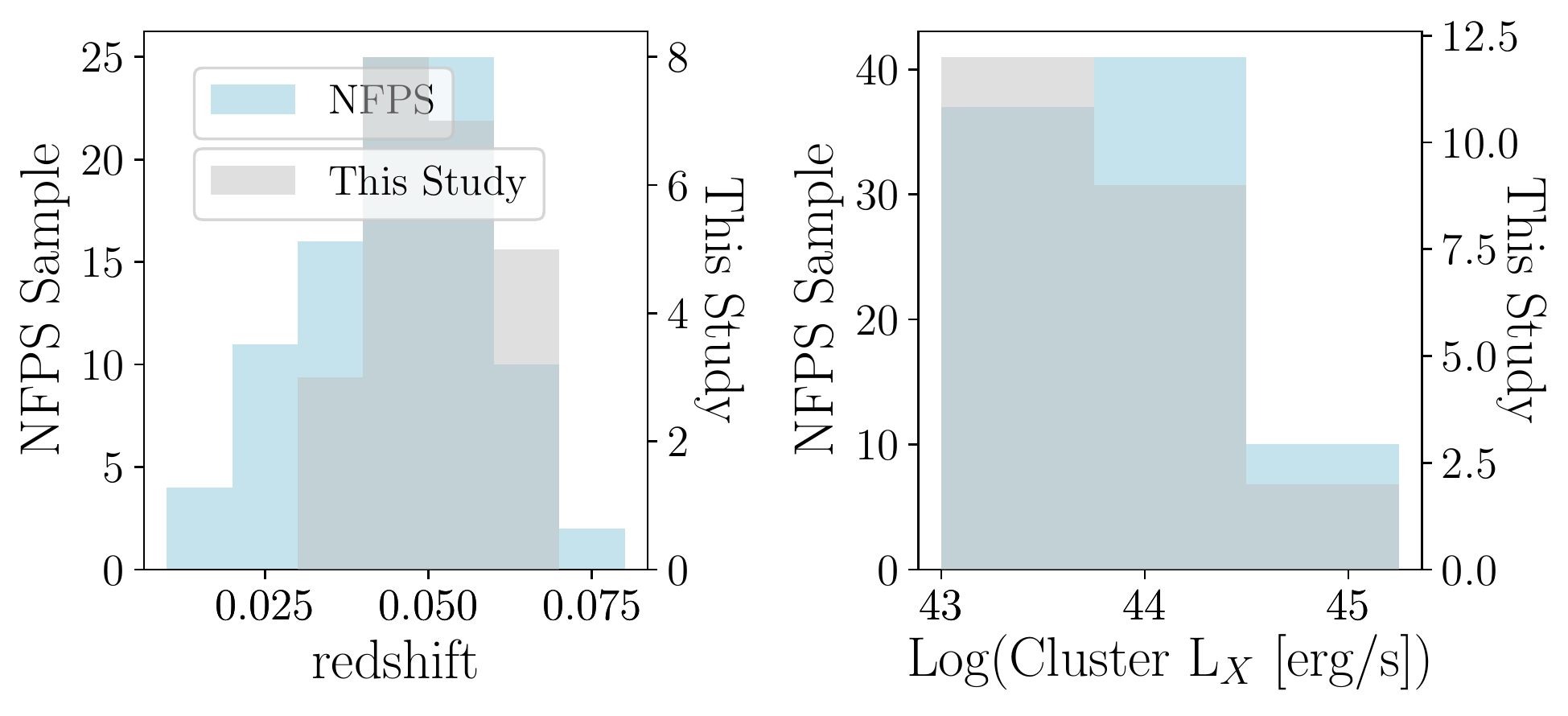}
\caption{{\it Characteristics of the sample cluster and BCG} Left: A histogram of the NFPS cluster redshifts (blue) and the subsample observed here with SparsePak  (grey). Right: Same but for cluster X-ray luminosity. }
\label{sample}
\end{figure}

\begin{figure*}
\includegraphics[height=8.5in]{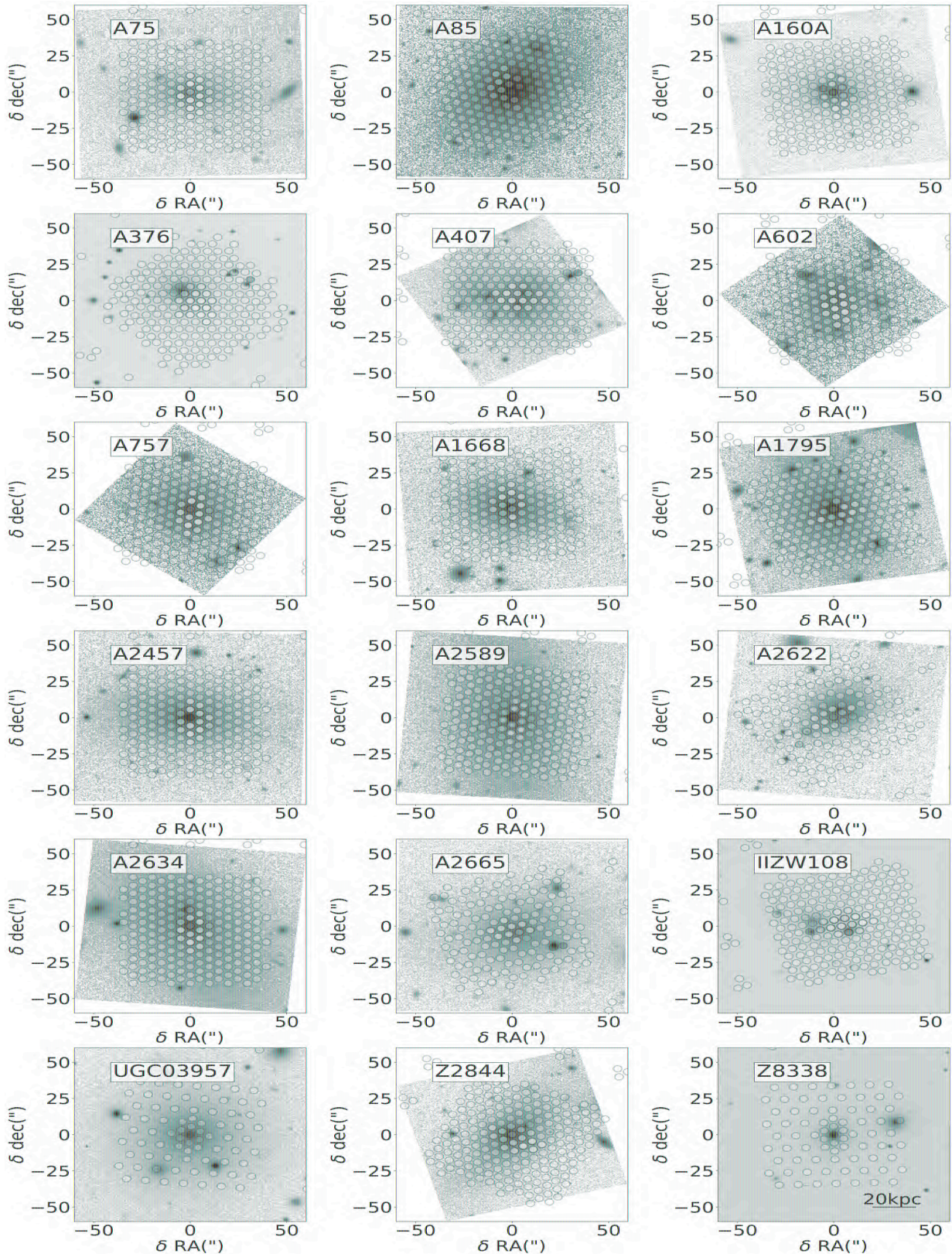}
\caption{{\it Image of the cluster centres and the SparsePak fibre positions.} The location of the SparsePak fibres, centred on the BCG, is illustrated for each of the 23 cluster cores in this sample.  The FOV is $120^{\prime\prime}\times120^{\prime\prime}$ for each cutout and each SparsePak fibre has a $5^{\prime\prime}$ diameter. All background images shown are SDSS r$^{\prime}$ images except for A376, IIZW108, Z8338 which are R-band images from the NFP survey, and UGC03957 which is a 2MASS K$_{s}$ image. The colour gradient within the SparsePak fibres shows the relative intensity of each spectrum summed over $5000\,$\AA~ and $6000\,$\AA, plotted with a linear scale. The data capture several resolution elements across the BCG.  North is up and East is to the left. The conversion from arcseconds to kpc for Z8338 is shown and that for each source is listed in Table~\ref{tab:ObjectData}.
}
\label{fig:sample}
\end{figure*}

The sample is selected from the NOAO Fundamental Plane (NFP) survey \citep[]{smi04}, which itself is based on ROSAT X-ray bright cluster catalogues \citep{ebe96,ebe98}. Figure~\ref{sample} shows the similarity in distributions of redshift and X-ray luminosity in the parent NOAO sample, with the subset of clusters observed in this program. No bias is assumed from looking only at the 61 northern targets. Twenty-five clusters have a redshift in the range $z=0.03-0.07$ and were considered for observation. This cut in redshift assures some level of uniformity across the sample.  The ICL of clusters at $z=0.03$ is likely more contaminated by the BCG outskirts than those at 0.06, but is within a factor of 2 to the physical size being compared. Similarly, the linear size difference covered in a single fibre between $z=0.03$ and $z=0.07$ is comparable. The redshift limit also assures the observations extend to nearby neighbours and the ICL. This allows for a statistical comment on local X-ray bright BCGs with a quoted 95\% confidence level and a 10\% margin of error for the main derived parameters.  BCG mass is known to correlate well with cluster mass \citep[]{edg91,hud97,bro05,lau14} so starting with X-ray bright clusters of the NFP survey ensures a fairly uniform population of BCGs. 

Twenty three clusters were ultimately observed, randomly chosen from the 25 possible, given limits of RA and airmass. These are shown in Figure~\ref{fig:sample}, which presents an image for each target in the sample, overlaid with the position of the SparsePak optical fibres.  The data capture several resolution elements across the BCG, and often include several high S/N projected companions. 

Table~\ref{tab:obdat} lists the 23 clusters that were observed with columns that list the cluster name, dates of observation, position and integration time of the central position, the rotator offset angle, the cluster redshift and extinction, as well as the dimensions of the observed area. The redshift and extinction values are taken from NED\footnote{https://ned.ipac.caltech.edu}.

\begin{table*}
  \centering
  \caption{Object data for the host cluster and BCG are compiled for each target. Columns two, three and four show the X-ray luminosity as measured in the $0.1-2.4\,$keV bandbass, the cluster velocity dispersion and the cluster $\rm r_{200}$, respectively, taken from the NFP survey. Next listed is the cool core status, as defined in the text. Strong cool cores are listed as having $\rm CC=2.0$, weak cool cores with $\rm CC=1.0$ and non-cool cores with $\rm CC=0.0$. The galaxy K-band absolute magnitude is calculated from the 2MASS apparent magnitude, and the effective radius is also from 2MASS. The central surface brightness and central Sersic index are taken from the \citet{don11} measurements. The presence of emission lines in the observed data are listed, as well as the separation between the BCG core and X-ray peak luminosity. The final column states the conversion from arcseconds to kpc.}
  \begin{tabular}{lccccccccccc}
    \hline \hline
Name &  $\rm L_{X}$&   $\sigma_{\rm cl}$&    r$_{200}$ & CC &M$_{\rm K}$&     $ r_{e}$ & $\mu_{\rm 0.5}$ & n & Lines  & X-ray Sep & 1 $^{\prime\prime}$  \\
               & ($10^{44}\rm erg/s$)              &      (km/s)    &      (Mpc)    &          &        &($^{\prime\prime}$)  &  &                  &     &  ($^{\prime\prime}$)  & (kpc)  \\
      \hline
A75      &    0.42 &     ... &     ... &     ... &  ... & 8.08 &  21.86   &4.74&    \checkmark &    118.6 &1.158\\
A85      &    4.92 &  736 &   1.274 & 2.0 & $-$26.0 &  15.21 &   21.43   &0.86&    \checkmark  &    3.0 &1.026\\
A160A  &    0.18 &  560 &   0.970 & 0.0 & $-$25.0 &  11.82 &  20.38  &1.04 &    $\times$  &   23.7 &0.865\\
A193     &    0.74 &  820 &  ...  & 1.0 & ... &   8.79 &  18.6    &4.13&    $\times$ &      1.5& 0.925\\
A376     &    0.68 &  975 &   1.689 & 0.0 & $-$24.4 &  14.13 &   25.76    &8.00&   $\times$ &    2.3& 0.907\\
A407     &    0.21 &  670 &   1.161 & 2.0 & $-$25.9 &  17.73 &  N/A   & N/A&   $\times$   &    4.8 &0.873\\
A602  &    0.53 &  675 &   1.169 & ... & $-$24.2 &   7.92 &21.18&3.30&     $\times$   &  110.8& 1.141\\
A671  &    0.44 &     ... &     ... &    0.0 & ... &  14.77 &    20.14   &2.21&    \checkmark &   77.6 &0.942\\
A757  &    0.46 &  381 &   0.660 & ... &  $-$24.5 &   5.60 &  20.80   &2.17& $\times$  &  216.0 &0.976\\
A1668    &    0.88 &  476 &  0.824 & 0.0 & $-$25.2 &   9.51 &  N/A     &N/A&   \checkmark   &    3.1& 1.158\\
A1795    &    6.33 &  725 &   1.256 &2.0 & $-$25.7 &  16.90 &   20.94  &1.88&    \checkmark   &    4.9& 1.141\\
A2199    &    1.90 &  647 &   1.120 & 2.0 & $-$25.7 &  18.13 & 21.18    &1.40&      \checkmark   &    3.9 &0.587\\
A2457    &    0.45 &    ... &   1.400 &  1.0 & ... &  10.95 &21.10 &2.60&     $\times$&     3.0 &1.092\\
A2589 &    0.89 &  583 &   1.010 & 1.0 & $-$25.4 &  15.65 &  28.54  &9.52&   $\times$ &    1.0& 0.786\\
A2622    &    0.51 &  942 &  ... &  ...& ... &   9.90 &  24.53 &6.13& $\times$ &     2.6 &1.141\\
A2626    &    0.97 &  681 &   ... &  1.0 &... &  16.98 &   24.10    &37.52&   \checkmark  &       2.2 &1.026\\
A2634 &    0.48 &  692 &   1.199 & 1.0 & $-$25.5 &  14.70 &      20.33    &3.16&  \checkmark  &       1.5 &0.606\\
A2665    &    0.93 &    ... &   1.500 &  0.0 & ... &  13.57 & 23.94    &4.15&       \checkmark  &       2.6 &1.043\\
IIZW108  &    1.09 &  399 &   0.692 & 0.0 & $-$25.6 &  29.54 & N/A&N/A&$\times$&     2.3& 0.925\\
MKW3     &    1.39 &  533 &   0.924 & 2.0 & $-$24.8 &   9.39 &    N/A  &N/A&     \checkmark  &    9.0 &0.838\\
UGC03957   &    0.48 &  511 &   0.885 & 1.0 & $-$25.3 &  10.55 & N/A  &N/A&  $\times$ &    3.1 &0.660\\
Z2844    &    0.30 &  462 &   0.800 & ... & $-$25.4 &   9.69 & N/A &N/A&     $\times$ &   2.1 &0.942\\
Z8338    &    0.45 &  533 &   0.922 & ... & $-$25.7 &  11.86 & N/A  &N/A&      \checkmark  &    1.0 &0.890\\

\hline
\label{tab:ObjectData}
\end{tabular}
 \end{table*}
 
Key physical properties of the BCG and its host cluster are listed in Table~\ref{tab:ObjectData}. Columns 1-4 list the cluster's name, X-ray luminosity, velocity dispersion, and $\rm r_{200}$. Column 5 lists is the cool core status, as defined in \citet{hud10} and calculated using the data from the ACCEPT \citep{cav09} database. The status for those clusters not in the database is taken from the existence of a central entropy dip, from a short cooling time, or a high mass deposition rate and referenced as follows: A1668 \citep{sal03}; A160, A407, A671 and A2626 \citep{whi97}; A193, A376, A2457, A2665 \citep{lac14}; A2589, A2634, IIZW108, UGC03957 \citep{che07}. The BCG K-band magnitude and effective radius, taken from 2MASS\footnote{https://irsa.ipac.caltech.edu/Missions/2mass.html}, are then listed along with central surface brightness and Sersic index measured by \citet{don11}. Whether emission lines are observed in the BCG is shown in column~10, and the separation between the BCG core and X-ray peak luminosity is given in column~11. The last column shows the arcseconds to kpc conversion.

\subsection{Observations}

The BCGs and associated standard stars were observed over the course of five observing runs from April~2013 -- December~2016. SparsePak is a sparsely packed quasi-integral field unit with 82 fibres of which 75 are arranged in a grid of dimensions $72^{\prime\prime}\times71.3^{\prime\prime}$ with a tightly-packed centre, and 7 are sky fibres located on the outside of two of the sides of the grid \citep{ber04}. As such, each source has an FOV between $43\,$kpc and $95\,$kpc. The data is thus expected to reach the ICL, as \citet{pre14} have found the ICL begins to dominate the light at $\sim40\,$kpc from the BCG core. This is further discussed in Paper~I.  In general, galaxies were observed in blocks of $600-900\,$s exposures three times at each position (Table~\ref{tab:obdat}) to help remove cosmic rays.  Each target was observed with three pointings so as to fully integrate the field, cover all close companions, and oversample the BCG. The total integration times are typically of order one hour such that the wide-field IFU observations allow for high spectra within the inner 15$^{\prime\prime}$ of the BCG. Stars and galaxies with fluxes 30 times fainter than the core can be distinguished and removed when necessary. The $2.1\,$\AA~($\sim 140\,$km/s) spectral resolution is well-matched to the $2.5\,$\AA~($\sim160\,$km/s) resolution of the template spectra that will be used to find the best-fit model of the stellar populations and kinematics.

The nights varied in their cloud cover and the seeing ranged from about $0.5^{\prime\prime}$ to $1.5^{\prime\prime}$. Cloudy conditions during the third night and part of the second night of the November 2013 observing run, resulted in some unusable data due to poor sky subtraction. During the October 2014 run many of these same objects were re-observed, but with an incomplete set of offset positions. Thus for clusters Abell~2665 and Abell~2622 data are presented for only two and one positions, respectively.  For the present study, this has the effect of reducing the $\rm S/N$ throughout the regions analysed. For Abell~376, the data were taken with two different rotator position offset angles, which results in a greater than average $\rm S/N$ for this system. For Abell~1795, the position offsets did not completely line up, thus only one of the offset positions in the maps for this cluster is used. This incompleteness is not expected to impact the main results of this paper which rely on stacked region spectra (discussed in Section~\ref{stackingspec}).

Abell~407 was observed on different nights during the same run. From comparing the sky-subtracted and flux calibrated spectra using standard stars, the stability of the flux calibration is estimated to be 4\%. IIZw108 was observed over different runs with slightly different positional and rotational offsets. In this worst-case scenario, the flux calibration is stable to 7\%.

\subsection{Data Reduction}

\begin{figure}
\includegraphics[width=3.3in]{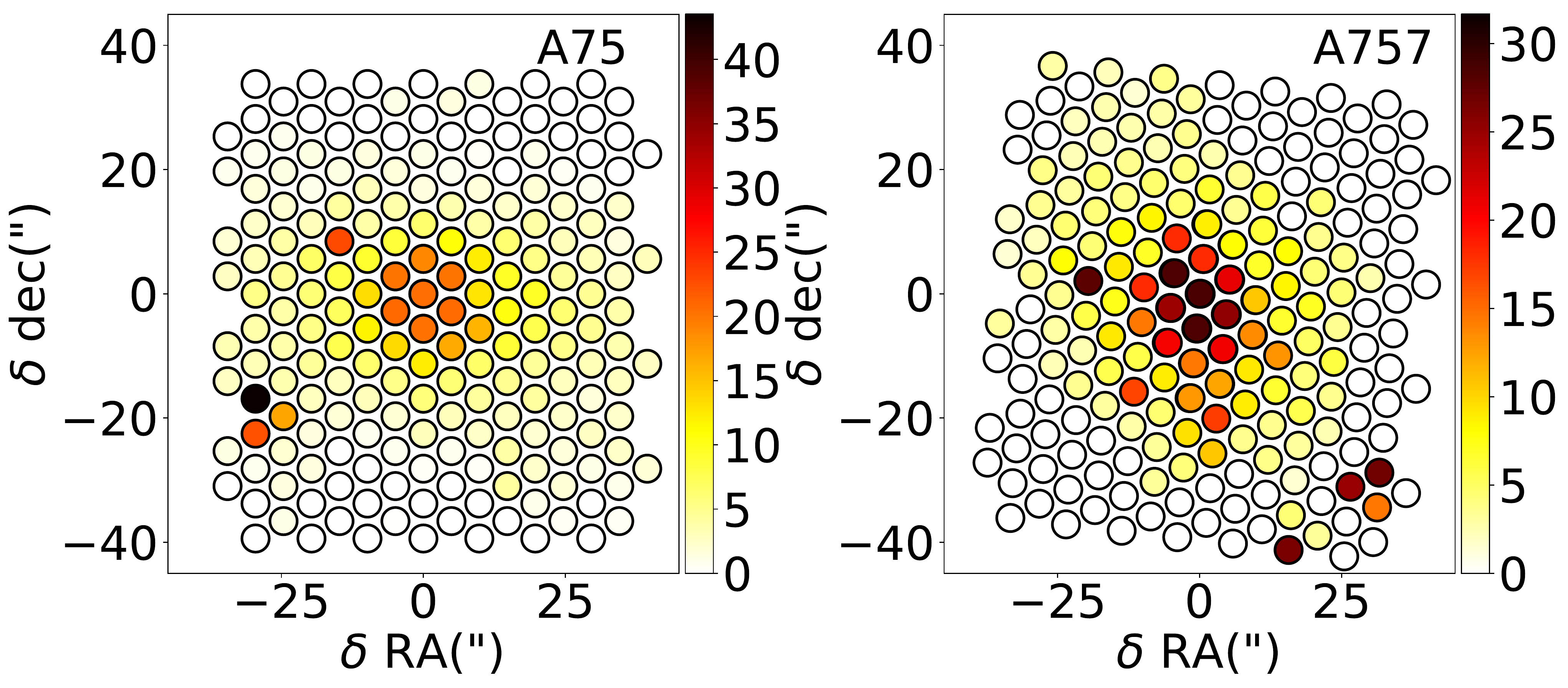}
\caption{{\it The fibre $\rm S/N$ ratio for representatives of the sample.} The colour of the SparsePak fibre shows the value of the $\rm S/N$ ratio for Abell~75 and Abell~757, two clusters that are typical in their integration times. A757 is at a lower redshift than A75, and the $\rm S/N$ for the BCG core is greater.  High in the galaxy cores, the $\rm S/N$ falls below 3 in the individual fibres of the outskirts and ICL. This motivates the use of stacked region spectra. The FOV is $90^{\prime\prime}\times90^{\prime\prime}$ for each cutout and each SparsePak fibre has a $5^{\prime\prime}$ diameter (6\,kpc for A75; 5\,kpc for A757).  North is up and East is to the left.}\label{fig:sn}
\end{figure}

The data reduction is detailed in Paper~I. Briefly, the images are bias, dark, and flat field corrected and sky subtracted. Cosmic rays and sky lines are removed and the spectra are flux-calibrated using the SDSS when available, and with standard stars taken during the run when no SDSS spectra exist as is the case for Abell~376. The spectra are all shifted to the rest frame wavelength.

\subsection{Identifying region spectra} \label{stackingspec}

\begin{figure}
\centering
\includegraphics[width=0.48\textwidth, angle=0]{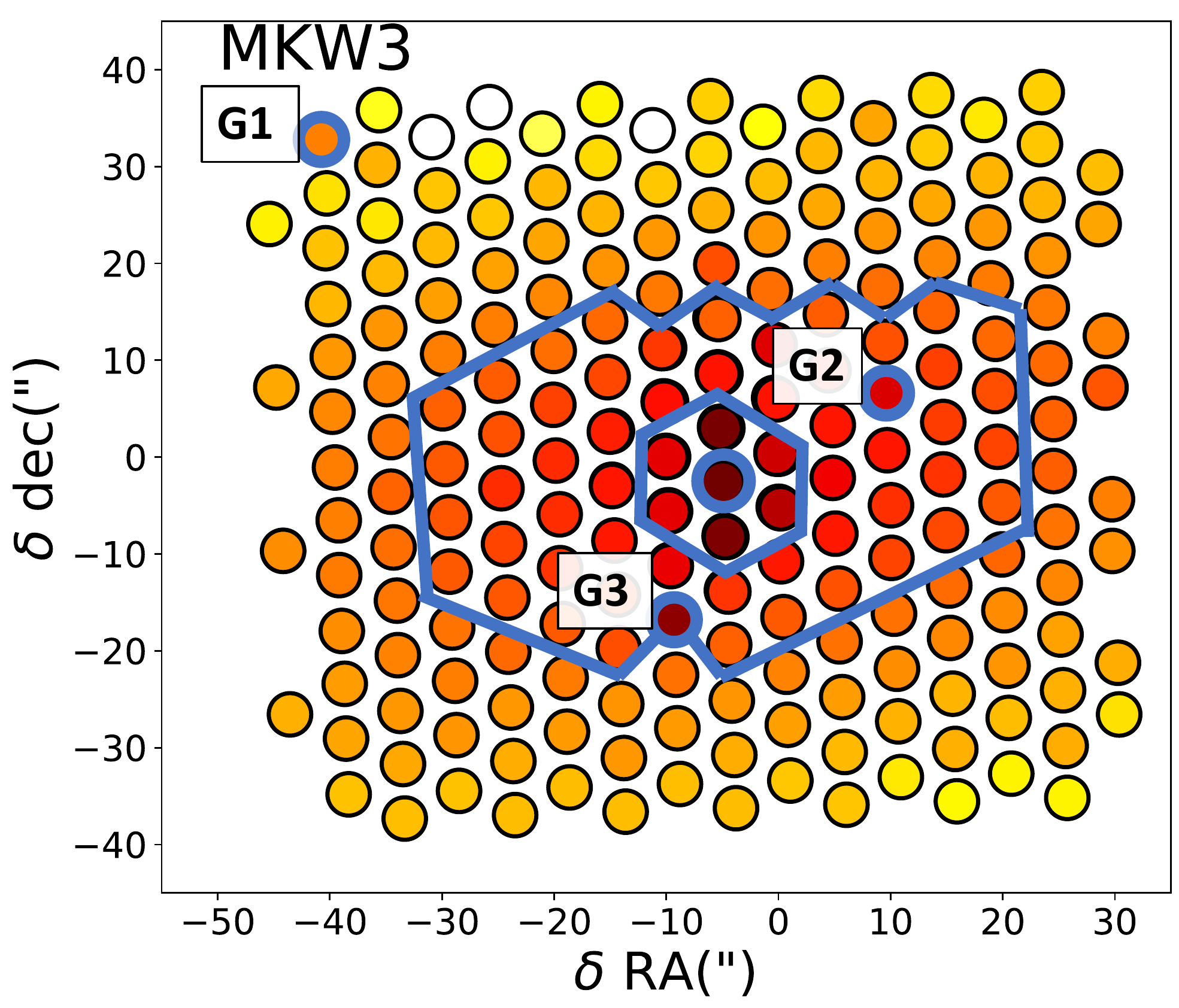}
\caption{{\it Identifying region spectra for MKW3.} The mean intensity, in arbitrary units, for each fibre is plotted with high intensity as red and low intensity as yellow.  Three possible companion galaxies are identified (G1, G2, and G3). The BCG regions are demarcated in blue. Here, there is a {\it core} region, surrounded by a concentric {\it centre} region of width $5^{\prime\prime}$. An {\it outer} region follows the general shape of the galaxy isophotes, and all remaining fibres are combined to form the {\it ICL} region.  The FOV is $90^{\prime\prime}\times90^{\prime\prime}$ and each SparsePak fibre has a $5^{\prime\prime}$ (4.2\,kpc) diameter.  North is up and East is to the left.}
\label{fig:reg}\label{fig:spec}
\end{figure}

Figure~\ref{fig:sn} shows the $\rm S/N$ per angstrom of the individual fibres for representative clusters Abell~75 and Abell~757. A value of $\sim20-30$ is measured at a 5$^{\prime\prime}$ spatial resolution for the inner 15$^{\prime\prime}$ of the BCG. But, the $\rm S/N$ drops down to $\sim 2$ at the outskirts of the field. The data are therefore binned to produce high $\rm S/N$ regions for the population synthesis analysis. Figure~\ref{fig:reg} illustrates the regions in the case of MKW3: The BCG core (central $5^{\prime\prime}$), the centre of the BCG (an annulus of the surrounding 5$^{\prime\prime}$), the outer region (where the galaxy intensity is still high), and the ICL region, which includes the remaining fibres (beyond $20-30^{\prime\prime}$ from the core). Stars and interloping galaxies are removed from the BCG and ICL regions. 

The spectra of these various regions are shown in Figure~\ref{fig:spec}. The highest $\rm S/N$ region of the combined BCG core and bright companions includes only the aperture size of one fibre, but as there are 3 pointings required to fully integrate the field, the BCG core $\rm S/N$ generally rises to $\sim 50$. The ICL is binned using only fibres with $\rm S/N\ge1$. Each cluster contains between 70-100 fibres within the ICL region, such that the $\rm S/N$ of the combined ICL spectrum is $\sim 20$, as shown in Figure~\ref{fig:spec}. It is certainly true that the majority of the ICL is beyond our FOV, however, the average stacked ICL spectrum that is constructed here matches well with the photometrically determined limit of $40\,$kpc, where the light begins to be dominated by the ICL \citep{pre14}. 

\begin{figure}
\centering
\hspace*{-0.8cm}\includegraphics[width=0.49\textwidth, angle=0]{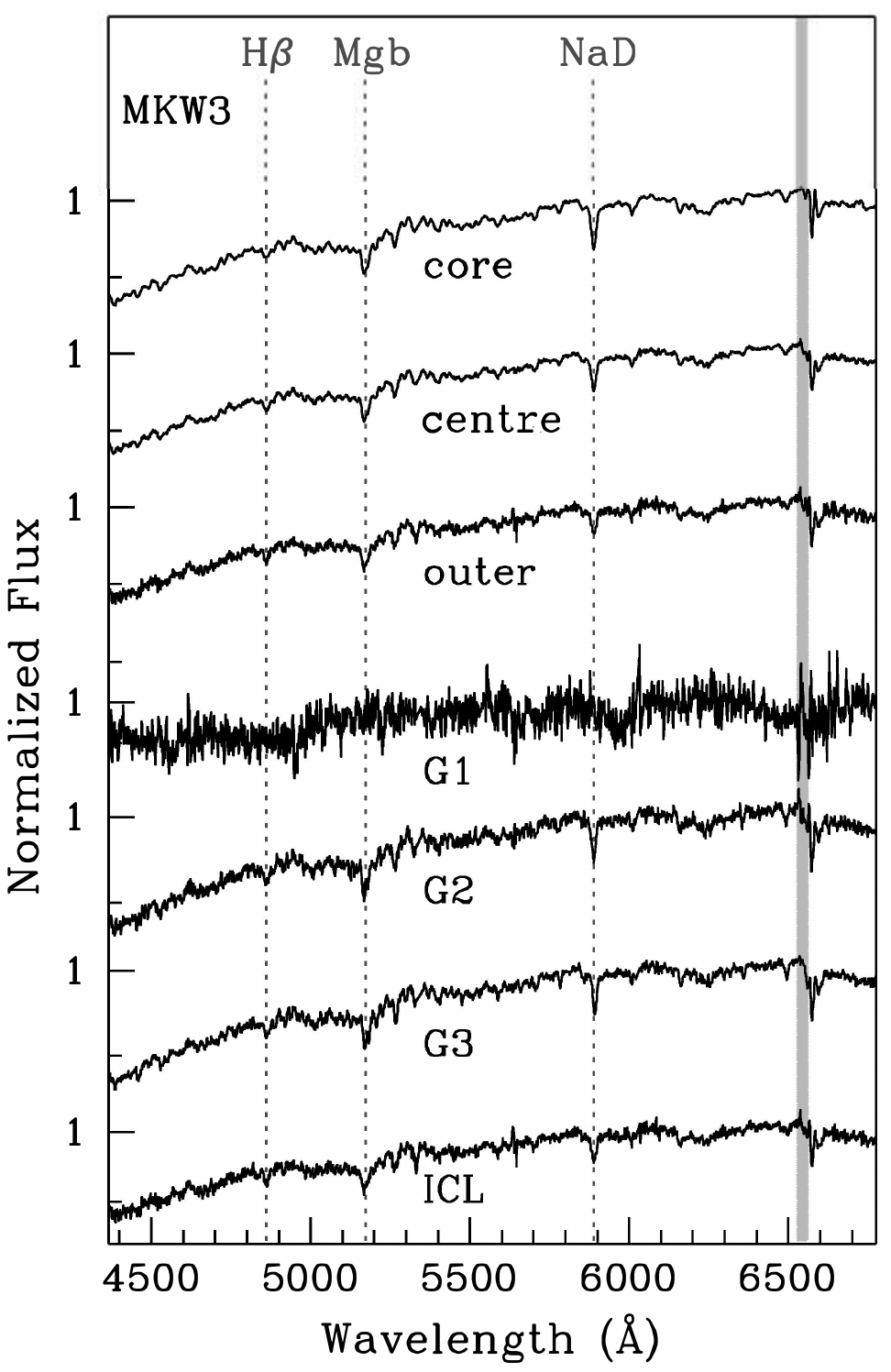}
\caption{{\it Region spectra for MKW3.} Spectra of the binned regions of Figure~\ref{fig:reg} are shown. The x-axis is the rest wavelength. Stacking the individual spectra of the regions results in high $\rm S/N$ spectra. The exception is G1, a faint source with very low $\rm S/N$. The individual fibres covering the ICL have $\rm S/N$ as low as 1, but their binned average spectrum is adequate for comparing to results from population synthesis models. The grey band to the far right indicates a portion of the spectrum affected by an instrumental artefact.}
\label{fig:spec}
\end{figure}

\subsection{Population Synthesis}

The STARLIGHT \citep{cid07} code, which employs a full-spectrum fitting algorithm, is used to find the best-fitting model spectra. Several methods for obtaining stellar populations and kinematics from the reduced region spectra are explored as various model input spectra are used, including those of \citet{bru03}, as well as models which include $\alpha$-enhancement from \citet{wal09}  and \citet{vaz15}. 

\subsubsection{Full Spectrum Fitting}
\begin{figure*}
\centering
\includegraphics[width=6.9in]{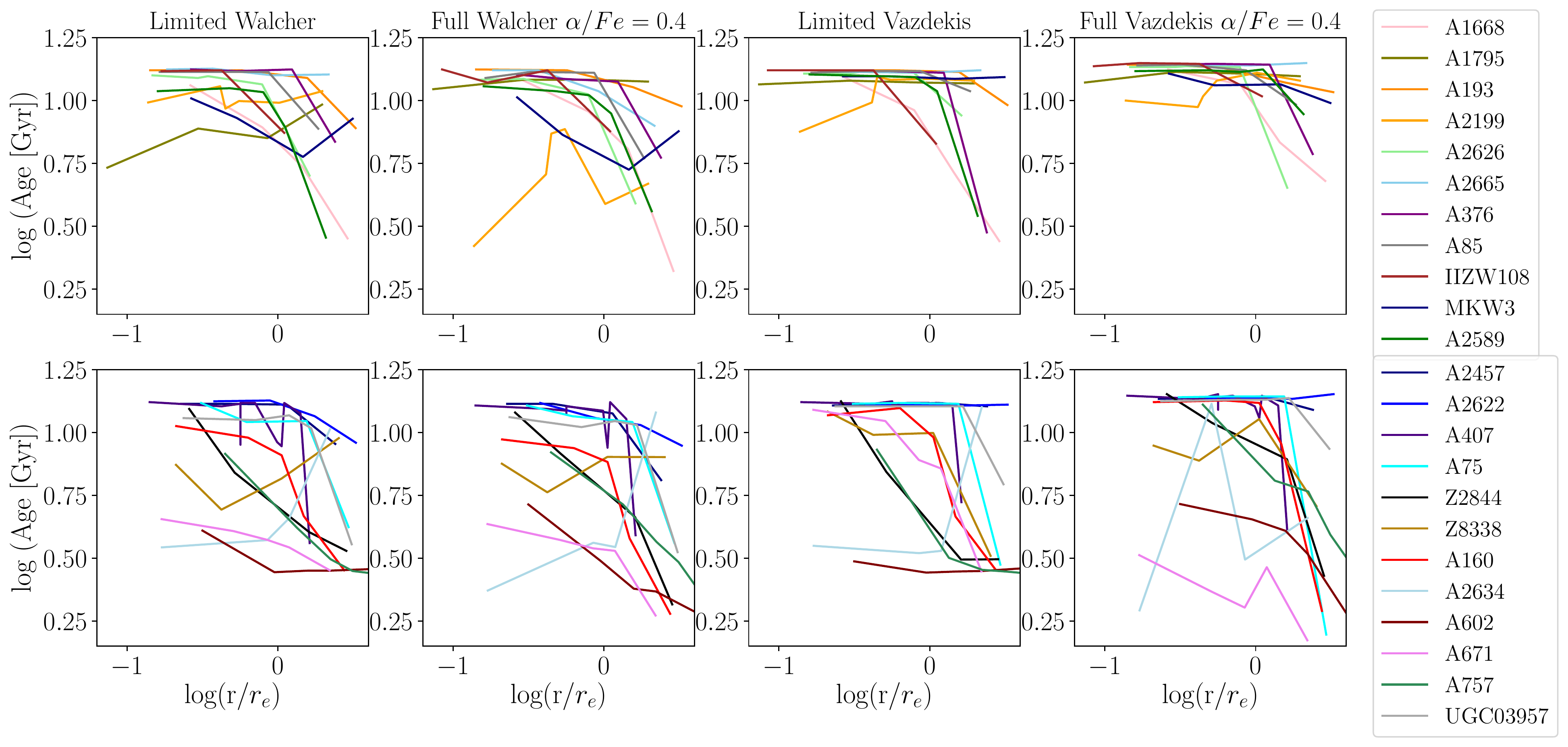}
\caption{{\it Effect of varying base spectra.} A comparison of the best-fitting average age for each region in of the cluster cores in the sample. The fitting is done using STARLIGHT with the base spectra varied. From left to right, the input base spectra include:  a limited set of 54 models from \citet{wal09}, all 0.4 $\alpha$-enhancement models from Walcher, the same limited and full set of $\alpha$-enhanced models from  \citet{vaz15}. Galaxies are split on the X-ray luminosity of the parent cluster. Systems with $\rm L_{X}>0.6 \times 10^{44}\rm erg/s$ are shown on the top, and those with $\rm L_{X} \le 0.6 \times 10^{44}\rm erg/s$ are on the bottom panels. Most of the sample shows the same general trend of negative age gradients no matter the set of base spectra.}
\label{fig:models}
\end{figure*}

Paper~I discusses the details of determining the best fit populations and kinematics using the full spectrum fitting code, STARLIGHT \citep{cid07} to determine the stellar populations by comparing to model spectra. There, a limited set of 54 models from \citet{wal09} were used to determine the stellar populations. These models include $\alpha$-enhancement based on synthetic spectra from \citet{coe07}. The models are based on a Chabrier IMF \citep{cha03}, include ages from $2-13\,$Gyr, metallicities in the range $-0.5 \le [{\rm Fe/H}] \le 0.2$, and $\alpha$-enhancements of $0.0-0.4$, and are run with the \citet{cal00} dust model. The stacked region spectra are run though STARLIGHT to find the best combination of single stellar populations that results in the lowest $\chi^{2}$ spectrum fit to the data. 

In this paper, the same base set is explored, in addition to including the full set of age ($30\,$Myr$ - 14.0\,$Gyr) and metallicity ($-2.27 \le [{\rm Fe/H}] \le +0.4$) spectra and limiting to only an $\alpha$-enhancement of 0.4 as well as the same full set of $\alpha$-enhanced models from the newer \citet[]{vaz15} models taken from the MILES database \footnote{http://www.iac.es/proyecto/miles/}. Figure~\ref{fig:models} illustrates how the results are sensitive to the number and variety of base spectra to which the observed spectra are matched. As expected, when exploring the results from different sets of base spectra, the individual values of age (and metallicity, not shown) can vary quite a bit. However, in the Limited Walcher, Full Walcher, Limited Vazdekis, and Full Vazdekis runs, core BCG ages of 10$\,$Gyr or more dominate the sample, and younger outskirt ages are common among all runs. High core metallicites and low outskirt metallicities (not shown) are also common throughout the sample in all test runs.

\subsubsection{Errors and Uncertainties in the Modelling}\label{obserrs}

Results of population synthesis codes are subject to several well-known limitations such as the treatment of AGB stars \citep{mar11} and the age-metallicity degeneracy \citep{wor94}. The results are also highly sensitive to the choice of input parameters such as the initial mass function \citep{con12} as well as choice of base spectra as discussed above and in \citet{coe09}.  Thus, any comparison to other samples should take into consideration these important limits and caveats. On the other hand, comparing the qualitative difference in the spectra of the different regions in each target within this sample should be straightforward. 

To estimate the errors in age, and metallicity from STARLIGHT that are caused by the data quality, the following procedure is used, as discussed in Paper~I: The program is run with the same initial conditions 5 times for each region, for each cluster. Each time, $5-10\%$ of the fibres are randomly removed from the median average stack. This results in errors of less than 10\% in age and 15\% in metallicity. The error on the velocity is 10$\,$km/s and on the velocity dispersion is 20$\,$km/s.  In \citet{cid05}, the impact of $\rm S/N$ of the observed spectrum on the STARLIGHT results were explored. They are able to recover mean stellar age and mean stellar metallicity to 0.1$\,$dex, when the input spectrum's $S/N\ge10$. This serves as another motivation for using the higher $\rm S/N$ stacked region spectra.

The results below are based on STARLIGHT runs using the full set of $\alpha$-enhanced models \citep{vaz15}. The average light-weighted age and metallicity for each region are reported in the figures of Section~\ref{res}.  The metallicities shown throughout represent $\rm[Fe/H] = log_{10}[(Fe/H)_{\rm region} / (Fe/H)_{\odot}] = 0 $ for solar values which corresponds to $\rm Z=0.02$ metal content by mass for the Sun. 

\section{Results} \label{res}

An important step toward identifying and discerning the origin of the BCG and ICL is to compare the stellar populations and kinematics of both, using a large dataset. A smooth transition of properties would be expected if the two components share an origin, whereas dramatic breaks in the stellar populations and kinematics in the galaxy outskirts would be expected if the two are unrelated. In the analysis, single power-law fits are used as a way to distinguish whether or not the properties of the inner and outer regions differ. The calculated slopes and errors for the age, metallicity and velocity dispersion profiles are listed in Table~\ref{tab:FitData}). This is chosen over a more complex fitting function so as to avoid over-fitting a few points with a more complex function. Also, the single power-law fits rely less on the exact definition of 'inner' and 'outer' as compared to simply calculating the difference of the properties of the inner and outer regions. 
 
 \begin{table}
  \centering
  \caption{Powerlaw fits to the data are catalogued. The slope and error on the profiles of log(age), metallicity and log(velocity dispersion) are listed, respectively. The number of bins and range in r ('') are also given.}
\begin{tabular}{lccccc}
\hline \hline
Cluster &  $d\,{\rm log(age)}/dr$  & $dZ/dr$ &  $d\,{\rm log(\sigma)}/dr$  & \# & $r$ \\ 
      \hline
A1668    &  $-$0.44 $\pm$ 0.11 &  $-$0.04 $\pm$  0.05 &   0.12 $\pm$  0.11 & 4 &28\\ 
A1795    &   0.04 $\pm$  0.02 &   0.08 $\pm$  0.05 &   0.24 $\pm$ 0.02 & 3 &14\\ 
A193     &  $-$0.08 $\pm$  0.02 &  $-$0.14 $\pm$ 0.04 &  $-$0.10 $\pm$  0.02 & 4 &29\\ 
A2199    &   0.08 $\pm$  0.08 &  $-$0.26 $\pm$ 0.11 &   0.17 $\pm$ 0.08 & 4 &36\\ 
A2457    &  $-$0.04 $\pm$  0.03 &  $-$0.44 $\pm$ 0.16 &  $-$0.27 $\pm$ 0.03 & 4 &26\\ 
A2622    &  $-$0.01 $\pm$  0.00 &  $-$0.13 $\pm$  0.01 &   0.15 $\pm$ 0.00 & 3 &17\\ 
A2626    &  $-$0.40 $\pm$ 0.25 &  $-$0.03 $\pm$ 0.08 &   0.29 $\pm$ 0.24 & 4 &28\\ 
A2665    &  $-$0.00 $\pm$  0.01 &  $-$0.08 $\pm$ 0.05 &   0.30 $\pm$  0.01 & 3 &13\\ 
A376     &  $-$0.00 $\pm$  0.01 &  $-$0.02 $\pm$ 0.01 &   0.10 $\pm$ 0.01 & 3 &18\\ 
A407     &  $-$0.07 $\pm$  0.07 &  $-$0.21 $\pm$ 0.13 &  $-$0.03 $\pm$ 0.07 & 4 &19\\ 
A75      &   0.00 $\pm$ 0.00 &  $-$0.09 $\pm$ 0.03 &   0.09 $\pm$ 0.00 & 3 &13\\ 
A85      &  $-$0.14 $\pm$ 0.06 &  $-$0.15 $\pm$  0.09 &  $-$0.02 $\pm$ 0.06 & 4 &28\\ 
IIZW108  &  $-$0.10 $\pm$  0.06 &  $-$0.19 $\pm$  0.09 &   0.05 $\pm$  0.06 & 4 &33\\
MKW3     &  $-$0.09 $\pm$  0.03 &  $-$0.28 $\pm$ 0.13 &   0.07 $\pm$  0.03 & 4 &29\\
Z2844    &  $-$0.32 $\pm$ 0.02 &  $-$0.08 $\pm$ 0.02 &  $-$0.19 $\pm$  0.02 & 3 &16\\
Z8338    &   0.17 $\pm$  0.17 &  $-$0.32 $\pm$ 0.14 &  $-$0.06 $\pm$  0.17 & 3 &13\\
A160     &  $-$0.14 $\pm$ 0.13 &  $-$0.10 $\pm$ 0.07 &   0.21 $\pm$  0.13 & 4 &18\\
A2589    &  $-$0.18 $\pm$  0.08 &  $-$0.21 $\pm$0.03 &   0.19 $\pm$  0.08 & 5 &33\\
A2634    &  $-$0.71 $\pm$   N/A &   0.07 $\pm$ N/A &  $-$0.54 $\pm$  N/A & 2 &33\\
A602     &  $-$0.21 $\pm$ 0.06 &  $-$0.18 $\pm$0.11 &   0.16 $\pm$  0.06 & 4 &18\\
A671     &  $-$0.24 $\pm$  0.12 &  $-$0.01 $\pm$ 0.00 &   0.09 $\pm$  0.12 & 5 &33\\
A757     &  $-$0.57 $\pm$ 0.08 &  $-$0.5 $\pm$ 0.06 &  $-$0.10 $\pm$  0.08 & 4 & 18\\
UGC03957 &  $-$0.13 $\pm$ 0.10 &  $-$0.33 $\pm$  0.06 &   0.23 $\pm$  0.10 & 5 &33\\
\hline
\label{tab:FitData}
\end{tabular}
 \end{table}
 
The stellar populations of regions within the BCG and ICL are explored, as are the dynamics of the BCG, ICL and host cluster. Unless in actively star-forming cool core clusters, all BCGs adjacent to their host cluster's X-ray peak have cores dominated by very old high-metallicity stars, and have negative radial gradients of average age and metallicity. The average gradient of velocity dispersion ($\sigma$) profile is positive, with the value of $\rm \sigma_{ICL}$ climbing toward the value of $\sigma_{\rm cl}$ in many systems.

 \begin{figure*}
\includegraphics[width=6.6in]{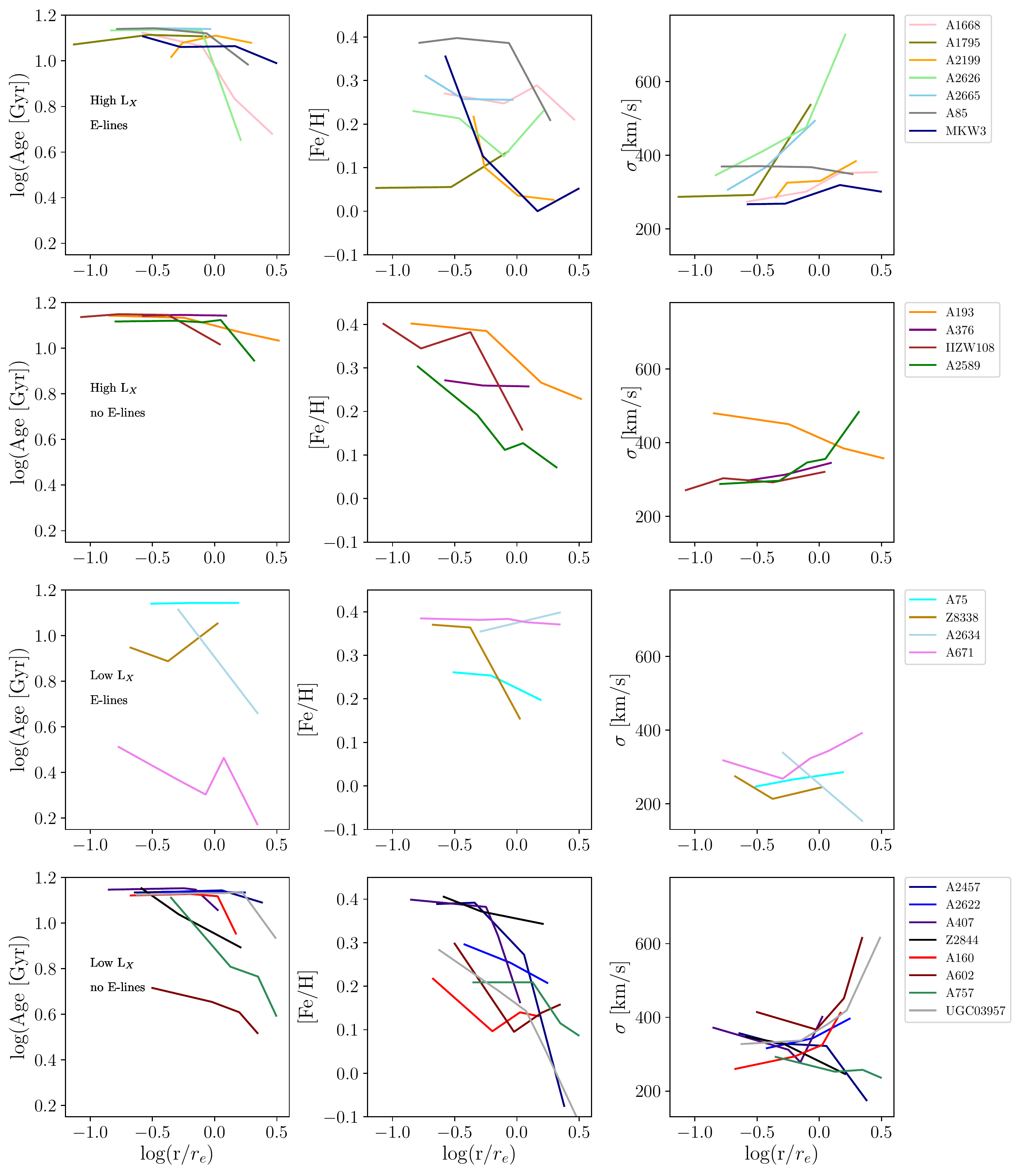}
\caption{{\it Profiles of age, [Fe/H] and velocity dispersion across the SparsePak FOV.} Left panels: The average light weighted age for each region is plotted as a function of the distance from the BCG core, log(r/$r_{e}$). Ages generally remain high in the BCG core, and fall past one effective radius. centre panels: The average light weighted [Fe/H] is plotted with log(r/$r_{e}$). Negative metallicity gradients are common within the sample.  Right panels: The stellar velocity dispersion for each region is plotted with  log(r/$r_{e}$). The velocity dispersion gradients can be described as flat, falling or rising. The sample is separated into galaxies whose host clusters have $\rm L{_X}>0.6\times10^{44}$erg/s (top), and those in clusters with $\rm L{_X}<0.6\times10^{44}$erg/s (bottom). Only regions with $\rm S/N\ge$10.0 are included.}
\label{ageLx}\label{ZLx}\label{vLx}\label{sigLx}
\end{figure*}

\subsection{Stellar Populations: Age and Metallicity Gradients}

Figure~\ref{ageLx} shows the best-fitting luminosity-weighted age, metallicity ([Fe/H]) and $\sigma$ from stellar population synthesis for all regions with a $\rm S/N\ge10$, in each galaxy, as a function of distance from the BCG core.  To better view the details for each cluster, the sample is separated by X-ray luminosity of the parent cluster with high values of $\rm L_{X}$  on the top two rows, and low-$\rm L_{X}$ on the bottom two rows. Negative age and metallicity gradients are common within the sample.

\subsubsection{Negative age gradients}

The left panels of Figure~\ref{ageLx} reveal that the majority of BCG cores are composed of a very old stellar population resulting in an average BCG core age of $\rm13.3\pm2.8\,$Gyr. All of the BCGs with core ages $<10\,$Gyr are in low-$\rm L_{X}$ clusters ($\rm L_{X}<$0.6$\times$10$^{44}\,$erg/s), however, there are only a small number of these younger BCGs, such that the average value of core age for the low- and high-$\rm L_{X}$ samples are consistent within errors ($\rm13.6\pm1.1\,Gyr$ compared to $\rm13.3\pm3.7\,Gyr$). 

\begin{figure*}
\includegraphics[width=6.6in]{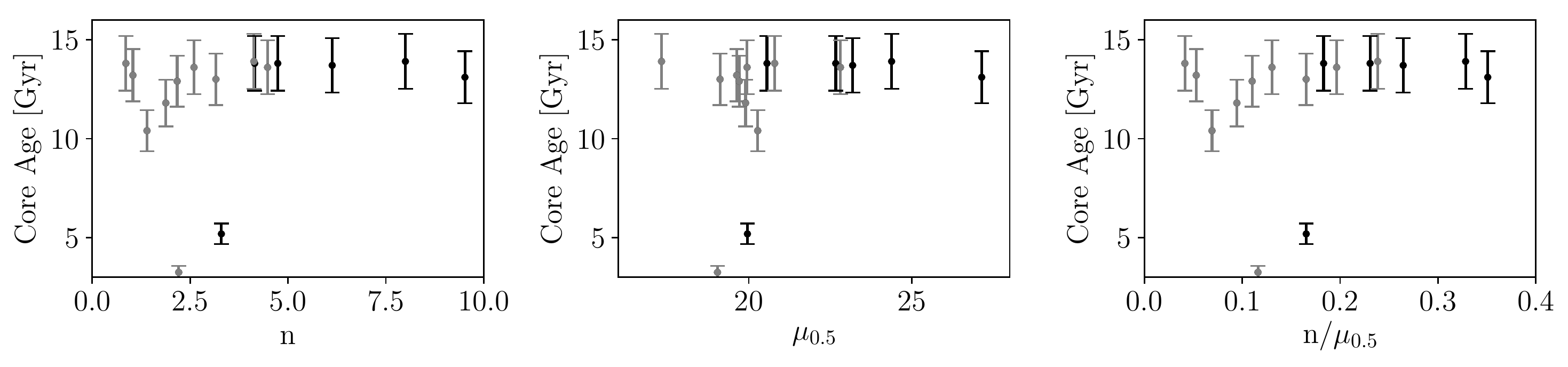}
\caption{{\it The core age is plotted against parameters from Sersic profile fits}.  From left to right, the central Sersic index, surface brightness within $0.5\,  r_{e}$, and ratio of the two are shown for 17 of the 23 BCGs in this sample that are also in the \citet{don11} study. The Sersic fits are taken from the  \citet{don11} paper and errors are as quoted in the text. Grey points are galaxies best fit with a double Sersic profile, and black points are best fit with a single Sersic profile. There is no evidence in this sample that the two BCGs with particularly young core ages (Abell~671 and Abell~602) exhibit extremes in their luminosity profiles.}
\label{sbage}
\end{figure*}

Figure~\ref{sbage} shows the core age of the BCG as a function of the inner Sersic profile and central surface brightness for 17 BCGs in our sample that were studied in \citet{don11}. They calculate best-fitting values using photometric data in the R-band with the KPNO $2.1\, m$ and CTIO $1.5\, m$ telescopes.   There is no evidence in this sample that the two BCGs with particularly young core ages (Abell~671 and Abell~602) exhibit extremes in their luminosity profiles. Rather, for the majority of this dataset, all core ages are old, regardless of central surface brightness or Sersic index.

\begin{figure*}
\includegraphics[width=6.9in]{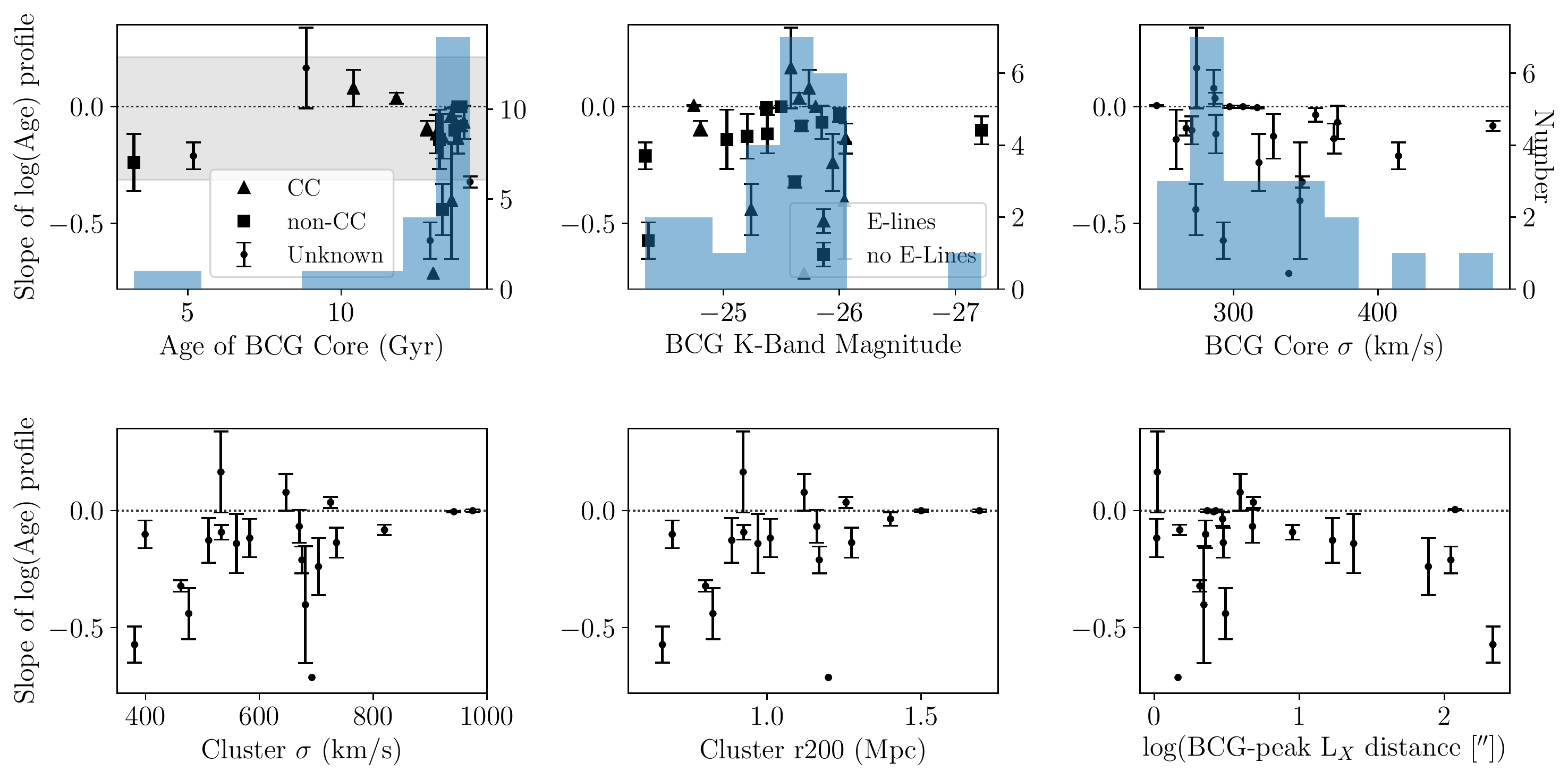}
\caption{ {\it The slope of the gradient in age is plotted against properties of the BCG and of the host cluster.} TOP: Properties of the BCG are explored. Left: The average age of the BCG core region is shown on the $x$-axis. BCGs in CC clusters are shown as triangles, and those in known non-CC clusters are shown as squares. The histogram guides the eye to the most common core ages. The grey region is $1\sigma$ around the average value of age gradient slope and the dotted line shows the zero-slope value. Errors are from Table~\ref{tab:FitData}. The most negative age gradients are found in the BCGs with the oldest core ages.  Centre:  The absolute K-band BCG magnitude is shown along the $x$-axis. BCGs with emission lines are shown as triangles, and those without are shown as squares. The histogram shows the distribution of BCG K-band magnitudes. Many of the bright BCGs with negative age gradients are also line-emitting galaxies. Right:  The velocity dispersion of the BCG core region is plotted.  BOTTOM: Properties of the host cluster are explored.  Left: The cluster velocity dispersion is shown on the $x$-axis. Centre:  The cluster $\rm r_{200}$ is shown on the $x$-axis. The steepest negative age gradients are found in small clusters with low $\sigma_{\rm cl}$. Right: Distance from the BCG to the peak of the cluster X-ray luminosity is shown on the $x$-axis. Sources with large offsets between the two peaks often have very steep age gradients. }
\label{ageCFENN}
\end{figure*}

The generally high values of core age are further illustrated in the histogram of the first panel of Figure~\ref{ageCFENN}.  Separating the galaxies in cool core from those in non-cool core systems has an insignificant effect on core age ($\rm13.4\pm1.1\,Gyr, 13.4\pm5.0\,Gyr$, respectively), and separating on presence of emission lines results in consistently old average core ages, to within errors ($\rm12.9\pm3.9\,Gyr, 13.7\pm2.6\,Gyr$, respectively). 

For 20/23 galaxies, the luminosity-weighted age remains above $\rm10\,Gyr$ within one effective radius, with 2/3 of the young cores existing in low-$\rm L_{X}$ clusters and 2/3 having emission lines. 

Overall, the ages continue to decrease to the outskirts and ICL (Figure~\ref{ageLx}, left panels). This decrease is parameterized with a linear fit to the logarithmic data at each region, for each galaxy. The majority of the gradients are negative, with an average slope of $\rm-0.14$ ($\rm-0.10$ and $\rm-0.25$, for the high and low-$\rm L_{X}$ samples, respectively). Figure~\ref{ageCFENN}, top-left panel shows that most systems have very old core populations. The slopes in age for these systems are either flat or decreasing. The four galaxies with increasing slopes all show emission lines (Figure~\ref{ageCFENN}, top-middle panel), and none are in known non-cool core clusters (Figure~\ref{ageCFENN}, top-left panel).

The gradient in age is plotted with several other properties of the BCG and cluster (Figure~\ref{ageCFENN}). This exercise finds that those BCGs with the steepest negative age gradients (slope$~<-0.2$) are found in small ($r_{200}<1.0\,$Mpc) clusters (bottom-middle panel) with low $\sigma_{\rm cl} < 600\,$km/s (bottom-left panel). The physical FOV of SparsePak covers $>27.5\,$kpc, so the physical coverage of these small systems is well beyond the galaxy $r_{e}$. The steep negative slopes can then be explained if a larger portion of the galaxy outskirts (and presumably more ICL) is observed.

\begin{figure}
\includegraphics[width=3.3in]{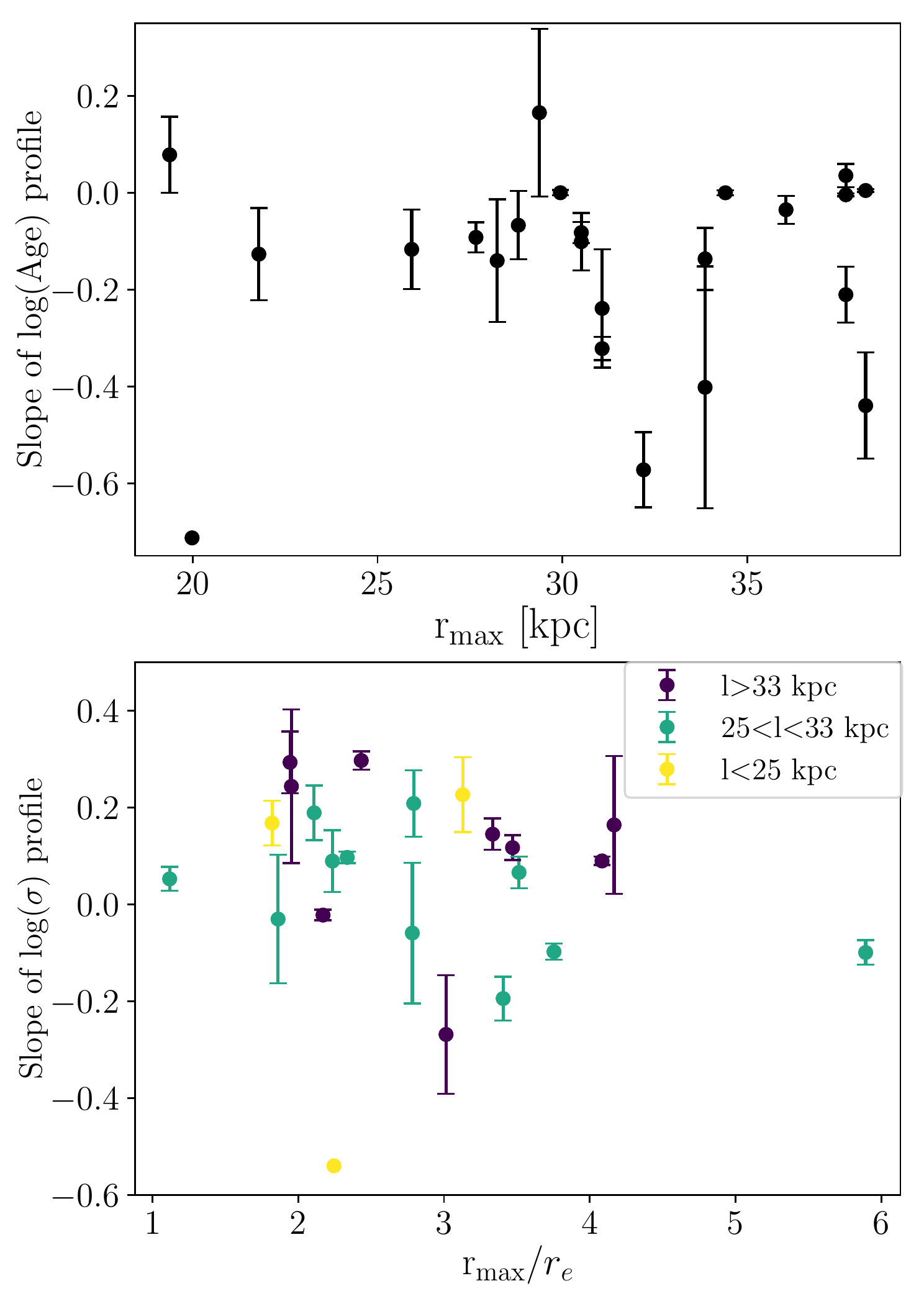}
\caption{{\it The physical extent of the observations}. Left: The slope and error of the age gradient is shown as a function of how far the observations extend beyond the BCG core (r$_{\rm max}$). Right: The slope and error of the velocity dispersion is shown as a function of how many effective radii the observations extend (r$_{\rm max}$/$r_{e}$). The largest FOV are cases plotted with purple, the moderate FOV with blue, and the small FOV cases with green.}
\label{ssfov}\label{ageslope}
\end{figure}

On the other hand, Figure~\ref{ageCFENN} shows that there are six systems that do not have negative gradients. These are mostly line-emitting galaxies co-spatial with the X-ray emission peak (bottom-right panel), in large, massive clusters with $r_{200}>1.1\,$Mpc and $\sigma_{\rm cl} > 650\,$km/s. The galaxies themselves have low core $\sigma$ (5/6 have $\sigma<340\,$km/s) but large K-band magnitudes $M_{K}$ (4/6 have $M_{K} <-25.5$; top-middle panel).  Abell~1795 and A2199 are in cool core clusters and show strong H$\alpha$ and H$\beta$ line emission. Abell~2665 is likewise in a high-$\rm L_{X}$ cluster and has emission lines, though it is not in a cool core cluster. A75 has emission lines. A1795 and A2199 have the highest positive age gradients, also have youngest average core ages, given they are in high mass clusters.  Optical emission lines and new central star formation are often \citep{cra99,edg02,cav08,tre15}, but not always \citep{edw09} associated with a cooling core cluster. If this activity provides a small amount of bright stars, it could make an otherwise negative gradient age gradient flat or positive. Circumnuclear discs \citep{tre15} and dust lanes of order $\sim$1$\,$kpc \citep{lai03} have been seen in BCGs. The SparsePak core fibre covers between 2-6$\,$kpc for these BCGs. Bright circumnuclear emission from a star forming ring may regulate the velocities and decrease the core velocity dispersion. In this scenario, one would expect that for systems with positive age gradients, their low core $\sigma$ would introduce a positive $\sigma$ gradient. Indeed, 4/6 of BCGs with positive or flat age gradients also have positive $\sigma$ gradients. Within the sample as a whole, there are 3 other BCGs in CC clusters that have emission lines - but those systems have negative age gradients. For each of these, the max FOV is over $27.5\,$kpc, so any starburst populations within the core are more polluted by quiescent regions, and perhaps a younger ICL component has evened things out, consistent with the large step observed in the ICL regions of MKW3 and A85. The FOV covered by the most extreme positive age gradient BCG is only $20\,$kpc, the smallest of the sample. This reaches out to $2\, r_{e}$ for the galaxies, but is likely too small to include a significant ICL component (Figure~\ref{ageslope}).  A2622 has insufficient information to be compared to the other BCGs as it has unknown cool core status.

The cluster redshift and galaxy core metallicity were also explored, but show little correlation with age gradient. 

\subsubsection{Negative metallicity gradients}

The core values of metallicity are strongly supersolar ($[{\rm Fe/H}] \ge 0.2$) for all BCGs except A1795 (Figure~\ref{ageLx}, middle panels). A1795 is a well-studied CC cluster with known star formation and complex morphological and spectral properties.  In particular, the X-ray emission extends in a southern tail \citep{fab01,cra05} and  a long H$\alpha$ emitting tail of gas also extends South from the core \citep{cow83,mcd09}. SparsePak can image this H$\alpha$ emitting tail, and in this analysis, the fibres that overlap with the central and outer regions which include the H$\alpha$ emitting tail have been removed. The core fibre shows strong H$\alpha$ emission.  The low metallicity of the core may indicate a population of stars made from more pristine gas may have been added recently. This is consistent with a CC origin, or the merging of a smaller galaxy, but not with early star formation in high-density peaks associated with most of the core stellar population. 

\begin{figure*}
\includegraphics[width=6.9in]{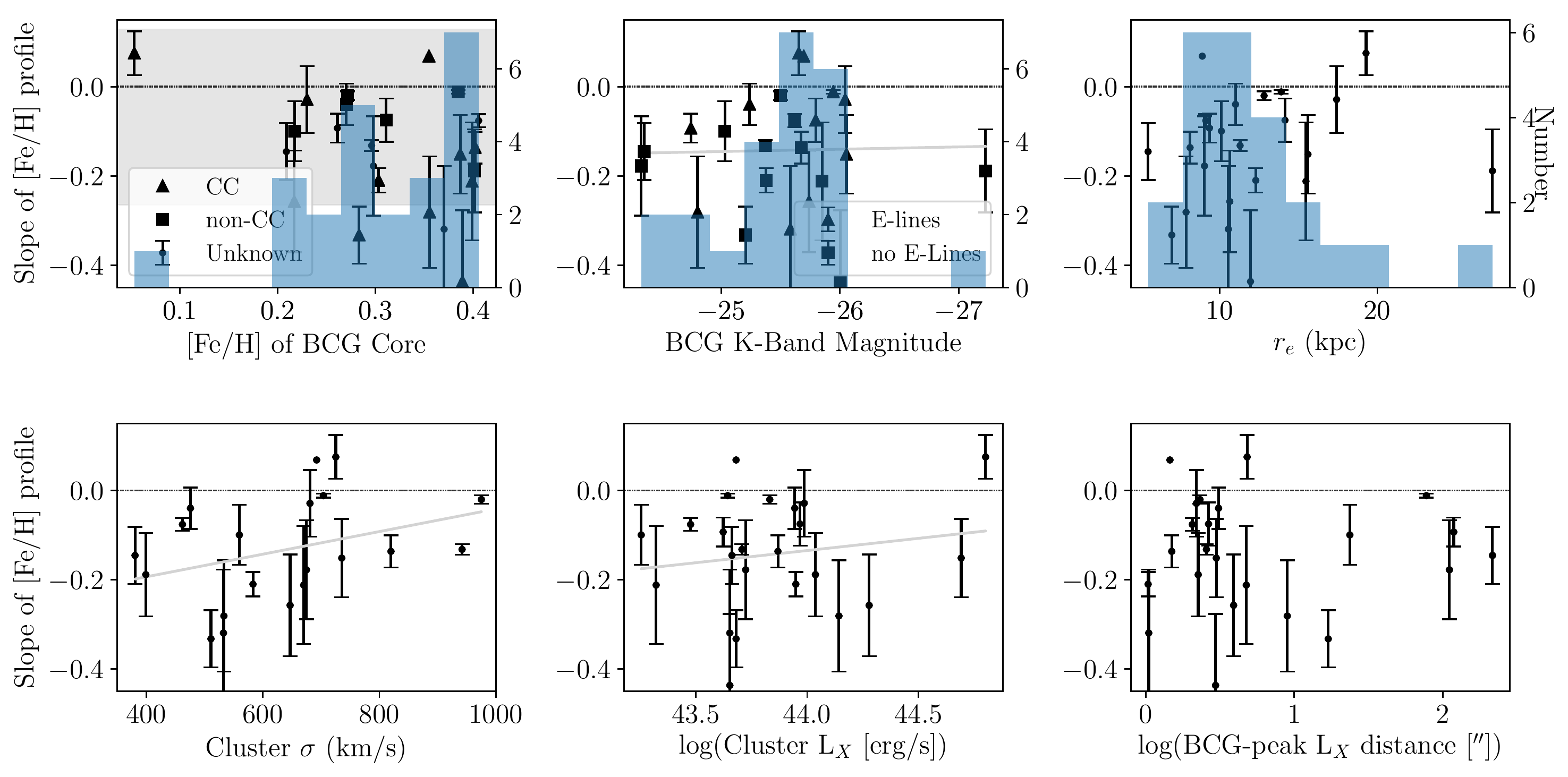}
\caption{{\it The slope of the gradient in metallicity is plotted against properties of the BCG and of the host cluster.} The symbols are as described in Figure~\ref{ageCFENN}. Errors are from Table~\ref{tab:FitData}. TOP: Properties of the BCG are explored. Left: The average metallicity of the BCG core region is shown on the $x$-axis. The grey region is $1\sigma$ around the average value of metallicity gradient slope and the dotted line shows the zero-slope value. The most negative gradients are found in the BCGs with the highest metallicity core ages.  Centre:  The absolute K-band BCG magnitude is shown along the $x$-axis.  Right:  The effective radius of the BCG is plotted. BOTTOM: Properties of the cluster are explored, including: The cluster velocity dispersion (left), the cluster X-ray luminosity (centre), and the distance from the BCG to the peak of the cluster X-ray luminosity (right). }
\label{ZCFENN}
\end{figure*}

There is an overall trend for the metallicities of the galaxies to decrease from the supersolar values found in the core, to solar toward the ICL (Figure~\ref{ZLx}, middle panels). However, in contrast with the decease in ages, a negative gradient tends to start before $1\, r_{e}$.  There are only two BCGs (A1795, A2634) with strong positive metallicity gradients, and they are again in CC systems (Figure~\ref{ZCFENN}, top-left panel) and line emitting galaxies (Figure~\ref{ZCFENN}, top-middle panel). A1795 has a particularly low value of central metallicity, where the new star formation could be coming from a reservoir of relatively pristine gas. All other systems have flat or falling metallicity slopes. 

The steepest negative metallicity gradients are in clusters with low-$\rm L_{X}$  and $\sigma_{\rm cl} < 700\,$km/s (Figure~\ref{ZCFENN}, bottom-left panel).There is little displacement of the BCG inside the cluster (bottom-right panel). The galaxies themselves are massive (Figure~\ref{ZCFENN}, top-middle panel) with small $ r_{e}$.  These are dense galaxies in low-$\rm L_{X}$ clusters, where new star formation does not occur, but a lot of time has passed for the metals to fall toward the gravitational potential and where no recent large-scale cluster mergers have upset the metallicity gradient. 

\begin{figure*}
\includegraphics[width=6.9in]{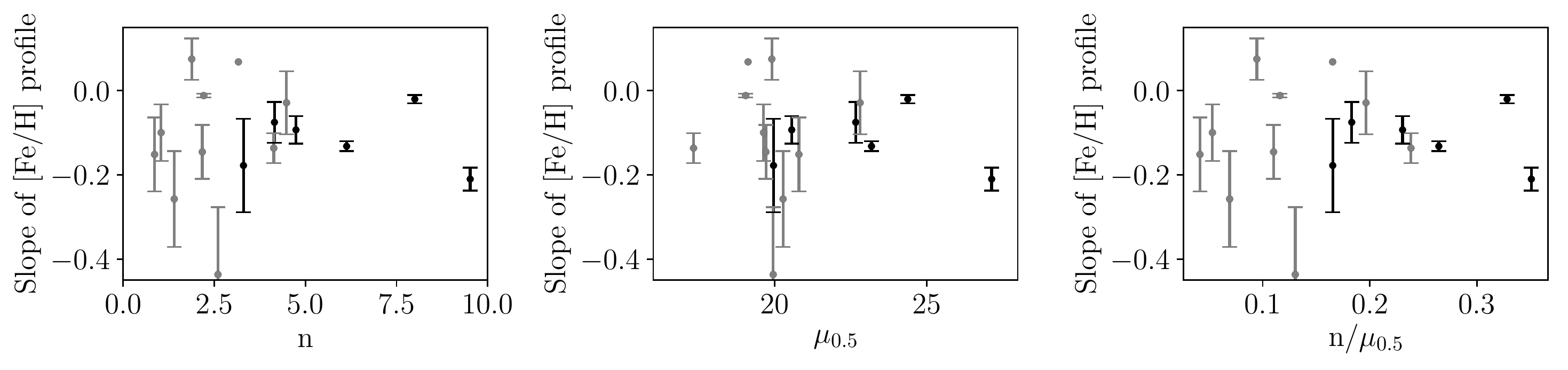}
\caption{{\it The slope and error of the metallicity gradient is plotted against parameters from Sersic profile fits}.  From left to right, the central Sersic index, surface brightness within $0.5\,  r_{e}$, and ratio of the two are shown for 17 of the 23 BCGs in this sample that are also in the \citet{don11} study. colours are the same as Figure~\ref{sbage}. }
\label{sbfeh}
\end{figure*}

Figure~\ref{sbfeh} shows the slope of the $[{\rm Fe/H}]$ profile against the inner Sersic profile and central surface brightness for 17 BCGs in our sample that were studied in \citet{don11}. We find no evidence in this dataset for a correlation between either the surface brightness slopes, parameterized by the Sersic index, n, or the central surface brightness with the gradient in metallicity. 

\begin{figure*}
\includegraphics[width=6.9in]{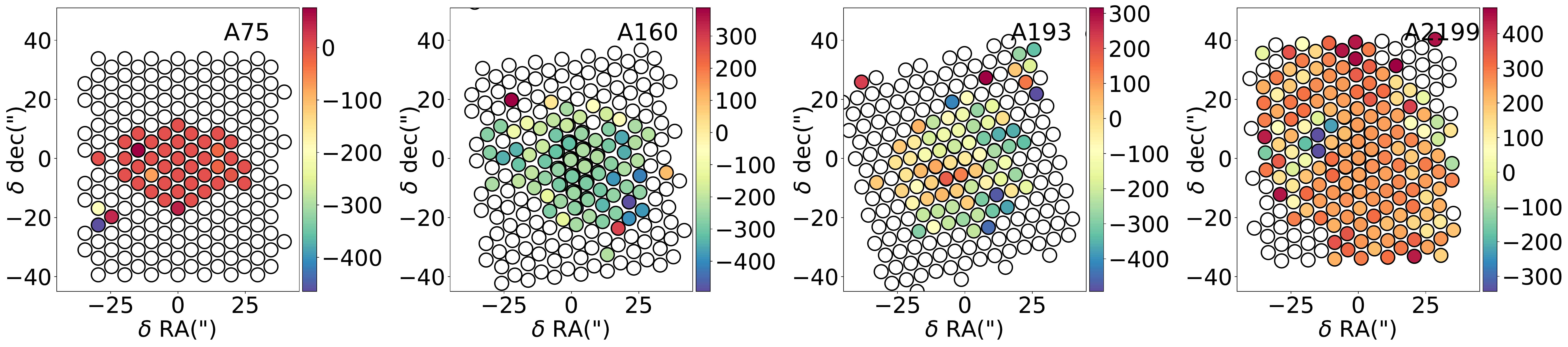}
\caption{{\it The velocity maps for a sample of the dataset.}  From left to right, A75, A160, A193, and A2199 are shown. These examples highlight systems with no measured rotation, rotations $\sim\pm 50 \, $km/s,  $\sim\pm 200 \,$km/s, and with multiple cores, respectively. Only fibres with $\rm S/N > 5$ are included.}
\label{fig:vmaps} 
\end{figure*}

\subsection{BCG and Cluster Dynamics}

Three dynamical measures are presented in this paper.  The first is maps of the peculiar velocity with respect to the cluster recessional velocity which are presented in Figure~\ref{fig:vmaps}. The second is the velocity dispersion within each BCG region, which is presented in Figure~\ref{sigLx}, above. A third measure attempts to describe how dynamically disturbed or relaxed the cluster may be, and is parameterized by calculating the distance between the BCG peak to the peak of the X-ray luminosity.  

\subsubsection{No large scale rotations in BCG sample}

The majority of BCGs are dispersion dominated, only A160, A602, A757, A2199, A2634, MKW3 and Z8338 show regions where the average velocity is at or near the value of $\sigma$. Furthermore, any large scale rotations measured from concentric rings around the BCG core, would be diluted. Thus, only the $\rm S/N>5$ spectra throughout the inner parts of the BCG are examined. These maps (Figure~\ref{fig:vmaps}) provide no evidence for rotation speeds greater than the minimum region velocity dispersion, within the inner $20^{\prime\prime}$ at the given resolution; therefore no correction for rotation is applied to the other dynamical properties studied. Jumps in the velocity are explained by the presence of nearby neighbours and multiple cores. For example, in A2199 a high $\rm S/N$ secondary core is moving with respect to the rest of the BCG at $\sim 500 \, $km/s (Figure~\ref{fig:vmaps}). Evidence for rotation speeds greater than $\sim100\,$km/s are seen only in six clusters (A193, A757, A2622, A2626,  IIZW108, and MKW3s). All of these have multiple cores or nearby galaxies in projection which could account for the velocity differences (Figure~\ref{fig:vmaps}). \citet{new13} likewise found little evidence for rotation in a set of seven BCGs. 

\subsubsection{Positive velocity dispersion profiles rise to cluster value}

\begin{figure*}
\includegraphics[width=6.6in]{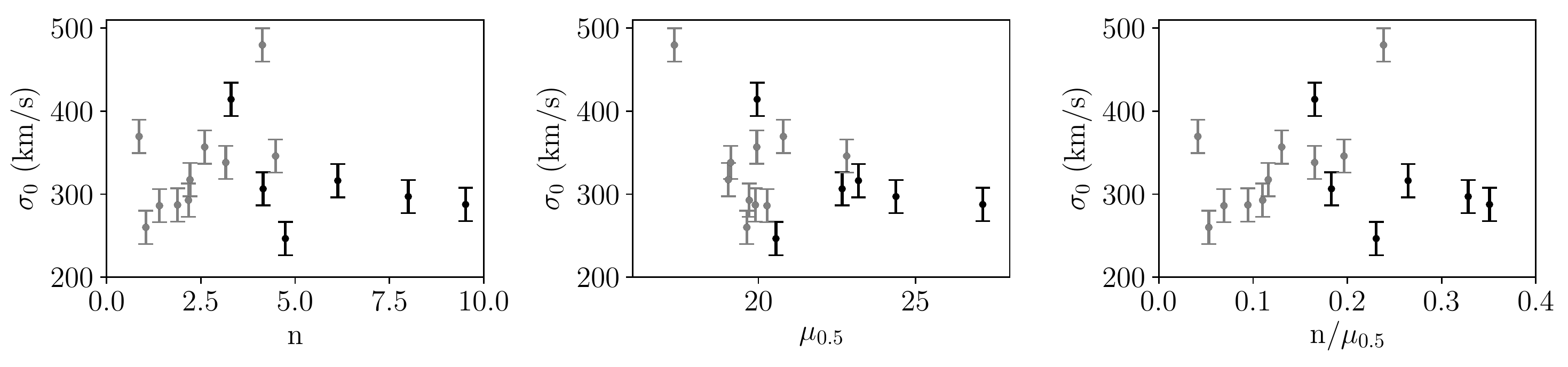}
\caption{{\it The core velocity dispersion, $\sigma_{0}$ is plotted against parameters from Sersic profile fits}.  From left to right, the central Sersic index and surface brightness within $0.5\,  r_{e}$ is shown for 17 of the 23 BCGs in this sample that are also in the \citet{don11} study.  colours are the same as Figure~\ref{sbage}.}
\label{sbcsig}
\end{figure*}

Figure~\ref{sbcsig} shows the central velocity dispersion as a function of Sersic parameters for the 17 BCGs studied in \citet{don11}. Values for most BCGs are $\sim 300 \, $km/s. There is no evidence in this sample that systems with very flat cores (low n), or low central surface brightnesses have either higher or lower central velocity dispersions than other BCGs.

\begin{figure*}
\includegraphics[width=6.6in]{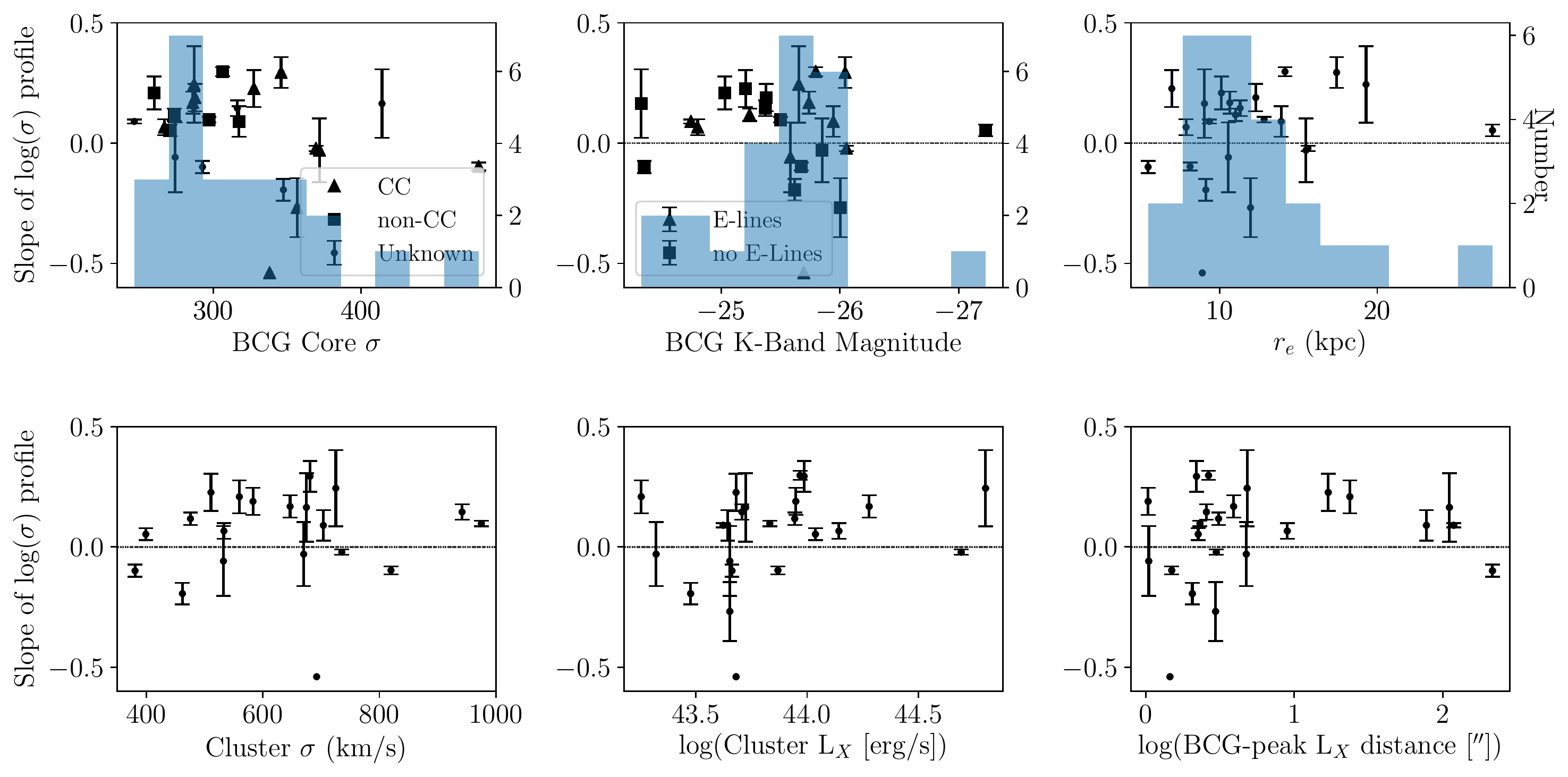}
\caption{{\it The slope of the velocity dispersion gradient is plotted against several properties of the BCG and its host cluster}. The symbols are as described in Figure~\ref{ageCFENN}. Errors are from Table~\ref{tab:FitData}.  TOP: Properties of the BCG are explored. Left: The average velocity dispersion of the BCG core region is plotted on the $x$-axis. The grey region is $1\sigma$ around the average value of the slope of the velocity dispersion and the dotted line shows the zero-slope value. Centre: The absolute K-band BCG magnitude is shown along the $x$-axis. Right: The effective radius measured in the K-band is plotted.  BOTTOM: Properties of the cluster are explored, including: The cluster velocity dispersion (left), the cluster X-ray luminosity 
(centre), and the distance from the BCG to the peak of the cluster X-ray luminosity (right). }
\label{sigCFENN}
\end{figure*}

\begin{figure*}
\includegraphics[width=6.6in]{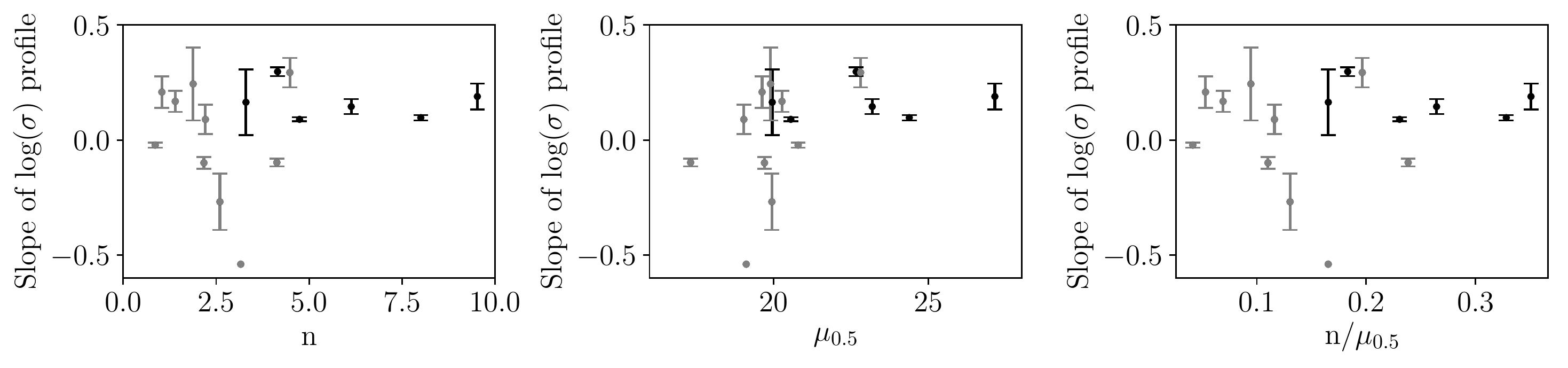}
\caption{{\it The slope and error of the velocity dispersion gradient is plotted against parameters from Sersic profile fits}.  From left to right, the central Sersic index, surface brightness within $0.5\,  r_{e}$, and ratio of the two are shown for 17 of the 23 BCGs in this sample that are also in the \citet{don11} study. The Sersic fits are taken from the  \citet{don11} paper. colours are the same as Figure~\ref{sbage}.}
\label{sbsig}
\end{figure*}

Most of the BCGs (17/23) have rising or flat velocity dispersion profiles, as can be seen in Figure~\ref{sigCFENN}  (see also Figure~\ref{sigLx}). The value of the slope is determined by the velocity dispersion of the ICL, as the core value is fairly constant throughout the subsample. Figure~\ref{sbsig} shows that a variety of Sersic parameters are seen throughout the sample. However, those with negative velocity dispersion slopes tend to have fainter central surface brightnesses and flatter cores.

\begin{figure}
\includegraphics[width=3.1in]{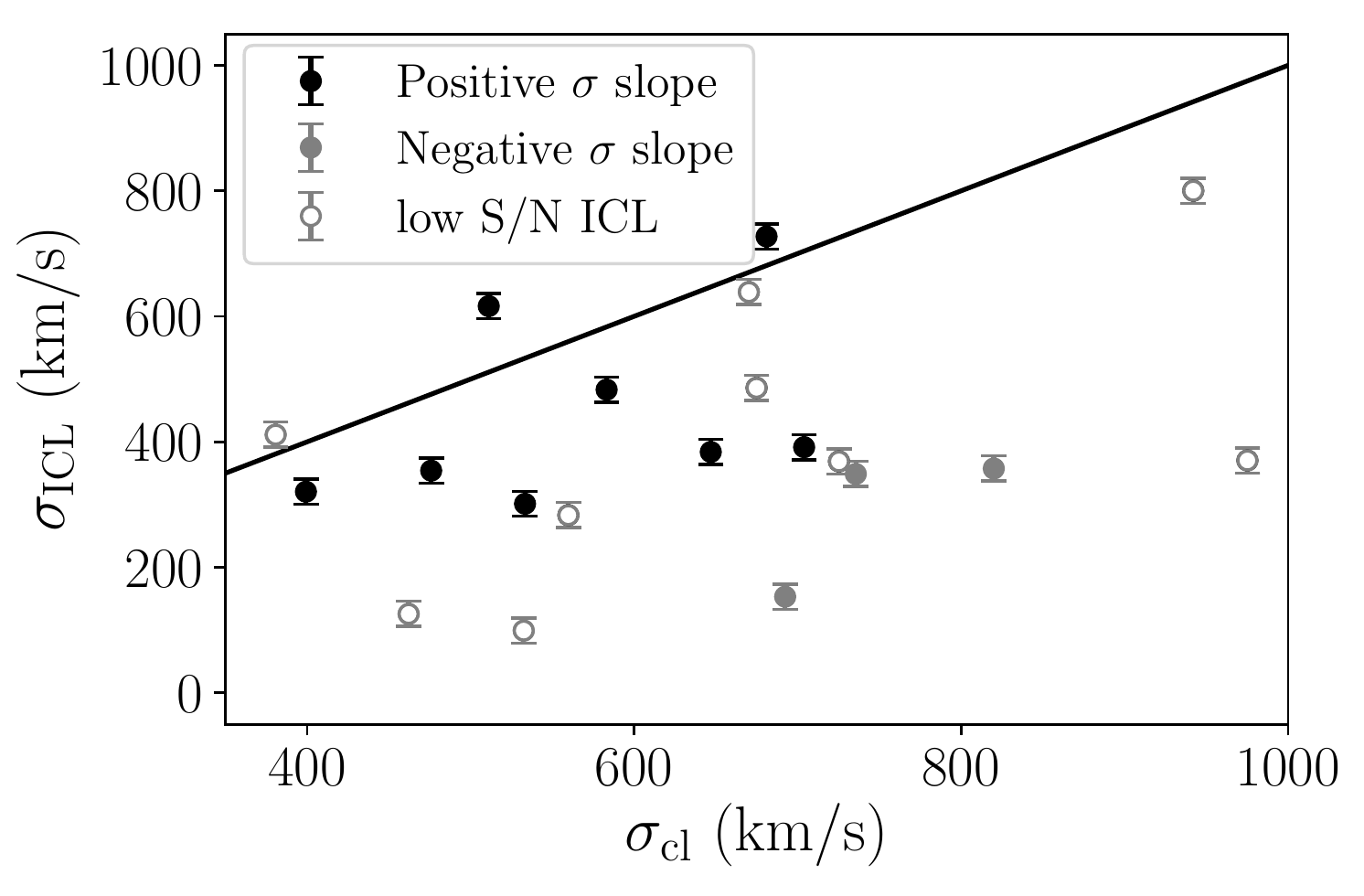}
\caption{{\it ICL vs. Cluster velocity dispersion.} BCGs with positive velocity dispersion gradients (black symbols) show that the $\rm \sigma_{ICL}$  approaches the value of the cluster velocity dispersion, whereas those with negative velocity dispersion gradients do not (grey symbols). BCGs with $\rm S/N_{ICL} < 10.0$ are shown as unfilled circles. The error bars are the 20$\,$km/s errors quoted in Section~\ref{obserrs}. A one-to-one correspondence is illustrated with a black line.}
\label{sigicl}
\end{figure}

For the BCGs with positive velocity dispersion slopes, the value of the $\rm \sigma_{ICL}$ rises toward that of the cluster as a whole, as can be seen in Figure~\ref{sigicl}. Assuming equilibrium, positive slopes are expected if the ICL follows the cluster potential rather than that of the galaxy. For a galaxy inside a more massive host cluster, one would expect to start to see rising velocity dispersions if the ICL stars follow the cluster potential, forming a population dynamically distinct from the BCG. The rising velocity dispersions are seen in all types of galaxies and host clusters (Figure~\ref{sigCFENN}). Cluster scaling relations would suggest that those with higher $\rm L_{X}$  are more massive clusters with deeper potential wells. There are only two high-$\rm L_{X}$  clusters that do not have rising $\sigma$. A85 is pretty close to flat. A193 has clear multiple cores which may be a signal of a recent upset in dynamical equilibrium. This illustrates the difficulty in disentangling the physical processes causing a variation in velocity dispersion. Projected companion galaxies are the subject of a forthcoming paper.

As for the low-$\rm L_{X}$  clusters, A407 has a flat $\sigma$ profile. This cluster is believed to be currently forming its BCG \citep{sch82,mac92,bij17} with 9 bright red galaxies occupying the geometric centre. There are two X-ray peaks on either side of the proto-BCG. It is unlikely in dynamic equilibrium. Z8338, A2457, AZ2844, A2634 have falling velocity dispersion profiles. Most of the spread in the velocity dispersion happens beyond $\rm 1\,r_{e}$.

For the six BCGs with falling velocity dispersions, most (4/6) are in low-$\rm L_{X}$ clusters, only A85 has $ r_{e} >15\,$kpc, and all are bright with $M_{K}<-25.3$. Further, none of them are in known non-cool core clusters, they all have high central metallicities ($[{\rm Fe/H}] \ge 0.35$), and 4/6 have high central ages ($>13.5\,$Gyr). Thus, the subset of BCGs with negative velocity dispersion gradients are massive galaxies in low X-ray luminosity clusters. This is expected if the galaxy mass profile is dominating the stellar motions, further discussed in Section~\ref{fallsigdisc}.  As they are in cool core clusters, these systems have had enough time without being disturbed by cluster-scale mergers to have come to equilibrium.

One potential criticism of a fixed FOV observational study is that the falling profiles may not be observed out to the physical scale of the ICL. To verify the fixed width FOV is not an issue for this result, the physical size of the BCGs and the FOV in kpc for the falling sample was compared to that of the BCGs with positive $\sigma$ slopes. The negative gradient BCGs are within the range of the positive gradient BCGs in terms of size ($\rm 5\,kpc<$$r_{e}<15\,$kpc), and how far the observations reach past the cluster core. For example, both sets cover at least out to $20\,$kpc from the core. Only one of the six negative $\sigma$ gradient BCGs is observed with an extent $l$ less than 25$\,$kpc on a side.  Together, this means that the field of view covers out to $2-3\, r_{e}$ for the negative gradient sample and $1-6\,  r_{e}$ for the full sample (Figure~\ref{ssfov}).

Figure~\ref{sigicl} shows that for the BCGs with flat and falling velocity dispersion slopes, the $\rm \sigma_{ICL}$ value stays low, hovering around the value of~$\sim350\,$km/s. The figure also shows that there are some clusters with large values of velocity dispersion, where the ICL region does not appear to follow the potential of the host cluster, even though they lie on the X-ray peak (A376, A193, A85, A1795, A2634. All of these except A1795 have flat or falling $\sigma$ slopes. One (A2634) has a smaller physical FOV, and so the observations may not reach the ICL. A2634, A1795, A85 are all massive, bright ($M_{K}<-25.5$) galaxies, so perhaps the host galaxy gravitational potential is dominant. Inspection of the velocity maps shows that A193 may have some rotation, so $\sigma$ may not be a good mass indicator in this case. A376 remains an outlier.  

\subsubsection{Cluster Environment} \label{bcgXsep}

Flat velocity dispersion profiles are found throughout the range of observed cluster velocity dispersions (Figure~\ref{sigCFENN}). There is no strong correlation with age, $\rm L_{X}$, or $\sigma$ slope and the distance between the BCG and the cluster centroid. However, most BCGs that are offset from the X-ray peak, have slightly negative metallicity gradients and moderately rising $\sigma$  profiles, consistent with that of the parent population. 

\begin{figure}
\includegraphics[width=3.1in]{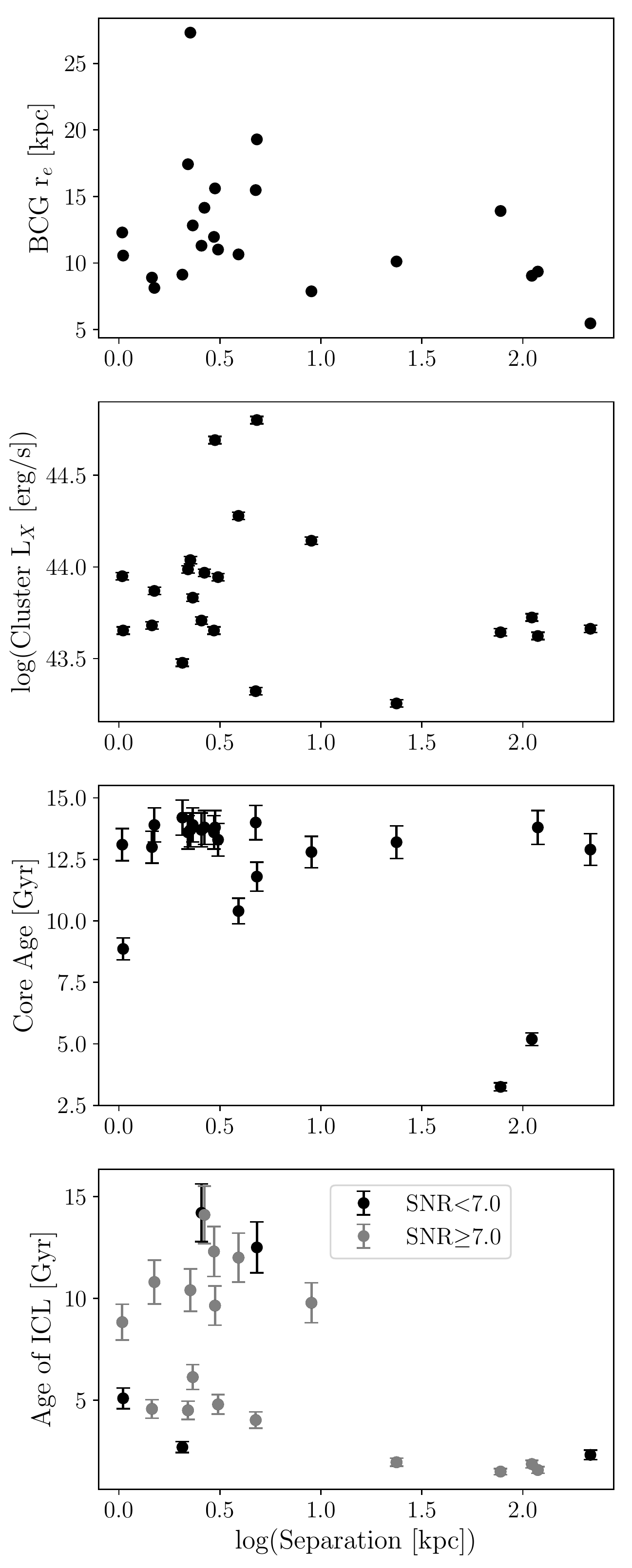}
\caption{{\it BCG and cluster properties with distance from X-ray peak.}From top to bottom: the BCG effective radius, cluster X-ray luminosity, BCG core age and ICL age are plotted as a function of the separation between the cluster X-ray peak and BCG core. The errors are as quoted in the text. The BCGs with large separations are smaller in size, found in less X-ray luminous clusters, and the ICL ages are younger. The grey points are those where the S/N of the ICL region spectrum is $< 7.$}
\label{seprekpc}
\end{figure}

Figure \ref{seprekpc} shows the separation of the BCG from the X-ray luminosity peak against several properties of the BCG and host cluster, including BCG size ($ r_{e}$), cluster X-ray luminosity, BCG core age, and ICL age. The few BCGs with large separations from the X-ray luminosity peak are smaller in size, are found in lower X-ray luminosity clusters, and have younger ICL age. They have a median effective radius of $9.2\,$kpc, compared with $11.6\,$kpc for BCGs close to the cluster centre. If the formation of BCG outskirts includes a component from minor merging, one would expect the observed result: that the large BCGs were more centrally located, where the galaxy density is highest and the rate of gravitational interactions is higher.  Furthermore, minor merging is thought to cause an increase in BCG size, without significantly increasing the mass.  

For the BCGs with large separations from the X-ray peak, the outermost region (labelled here as the ICL,) is likely more heavily weighted by the galaxy's own outskirts, as the ICL is usually more concentrated in the core of the cluster. Thus, the especially young ages being measured, are likely those of the outskirts of these galaxies, which may be new arrivals into the cluster.

\section{Discussion}\label{disc}
 
An analysis of the stellar populations and dynamics for the largest sample BCGs observed with integral field spectroscopy beyond $\rm 1\,$$r_{e}$, are presented above. A discussion of the ages, metallicities and velocity dispersions of these 23 BCGs in X-ray luminous clusters follows. 

\subsection{The ICL is younger than the BCG}
\subsubsection{Age of the BCG core}

This study finds that the BCG cores are very old, with 70\% of the population having average core ages of $13-14\,$Gyr, and 87\% having core ages above $10\,$Gyr, out to $\rm 1\,$$r_{e}$, as has been seen in previous studies  \citep[]{pos95,bro07b,bil08,lou09,coc10b,ben15}. Old stars in massive galaxies are expected in the downsizing scenario \citep{tre05}, where the largest galaxies are formed first. The semi-analytic models of \citet{del07} predict that most stars of today's BCGs formed by $z=3$ ($11.5\,$Gyr). This picture is supported by these data. 
 
The \citet{oli15} study finds that 6/9 of their BCGs have intermediate age cores ($\sim 7\,$Gyr). Their population is at a slightly higher mean redshift of $z\sim 0.1$ and they define their central region as less than $0.2 \,$$r_{e}$, which encompasses the core and central regions of the present sample. Even considering a 1.3$\,$Gyr evolutionary correction and the core and central regions together,  only 4/23 of the present sample have intermediate core ages, a much smaller fraction.  One unknown in the \citet{oli15} sample is whether the systems are in high or low-$\rm L_{X}$ clusters. For low-$\rm L_{X}$ clusters in this study, the fraction of intermediate ages is higher (3/12). Furthermore, the youngest BCG cores are mostly line-emitting galaxies in CC clusters, as previously seen by \citet{lou09} and \citet{bil08}. Perhaps the \citet{oli15} sample is dominated by cool cores clusters. 

\citet{lou12} and \citet{oli15} measure flat or slightly rising age gradients. This is also consistent with the present study. When constraining the SparsePak data to within $\rm 1\,$$r_{e}$ most of the age gradients are flat; they do not tend to decrease until beyond $\rm 1\,$$r_{e}$. For example, A376 is a common galaxy to both this study and \citet{lou12}. While its age stays constant to $\rm 1\,$$r_{e}$, the picture is quite different beyond $\rm 1\,$$r_{e}$.
  
\begin{figure}
\includegraphics[width=3.4in]{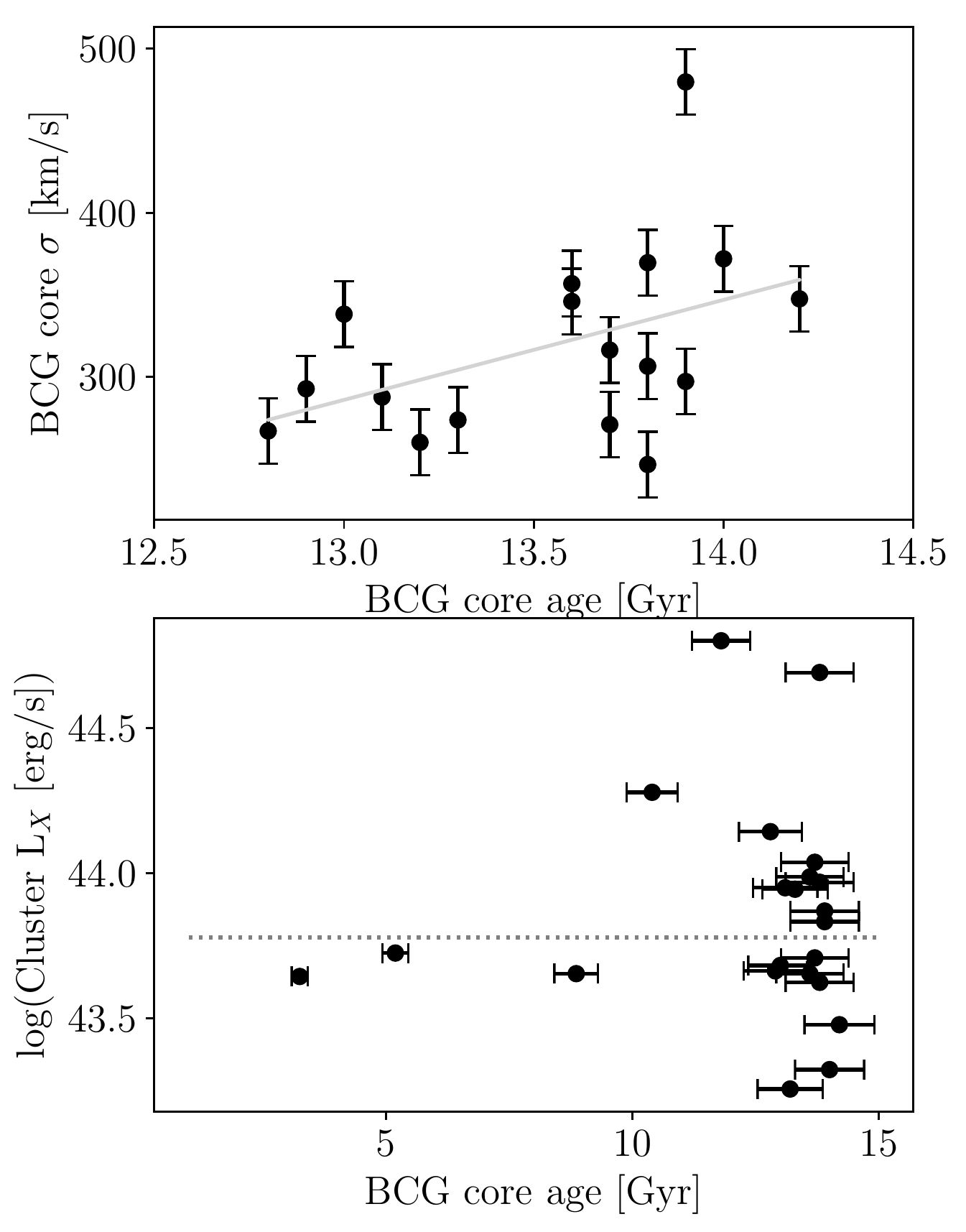}
\caption{{\it BCG and cluster mass with BCG age.} Top - The BCG core velocity dispersion as a function of BCG core age. Only BCGs with core ages greater than 12$\,$Gyr are shown. The errors are as quoted in the text. A weak trend (slope$\rm = 59\,km/s/Gyr$, std err $\rm = 30\,km/s/Gyr$ ) is measured. Bottom - The cluster X-ray luminosity as a function of the BCG core age shows no trends. }
\label{age0gal}
\end{figure}
  
\subsubsection{BCG age beyond $\rm 1\,$$r_{e}$}
  
This is the first study to present a large sample of spectroscopically derived stellar populations of BCGs to beyond $\rm 1\,$$r_{e}$. In this sample of X-ray selected BCGs, negative age gradients are common: 6/23 show strong negative age gradients and 11/23 show weak negative age gradients. They likewise take hold sharply beyond $\rm 1\,$$r_{e}$, consistent with being a separate component from the BCG core. Exquisite IFU measurements of the BCG in Hydra~I found negative age gradients beyond $\rm 1\,$$r_{e}$ \citep{bar16}. The present study shows this is a common occurrence for  local X-ray bright BCGs.

The three BCGs studied in \citet{bro07b} were all found to have uniformly old populations out to $3\,  r_{e}$. This is not at odds with the present findings, as six of the BCGs in the present sample show flat and positive age gradients beyond $\rm 1\,$$r_{e}$. The three positive age gradients of this sample are associated with central emission lines and CC clusters. The \citet{bro07b} study also found a correlation in age with galaxy mass and with cluster $\rm L_{X}$. Considering only BCGs with central ages above $12\,$Gyr, Figure~\ref{age0gal} (top panel) also presents a trend in central age with galaxy mass, in this case indicated by central velocity dispersion. But, the trend does not exist over the entire sample. The core age of BCGs is higher in the high-$\rm L_{X}$ sample, but this is only because all of the core ages below $10\,$Gyr belong to the low-$\rm L_{X}$ subsample (Figure~\ref{age0gal}, bottom panel). 

\subsubsection{Age of the ICL}

The average age of the ICL in this study is $9.2\pm3.5\,$Gyr, compared to $13.3\pm2.8\,$Gyr for the BCG core. This is in line with colour-based ages of the photometric studies of BCGs using HST data. \citet{mon18} find negative colour gradients beyond 10$\,$kpc from the galaxy core with age differentials of $2-4\,$Gyr at $40\,$kpc. This is in great agreement with the present study. The photometrically determined ages continue to drop, beyond what the SparsePak FOV covers, out to $120\,$kpc from the BCG core, where the ICL ages ultimately become $2-6\,$Gyr less than those of the cluster galaxies. 

The stars of the ICL formed more recently than those that make up the BCG, and according to downsizing, this would happen in lower-mass galaxy haloes. While the average value of the ICL is young compared to its host BCG, the ICL ages measured here vary widely from very old ($12.3\,$Gyr; A2457) to very young ($1.5\,$Gyr; A671), implying the build up of the ICL is an ongoing process. An ICL dominated by old stars reflects that of the host BCG observed by \citet{mel12}. While an old ICL is not a common case in the present sample, it is occasionally seen (for example, in A2457).

Note that the young ICL component is not associated with CCs as is the case for BCG cores.

\subsection{BCG formation is a two-phase process}

Negative metallicity gradients have been interpreted as supporting the two-phase build up, with low metallicity outskirts signaling stars that formed in shallower potentials, subsequently joining the main galaxy though mergers. The change in the metallicity from the BCG core to outskirts is discussed in this context.

\subsubsection{BCG metallicities}

In this sample, 91\% of the core metallicities are above 0.2. Such high central metallicities are ubiquitous throughout the quoted studies above, as expected for massive galaxies. In contrast to the ages, which remain high out to $\rm 1\, $$r_{e}$, the metallicity gradients are set up well within $\rm 1\,$$r_{e}$. Eleven of the sample have steep negative gradients $<-0.15$, ten have moderately negative gradients, and two have positive gradients. Smaller samples of the metallicity gradients in BCGs have shown shallow \citep{oli15} and steep metallicity gradients \citep{bro07b}. The \citet{lou12} sample is also dominated by negative metallicity gradients. The two samples share a similar range in core $\sigma$ and metallicity gradients. A correlation between the two is found in \citet{lou12} and for the 3 BCGs of \citet{bro07b}, but it is not recovered here. \citet{lou12} also measure a weak correlation with the K-band magnitude of the galaxy (which was not seen by \citet{bro07b}). K-band magnitude, cluster $\rm L_{X}$, and  cluster velocity dispersion were explored in Figure~\ref{ZCFENN}. This dataset shows no evidence for a correlation with K-band magnitude ($\rm slope = -0.007, std$ $\rm err = 0.040$). \citet{bro07b} find a correlation with $\rm L_{X}$. This data is unable to support or refute the correlation with $\rm L_{X}$ as the trend is unreliable ($\rm slope=0.025, std$ $\rm err=0.018$), being dominated by one exceptionally high luminosity point. However, a weak trend is measured with the cluster velocity dispersion ($\rm slope=0.00024, std$ $\rm err=0.00016$). 

Cosmological simulations predict the formation of steep radial metallicity gradients ($-0.4$ to $-0.3$) when galaxies are formed through dissipative collapse \citep{kob04,hir15}, and a flattening of the gradients governed by merging history \citep{hop09}. This naturally accounts for a diversity in metallicity gradients, and in this picture, the galaxies with more recent merging histories would have shallower metallicity ~gradients. All the galaxies with large separations from the X-ray peak have shallow metallicity gradients. \citet{lou12} find a correlation with separation between the X-ray peak and the BCG which is not recovered in this study. A major merger between clusters would bring a large number of galaxies within the vicinity of the cluster core, increasing the possibility for merging and encounters with the BCG. The steepest metallicity gradients are in clusters with low-$\rm L_{X}$ and low $\rm \sigma_{cl}$. These may have just formed -- allowing for less time to encounter a major merger. They are also in cool cores and close to the X-ray peak, evidence that no cluster-scale merging has happened recently. The large scatter in metallicity gradients reflects the diversity in the individual cluster's merging history.

\subsubsection{Metallicity of the ICL}

The ICL stars formed more recently, and from gas that has a range of metallicity.  The metallicity of our ICL regions range from $-0.1$ to $0.4$. A broad range in metallicity could be explained from gas or stars that join the ICL through gravitational interactions with the BCG, providing lower mass galaxies with lower metallicity gas or stars. But this would have happened after the core and central parts of the BCG were formed. This older, more metal-rich BCG, compared to ICL is also found in cosmological simulations analysed by \citet{cui14}.

The average $[{\rm Fe/H}]$~is $0.18\pm0.16$, which is much lower than the average $[{\rm Fe/H}]$~of the core ($0.30\pm0.09$). Within the context of hierarchical building, the mass-metallicity relationship \citep{gal05} can help identify progenitors of the ICL. At 40$\,$kpc, it suggests they are massive galaxies with $\rm M_{*}\sim 10^{11}\, \rm M_{\odot}$. Following the measured average gradient for local BCGs, at $100\,$kpc from the core, the metallicity would be about $-1$ and the corresponding progenitor mass would be $\sim \rm 10^{9-10}\,M_{\odot}$. 

The photometric studies of ICL from \citet{mon18} also find negative metallicity gradients in 4/6 of the Frontier Fields. The metallicites measured at 40$\,$kpc ($\rm -0.4 < [Fe/H]~< 0.2$) are in line with this survey's results, and continue to much lower values at $100\,$kpc from the BCG core ($\rm -0.5 < Fe/H] < -0.3$), as expected given the strong radial gradients.  At $100\,$kpc, the ICL metallicity is similar to the Milky Way outskirts, and plausible ICL progenitors would be $\rm L_{*}$ galaxies. Coma and Hydra~I haloes show low metallicity, but old stars dominate the ICL, such that stripping from massive early type galaxies are likely sources \citep{coc10b,bar16}.  Note that merging today with an $\rm L_{*}$ galaxy would constitute a minor merger with the BCG, normally considered to be a mass ratio of 4:1 or greater.

\subsubsection{Population synthesis - beyond the average age and metallicity}

\begin{figure}
\includegraphics[width=3.4in]{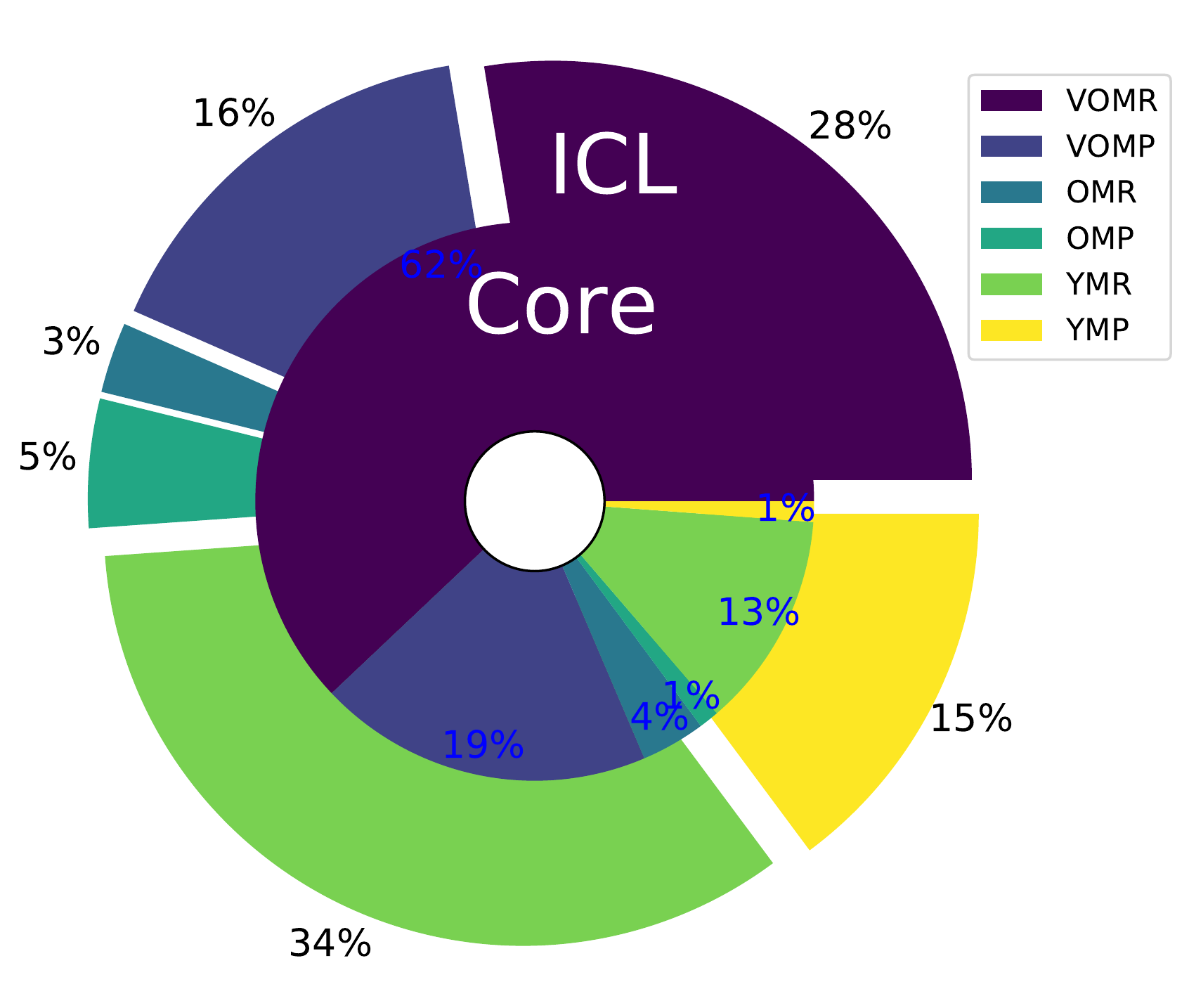}
\caption{{\it Breakdown of core and ICL population.} The fraction of each stellar population required for the best-fitting match is calculated from stellar population synthesis. The core populations are given in the centre, and that for the ICL in the outer pie chart. Populations are grouped into Very Old metal-rich (VOMR), Very Old metal-poor (VOMP), Old and metal-poor (OMP), Old and metal-rich (OMR), Young and metal-rich (YMR) and Young and metal-poor (YMP). Rich and poor metallicity populations are split at a value of $\rm [Fe/H] = 0.2$. Very old, old, and young populations have $\rm ages >12\,Gyr, ages$ between 8 and 12$\rm\,Gyr$, and $\rm ages <8\,Gyr$, respectively.}
\label{piepopsave}
\end{figure}

The combined spectra average out any inhomogeneous elements to the ICL populations at physical scales smaller than the analysed regions. Thus, although the physical location of separate populations cannot be garnered in this analysis, stellar population synthesis does return the fractions of the single stellar populations required for best-fitting. Figure~\ref{piepopsave} shows the percentage of individual populations that make up the average populations. The make-up of the average core and ICL is very different, although all populations are found in both regions. The ICL has $\sim30\%$ very old metal-rich (VOMR) populations, which is the dominant core population, and suggesting some shared formation history with the BCG.

This information from population synthesis can be used to test some of the predictions from cosmological models. In the models of \citet{mur07}, 50\% of the ICL stars come from the progenitors that form the BCG, and another 25\% from the progenitors that form the other large red galaxies in the cluster. Even beyond 100$\,$kpc, the ICL stars are mostly sourced from what are today the other massive cluster galaxies.  Figure~\ref{piepopsave} does show that $\sim 55\%$~ of the populations of the observed BCG cores are shared with the ICL, however, the much younger ages and lower metallicities of the ICL suggest a source other than cluster LRGs. On the other hand, the models of \citet{mar12} and \citet{con18} find that the majority of the ICL is from $\rm L_{*}$ galaxies, consistent with the observation that the average metallicity of the ICL is lower than that of the BCG. Figure~\ref{piepopsave} shows the ICL contains 36\% metal-poor (MP) populations, compared to 21\% metal-poor (MP) populations in the BCG core. 

Note that the comparisons above are first-order estimates. Extracting information for simulations that can then be compared to observations is not trivial. The basic approach consists of applying to simulation results the same methods that are used with actual observations. Several studies have addressed this issue, and shown that results can be quite sensitive to the observational parameters used \citep{dol10,tan18}.

\subsection{Some local BCGs are still assembling}

\citet[]{jim13} find a diverse merging history for four BCGs. However, with this larger sample of uniformly massive BCGs in massive clusters, we find little evidence for rotation greater than $100\,$km/s within $\rm 1\,$$r_{e}$, consistent with other recent studies of larger samples \citep{new13,vea18,lou18}, supporting the idea that any mass added to BCGs in the current epoch generally builds through minor mergers. 

In their large photometric study of BCG+ICL, \citet{klu19} find the brighter systems have a larger fraction of faint light, are in richer, more massive and larger clusters, suggesting the ICL grows with cluster growth. On the other hand, \citet{bro02} found that BCGs in high-$\rm L_{X}$ clusters ($\rm L_{X} > 1.9\times10^{44}erg/s$; A1795 and A85 in this sample) have uniform K-band magnitudes from redshifts 0.02 to 0.8, consistent with not experiencing significant stellar mass evolution since $z\sim 1$. In low-$\rm L_{X}$ clusters, the K-band magnitude of the BCGs are scattered, suggesting they are still increasing in mass. The high-$\rm L_{X}$ BCGs are larger and have fainter mean surface brightnesses than the low-$\rm L_{X}$ counterparts \citep{bro05}, signaling they have undergone more accretion and assembled their mass at $z>1$. The BCGs in low-$\rm L_{X}$ clusters are still assembling. Two possible cases are present in this sample: IIZW108 and A407. In these two clusters, the large number of similar luminosity and colour galaxies appear to still be undergoing BCG formation by major mergers, as must have been common during the formation of galaxies in the early universe \citep{con18}.

\subsection{The ICL is a distinct kinematic component, separate from the BCG}

The kinematics of planetary nebulae in galaxies in the Virgo cluster have been studied by many authors \citep{lon13,lon15,har18,lon18}. These studies have found that in M87 and M49 the planetary nebulae of the ICL have different velocity and spatial distributions than those in the BCG halo. Additionally, the planetary nebulae in the BCG are found to be older and more metal-rich than those in the ICL. This is in line with the different stellar populations we have discussed for the BCG and ICL stars. Below, we discuss kinematic evidence in our dataset that also suggests the BCG and ICL are different components. 

\subsubsection{Rising velocity dispersion profiles are common among BCGs, when observed beyond $\rm 1\,$$r_{e}$}

Two-thirds of the BCGs of this X-ray selected sample exhibit positive velocity dispersion gradients. 

On the other hand, this is uncommon in the general population of elliptical galaxies \citep{car95,ger01,pad04} where $\rm \Delta Log (\sigma) / \Delta Log (r) = -0.05 \pm0.05$ \citep{gon93}. None the less, rising velocity dispersions have been observed in BCGs (often cD galaxies) since  \citet{dre79} observed the BCG in A2029. A2029, however, remained essentially an outlier in the following decades, when data was not observed much beyond $\rm 1\,$$r_{e}$   
\citep{ton83,fis95a,lou08} or samples remained small \citep{bro07b}. Several of the galaxies observed in this paper had already been observed out to $\rm 1\,$$r_{e}$, and were determined to have falling profiles (A85, A2634, A193, A376, A2199, A2589, A2622). Only A85 and A2634 remain falling beyond $\rm 1\,$$r_{e}$, all other profiles are seen in the right panels of Figure~\ref{sigLx} to rise past this point.

\citet{new13} observe 7 BCGs at $z=0.2-0.3$ to $1-2 r_{e}$ finding all 7 have rising velocity dispersion profiles, and recent spectroscopic surveys of the MASSIVE team \citep{vea17,vea18} and \citet{lou18} have likewise found a significant fraction of rising velocity dispersion profiles for BCGs. The MASSIVE survey is an IFU survey of massive red galaxies that covers a similar extent at a similar spatial resolution as the present study, but only includes a few BCGs. The \citet{lou18} long-slit study observes 35 BCGs at $z=0.05-0.3$ with twice the spatial resolution as the present study, but only out to $15\,$kpc. They find a similar fraction of BCGs with positive velocity dispersion slopes (69\%).

\subsubsection{Signature of the host cluster halo}

One interpretation is that the large velocity dispersions are signaling the large total mass-to-light ratio of the host cluster. If the ICL stars are kinematically connected to the underlying massive cluster potential, rather than to the BCG,  positive velocity dispersion slopes would be expected, where the velocity dispersion of the ICL eventually reaches that of the host cluster. This was seen in \citet{mis11} and \citet{rit11} for Hydra~I using globular clusters, as well as for  A2199 in \citet{ben15} with the diffuse stellar light. There are 7 BCGs in the present sample with ICL region spectra that have $\rm S/N\ge10$, which indeed show a trend of the ICL velocity dispersion increasing to the value of the cluster velocity dispersion, illustrated in Figure~\ref{sigicl}. \citet{ben15}  concluded that the outer diffuse stellar component is recognizable not from the photometry, but rather the kinematics. As in this study, they find A2199 to have old, and slightly metal-rich populations. When looking at the present sample of BCGs as a whole, the ages and metallicities also appear to define the outer halo as a separate component, apart from the BCG.

\subsubsection{Mass of the galaxy}

\begin{figure}
\includegraphics[width=3.4in]{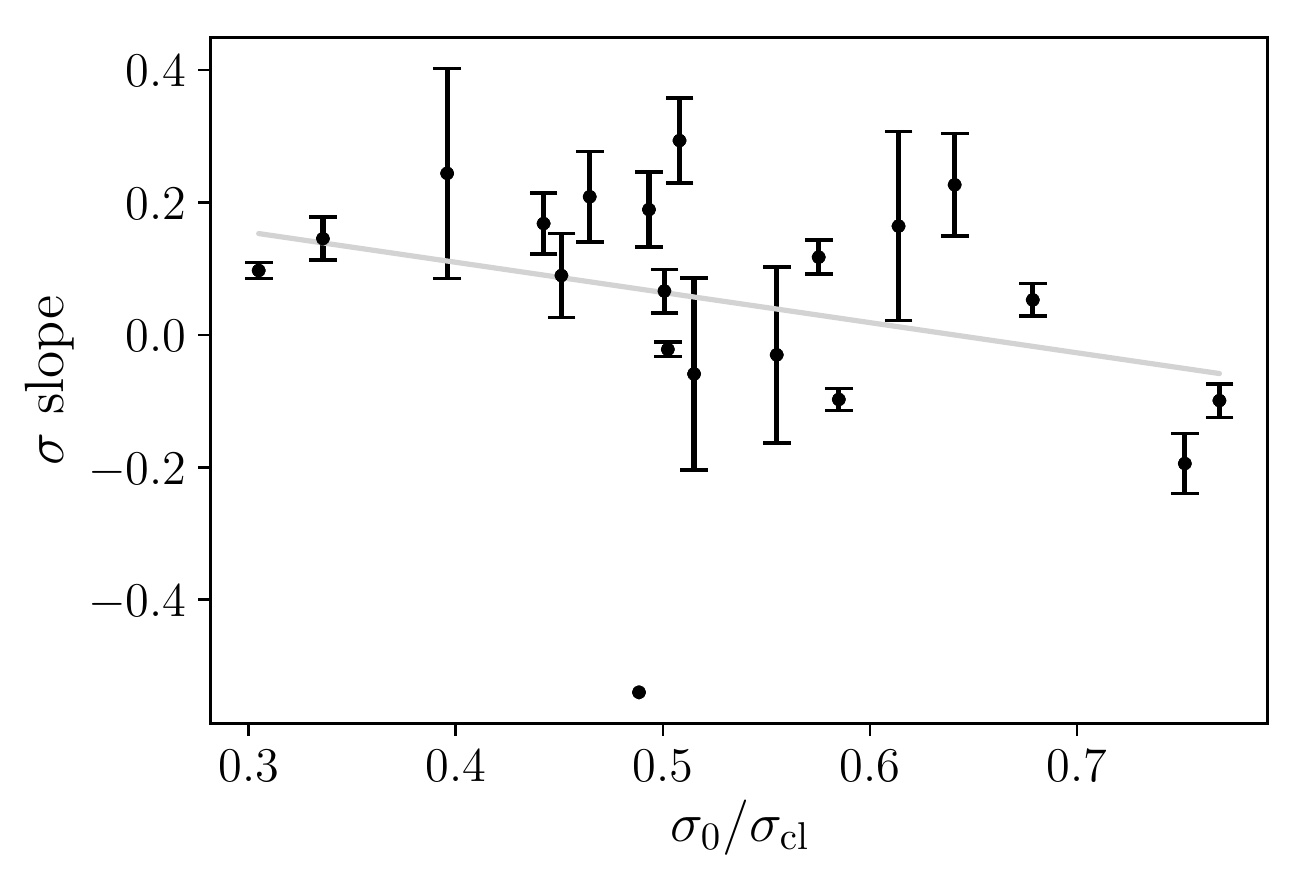}
\caption{{\it Influence of the BCG's gravitational potential}. A trend in the velocity dispersion slope is seen, where its value decreases as the ratio of BCG to cluster velocity dispersion increases. Falling velocity dispersion profiles are only seen when this ratio is $\gtrapprox 0.5$, suggesting the mass of the BCG becomes important. Errors are from Table~\ref{tab:FitData}. }
\label{potwell}
\end{figure}

Complicating this simple picture are the isolated, satellite and brightest group (BGG) galaxies observed by \citet{vea17,vea18} out to $20-40\,$kpc from the galaxy core. All types of galaxies are found to contain some fraction of rising, falling and flat velocity dispersion profiles. Surprisingly, the highest velocity dispersion slopes are the satellite galaxies. It would be interesting to determine whether the velocity dispersion of satellite galaxies rises all the way to the cluster value. \citet{vea18} discuss the possibility that the physical mechanism behind the rising velocity dispersions in central BCGs and satellite galaxies may differ, where subhalo masses and velocity anisotropies may play a role. They find a strong trend with galaxy mass, where their highest mass bin ($\rm -26.7<M_{K}<-26$) has the highest fraction of positive velocity dispersion slopes (8/11). This is in contrast to this sample of only BCGs, where no trend with galaxy mass is observed (either measured by $\rm M_{K}$ or $\sigma_{0}$). Perhaps the mass dependency is coming from the satellite galaxies. Note that the single power-law slopes calculated for the present sample are not exactly comparable to those of \citet{vea17}, who use a broken power-law fit. The more complex fit is possible because their sample is more local, translating into a higher physical spatial resolution. For the BCGs in this paper, the $5\,$kpc break radius defined by \citet{vea17} is generally within the first SparsePak resolution element (ie. core region). Thus, the total slopes are moderated by the inner core and central regions. However, the sign of the velocity dispersion slope is usually independent of the core and central values. The fits in this paper are to single power-laws, in the same way as \citet{lou18}. 

\citet{lou18}, who have a larger baseline of stellar mass than the present sample, likewise observe a correlation with galaxy mass, as measured by central velocity dispersion. No strong trend is measured in the present data, likely because the current sample does not probe the BGG range. None the less, the overall results do overlap with those of \citet{lou18}. Their Figure 3 shows that high-mass BCGs ($\rm 2.4 < Log\, \sigma_{0} < 2.6$) show a wide range of velocity dispersion slopes ($-0.2$ to $0.2$). Over the same mass range, the slopes observed here show values of $-0.4$ to $0.4$.   Furthermore, the highest values of velocity dispersion slope are in galaxies with large K-band magnitude and large values of $\sigma_{0}$.  We note that many of the systems in the present study have different slopes within $15\,$kpc and beyond $15\,$kpc (like the u-shaped profiles of \citet{vea17}). Thus the \citet{lou18} results are best compared with the inner slope results of \citet{vea17} and the core and central regions here. For example, A1795 is shared in both surveys. In \citet{lou18}, $\sigma_{0} = 268\,$km/s, $\sigma$ at $10^{\prime\prime}$ is $350\,$km/s, and the slope is 0.037. In this study, $\sigma_{0} = 287\,$km/s,  $\sigma$ at 5$^{\prime\prime} = 292\,$km/s and $\sigma$ at $15^{\prime\prime} =536\,$km/s. That is, the velocity dispersion slope is flat within the central $15\,$kpc before it rises. 

\subsubsection{The host cluster mass}

ICL from cosmological simulations was characterized by \citet{cui14}. The velocity dispersion was found to rise steeply as a function of the cluster M$_{200}$. \citet{vea18} sample a wide range of environmental density and host halo mass, and also measured a correlation. The present sample shows no correlation, which is not surprising as it is dominated by rich, X-ray bright clusters. 

\subsubsection{BCGs with falling velocity dispersion slopes}\label{fallsigdisc}

What about the BCGs that do not show rising velocity dispersions? Figure~\ref{ssfov} illustrates that this is not because of a small physical field of view, and Figure~\ref{sigCFENN} shows it is not the distance from the X-ray peak. Beyond velocity anisotropies, it could be that the BCG dark matter or stellar core is large, dominating the stellar dynamics beyond.  As an example, A85 has a flat velocity dispersion profile, and is known to have a particularly large core \citep{lop14}. Figure~\ref{potwell} supports the idea of a BCG core occasionally dominating the dynamics. The velocity dispersion slope is negative only when the ratio of BCG to cluster velocity dispersion is  $\gtrapprox 0.5$. It also shows a slight trend where the velocity dispersion slope falls as the ratio of the BCG to cluster mass increases ($\rm slope = -0.46, std err = 0.35;  slope = -0.55, std err = 0.22,$ when the very negative slope point is removed). 

\section{Summary and Conclusions} \label{con}

The main findings and conclusions of the paper are:
\begin{itemize}

\item The stars in the BCG core were the first in the galaxy to form and did so rapidly in deep potential wells long ago. The ages of the BCG cores are overwhelmingly very old with 70\% of the population having average core ages of $13-14\,$Gyr, and 87\% having core ages above $10\,$Gyr, out to $\rm 1\,$$r_{e}$. The highest metallicities are seen throughout $1\, r_{e}$, and the spectra are fit well by $\alpha$-enhanced stellar population models.  This is expected in the downsizing picture, where the largest galaxies are formed from the oldest stars, and it is consistent with the \citet{tof14}, \citet{van16} scenarios where the BCG cores have been quiescent since $z\sim2$. The consistent old ages within $\sim 5-10\,$kpc echo the size of the most massive, compact galaxies observed at high $z$~ \citep{tru06,van08}. 

\item The BCG cores in X-ray selected clusters are similar to each other in population and mass. Similar core stellar populations are seen throughout the sample and most core velocity dispersions are around $300-400\,$km/s. Any emission lines are associated with BCGs adjacent to the X-ray peak of cool core clusters and young stellar populations form a relatively minor contribution.  The stellar populations and velocity dispersion within the core and centre are similar for the galaxies in the sample, suggesting that there has been a long time for populations to mix.

\item The ages and metallicities then decrease with distance from the core, as expected if minor mergers continue to add mass to the BCG outskirts. Isotropic, dry minor mergers are expected to remove most of the initial angular momentum of BCGs, resulting in the slowly rotating galaxies of this local sample. The few BCGs with positive age gradients also host the youngest average core age.  New central star formation, again associated with an X-ray cooling core \citep{cra99,edg02,edw07,bil08}, could provide a small amount of bright stars that would make an otherwise negative gradient age gradient become flat or positive. 

\item The most extended BCGs are more centrally located in the cluster, consistent with a large number of minor mergers increasing the size of the galaxy without much increasing the mass.

\item This assembly of the BCG outskirts happened after core assembly, and the ICL stars formed more recently from gas that has a range of metallicity. This broad range in metallicity could be explained by gravitational encounters with the BCG. High-metallicity material may originate from the BCG core, while lower-mass galaxies could provide the lower-metallicity gas or stars. This would have happened after the core and central parts of the BCG were formed. 
 
\item The ICL stars are part of the cluster, rather than the BCG. Both the kinematics and stellar populations support this scenario. In the majority of cases, the velocity dispersion rises from the BCG core to the outskirts, often reaching the cluster velocity dispersion value. The average age and metallicity of the ICL are both significantly lower than in the BCG. A small subset of BCGs with negative velocity dispersion gradients are observed. This is expected if the galaxy mass profile is dominating the stellar motions. Indeed, these galaxies are massive, but in low X-ray luminosity clusters such that the ratio of BCG to cluster velocity dispersion is $\lessapprox 0.5$.  Many are in cool core clusters. Such BCGs would have had enough time without being disturbed by cluster-scale mergers to have come to equilibrium.

\end{itemize}

The formation of local BCGs is a long process. The stars observed in the core of these systems are consistent with the oldest possible ages in stellar population models. Yet, their outskirts formed over billions of years through the merging and interactions with other galaxies in the cluster. Even today, some large red galaxies in massive clusters are undergoing major mergers, forming what will be even brighter cluster galaxies of tomorrow. In future papers we will quantify the amount of merging currently taking place through a study of the close companions identified here. We will also more closely examine the emission line systems. 

These findings support efforts to obtain spectroscopic observations deep into the ICL so that stellar populations and kinematics together can be determined. As the endpoint of hierarchical merging, brightest cluster galaxies provide strong constraints for cosmological simulations. This paper presents constraints that cosmological evolution models should be able to reproduce: the gradients in metallicity and velocity dispersion for BCGs, and the lower stellar ages past $\sim1 r_{e}$. 

\section*{Acknowledgements}

We thank the anonymous referee for helpful and insightful comments. This study uses data gathered with SparsePak on the WIYN Telescope at Kitt Peak National Observatory, through time generously granted by the Yale University Astronomy department. We would like to thank instrument scientist Jenny Power and telescope operators Krissy Reetz and Dave Summers for their valuable assistance in obtaining observations. We thank Eric Wilcots for securing the final six cluster members through UW Madison, and Malanka Riabokin for performing the observations the evening Willow Edwards was born. Many undergraduate students were involved in this project and their funding sources were from: the Yale STARS and the CT Space Grant Consortium, the Cal Poly BEACON research, an NSF LSAMP grant (number HRD-1302873), as well as Frost Research Scholars supported by The Bill and Linda Frost Fund. This work is supported by the NSF AAG Award Number 1814375. PC acknowledges support from FAPESP project 2018/05392-8 and CNPq 305066/2015-3. HM acknowledges support from NSERC. We would also like to thank Anthony Festa, Vasilije Dobrosavljevic, Tara Abraham, Victoria Beizer, and Emily Lucas for their contributions to the project. This project also makes extensive use of data from several publicly accessible astronomical databases.  Funding for the Sloan Digital Sky Survey IV has been provided by the Alfred P. Sloan Foundation, the U.S. Department of Energy Office of Science, and the Participating Institutions. SDSS-IV acknowledges support and resources from the center for High-Performance Computing at the University of Utah. The Two Micron All Sky Survey,  is a joint project of the University of Massachusetts and the Infrared Processing and Analysis center/California Institute of Technology, funded by the National Aeronautics and Space Administration and the National Science Foundation. The NASA/IPAC Extragalactic Database (NED), which is operated by the Jet Propulsion Laboratory, California Institute of Technology, under contract with the National Aeronautics and Space Administration. Data from the Chandra Data Archive and the Chandra Source Catalog, and software provided by the Chandra X-ray center (CXC) in the application packages CIAO, ChIPS, and Sherpa was also used. The STARLIGHT project is supported by the Brazilian agencies CNPq, CAPES and FAPESP and by the France-Brazil CAPES/Cofecub program.

\bibliography{le1}

\newcommand{\noop}[1]{}
\begin{thebibliography}{133}
\expandafter\ifx\csname natexlab\endcsname\relax\def\natexlab#1{#1}\fi

\bibitem[{{Barai}, {Brito} \& {Martel}(2009){Barai}, {Brito}, \&
  {Martel}}]{bar09}
{Barai} P., {Brito} W., {Martel} H., 2009, Journal of Astrophysics and
  Astronomy, 30, 1

\bibitem[{{Barbosa} {et~al}\mbox{.}(2016){Barbosa}, {Arnaboldi}, {Coccato},
  {Hilker}, {Mendes de Oliveira}, \& {Richtler}}]{bar16}
{Barbosa} C.~E., {Arnaboldi} M., {Coccato} L., {Hilker} M., {Mendes de
  Oliveira} C., {Richtler} T., 2016, A\&A, 589, A139

\bibitem[{{Bender} {et~al}\mbox{.}(2015){Bender}, {Kormendy}, {Cornell}, \&
  {Fisher}}]{ben15}
{Bender} R., {Kormendy} J., {Cornell} M.~E., {Fisher} D.~B., 2015, ApJ, 807, 56

\bibitem[{{Bershady} {et~al}\mbox{.}(2004){Bershady}, {Anderson}, {Harker},
  {Ramsey}, \& {Verheijen}}]{ber04}
{Bershady} M., {Anderson} M., {Harker} J., {Ramsey} L., {Verheijen} M., 2004,
  PASP, 116, 565

\bibitem[{{Bezanson} {et~al}\mbox{.}(2009){Bezanson}, {van Dokkum}, {Tal},
  {Marchesini}, {Kriek}, {Franx}, \& {Coppi}}]{bez09}
{Bezanson} R., {van Dokkum} P.~G., {Tal} T., {Marchesini} D., {Kriek} M.,
  {Franx} M., {Coppi} P., 2009, ApJ, 697, 1290

\bibitem[{{Biju} {et~al}\mbox{.}(2017){Biju}, {Bagchi}, {Ishwara-Chandra},
  {Pandey-Pommier}, {Jacob}, {Patil}, {Kumar}, {Pandge}, {Dabhade}, {Gaikwad},
  {Dhurde}, {Abraham}, {Vivek}, {Mahabal}, \& {Djorgovski}}]{bij17}
{Biju} K.~G. {et~al.}, 2017, MNRAS, 471, 617

\bibitem[{{Bildfell} {et~al}\mbox{.}(2008){Bildfell}, {Hoekstra}, {Babul}, \&
  {Mahdavi}}]{bil08}
{Bildfell} C., {Hoekstra} H., {Babul} A., {Mahdavi} A., 2008, MNRAS, 389, 1637

\bibitem[{{Brough} {et~al}\mbox{.}(2005){Brough}, {Collins}, {Burke}, {Lynam},
  \& {Mann}}]{bro05}
{Brough} S., {Collins} C.~A., {Burke} D.~J., {Lynam} P.~D., {Mann} R.~G., 2005,
  MNRAS, 364, 1354

\bibitem[{{Brough} {et~al}\mbox{.}(2002){Brough}, {Collins}, {Burke}, {Mann},
  \& {Lynam}}]{bro02}
{Brough} S., {Collins} C.~A., {Burke} D.~J., {Mann} R.~G., {Lynam} P.~D., 2002,
  MNRAS, 329, L53

\bibitem[{{Brough} {et~al}\mbox{.}(2007){Brough}, {Proctor}, {Forbes}, {Couch},
  {Collins}, {Burke}, \& {Mann}}]{bro07b}
{Brough} S., {Proctor} R., {Forbes} D.~A., {Couch} W.~J., {Collins} C.~A.,
  {Burke} D.~J., {Mann} R.~G., 2007, MNRAS, 378, 1507

\bibitem[{{Bruzual} \& {Charlot}(2003)}]{bru03}
{Bruzual} G., {Charlot} S., 2003, MNRAS, 344, 1000

\bibitem[{{Calzetti} {et~al}\mbox{.}(2000){Calzetti}, {Armus}, {Bohlin},
  {Kinney}, {Koornneef}, \& {Storchi-Bergmann}}]{cal00}
{Calzetti} D., {Armus} L., {Bohlin} R.~C., {Kinney} A.~L., {Koornneef} J.,
  {Storchi-Bergmann} T., 2000, ApJ, 533, 682

\bibitem[{{Carollo} {et~al}\mbox{.}(1995){Carollo}, {de Zeeuw}, {van der
  Marel}, {Danziger}, \& {Qian}}]{car95}
{Carollo} C.~M., {de Zeeuw} P.~T., {van der Marel} R.~P., {Danziger} I.~J.,
  {Qian} E.~E., 1995, ApJL, 441, L25

\bibitem[{{Cavagnolo} {et~al}\mbox{.}(2008){Cavagnolo}, {Donahue}, {Voit}, \&
  {Sun}}]{cav08}
{Cavagnolo} K.~W., {Donahue} M., {Voit} G.~M., {Sun} M., 2008, ApJL, 683, L107

\bibitem[{{Cavagnolo} {et~al}\mbox{.}(2009){Cavagnolo}, {Donahue}, {Voit}, \&
  {Sun}}]{cav09}
{Cavagnolo} K.~W., {Donahue} M., {Voit} G.~M., {Sun} M., 2009, ApJS, 182, 12

\bibitem[{{Chabrier}(2003)}]{cha03}
{Chabrier} G., 2003, ApJL, 586, L133

\bibitem[{{Chen} {et~al}\mbox{.}(2007){Chen}, {Reiprich}, {B{\"o}hringer},
  {Ikebe}, \& {Zhang}}]{che07}
{Chen} Y., {Reiprich} T.~H., {B{\"o}hringer} H., {Ikebe} Y., {Zhang} Y.~.,
  2007, ArXiv Astrophysics e-prints

\bibitem[{{Cid Fernandes} {et~al}\mbox{.}(2007){Cid Fernandes}, {Asari},
  {Sodr{\'e}}, {Stasi{\'n}ska}, {Mateus}, {Torres-Papaqui}, \&
  {Schoenell}}]{cid07}
{Cid Fernandes} R., {Asari} N.~V., {Sodr{\'e}} L., {Stasi{\'n}ska} G., {Mateus}
  A., {Torres-Papaqui} J.~P., {Schoenell} W., 2007, MNRAS, 375, L16

\bibitem[{{Cid Fernandes} {et~al}\mbox{.}(2005){Cid Fernandes}, {Mateus},
  {Sodr{\'e}}, {Stasi{\'n}ska}, \& {Gomes}}]{cid05}
{Cid Fernandes} R., {Mateus} A., {Sodr{\'e}} L., {Stasi{\'n}ska} G., {Gomes}
  J.~M., 2005, MNRAS, 358, 363

\bibitem[{{Coccato} {et~al}\mbox{.}(2010){Coccato}, {Arnaboldi}, {Gerhard},
  {Freeman}, {Ventimiglia}, \& {Yasuda}}]{coc10a}
{Coccato} L., {Arnaboldi} M., {Gerhard} O., {Freeman} K.~C., {Ventimiglia} G.,
  {Yasuda} N., 2010, A\&A, 519, A95

\bibitem[{{Coccato}, {Gerhard} \& {Arnaboldi}(2010){Coccato}, {Gerhard}, \&
  {Arnaboldi}}]{coc10b}
{Coccato} L., {Gerhard} O., {Arnaboldi} M., 2010, MNRAS, 407, L26

\bibitem[{{Coelho} {et~al}\mbox{.}(2007){Coelho}, {Bruzual}, {Charlot},
  {Weiss}, {Barbuy}, \& {Ferguson}}]{coe07}
{Coelho} P., {Bruzual} G., {Charlot} S., {Weiss} A., {Barbuy} B., {Ferguson}
  J.~W., 2007, MNRAS, 382, 498

\bibitem[{{Coelho}, {Mendes de Oliveira} \& {Cid Fernandes}(2009){Coelho},
  {Mendes de Oliveira}, \& {Cid Fernandes}}]{coe09}
{Coelho} P., {Mendes de Oliveira} C., {Cid Fernandes} R., 2009, \hbox {MNRAS},
  396, 624

\bibitem[{{Conroy} \& {van Dokkum}(2012)}]{con12}
{Conroy} C., {van Dokkum} P.~G., 2012, ApJ, 760, 71

\bibitem[{{Contini} {et~al}\mbox{.}(2014){Contini}, {De Lucia}, {Villalobos},
  \& {Borgani}}]{con14}
{Contini} E., {De Lucia} G., {Villalobos} {\'A}., {Borgani} S., 2014, MNRAS,
  437, 3787

\bibitem[{{Contini}, {Yi} \& {Kang}(2018){Contini}, {Yi}, \& {Kang}}]{con18}
{Contini} E., {Yi} S.~K., {Kang} X., 2018, MNRAS, 479, 932

\bibitem[{{Cooke} {et~al}\mbox{.}(2019){Cooke}, {Kartaltepe}, {Tyler},
  {Darvish}, {Casey}, {Le F{\`e}vre}, {Salvato}, \& {Scoville}}]{coo19}
{Cooke} K.~C., {Kartaltepe} J.~S., {Tyler} K.~D., {Darvish} B., {Casey} C.~M.,
  {Le F{\`e}vre} O., {Salvato} M., {Scoville} N., 2019, ApJ, 881, 150

\bibitem[{{Cowie} {et~al}\mbox{.}(1983){Cowie}, {Hu}, {Jenkins}, \&
  {York}}]{cow83}
{Cowie} L.~L., {Hu} E.~M., {Jenkins} E.~B., {York} D.~G., 1983, ApJ, 272, 29

\bibitem[{{Coziol} {et~al}\mbox{.}(2009){Coziol}, {Andernach}, {Caretta},
  {Alamo-Mart{\'{\i}}nez}, \& {Tago}}]{coz09}
{Coziol} R., {Andernach} H., {Caretta} C.~A., {Alamo-Mart{\'{\i}}nez} K.~A.,
  {Tago} E., 2009, AJ, 137, 4795

\bibitem[{{Crawford} {et~al}\mbox{.}(1999){Crawford}, {Allen}, {Ebeling},
  {Edge}, \& {Fabian}}]{cra99}
{Crawford} C.~S., {Allen} S.~W., {Ebeling} H., {Edge} A.~C., {Fabian} A.~C.,
  1999, MNRAS, 306, 857

\bibitem[{{Crawford}, {Sanders} \& {Fabian}(2005){Crawford}, {Sanders}, \&
  {Fabian}}]{cra05}
{Crawford} C.~S., {Sanders} J.~S., {Fabian} A.~C., 2005, MNRAS, 361, 17

\bibitem[{{Croton} {et~al}\mbox{.}(2006){Croton}, {Springel}, {White}, {De
  Lucia}, {Frenk}, {Gao}, {Jenkins}, {Kauffmann}, {Navarro}, \&
  {Yoshida}}]{cro06}
{Croton} D.~J. {et~al.}, 2006, MNRAS, 365, 11

\bibitem[{{Cui} {et~al}\mbox{.}(2014){Cui}, {Murante}, {Monaco}, {Borgani},
  {Granato}, {Killedar}, {De Lucia}, {Presotto}, \& {Dolag}}]{cui14}
{Cui} W. {et~al.}, 2014, MNRAS, 437, 816

\bibitem[{{De Lucia} \& {Blaizot}(2007)}]{del07}
{De Lucia} G., {Blaizot} J., 2007, MNRAS, 375, 2

\bibitem[{{De Lucia}, {Kauffmann} \& {White}(2004){De Lucia}, {Kauffmann}, \&
  {White}}]{del04}
{De Lucia} G., {Kauffmann} G., {White} S.~D.~M., 2004, MNRAS, 349, 1101

\bibitem[{{De Maio} {et~al}\mbox{.}(2015){De Maio}, {Gonzalez}, {Zabludoff},
  {Zaritsky}, \& {Bradac}}]{dem15}
{De Maio} T., {Gonzalez} A., {Zabludoff} A.~I., {Zaritsky} D., {Bradac} M.,
  2015, MNRAS, 448, 1162

\bibitem[{{Dolag}, {Murante} \& {Borgani}(2010){Dolag}, {Murante}, \&
  {Borgani}}]{dol10}
{Dolag} K., {Murante} G., {Borgani} S., 2010, MNRAS, 405, 1544

\bibitem[{{Donzelli}, {Muriel} \& {Madrid}(2011){Donzelli}, {Muriel}, \&
  {Madrid}}]{don11}
{Donzelli} C.~J., {Muriel} H., {Madrid} J.~P., 2011, ApJS, 195, 15

\bibitem[{{Dressler}(1979)}]{dre79}
{Dressler} A., 1979, ApJ, 231, 659

\bibitem[{{Ebeling} {et~al}\mbox{.}(1998){Ebeling}, {Edge}, {Bohringer},
  {Allen}, {Crawford}, {Fabian}, {Voges}, \& {Huchra}}]{ebe98}
{Ebeling} H., {Edge} A.~C., {Bohringer} H., {Allen} S.~W., {Crawford} C.~S.,
  {Fabian} A.~C., {Voges} W., {Huchra} J.~P., 1998, MNRAS, 301, 881

\bibitem[{{Ebeling} {et~al}\mbox{.}(1996){Ebeling}, {Voges}, {Bohringer},
  {Edge}, {Huchra}, \& {Briel}}]{ebe96}
{Ebeling} H., {Voges} W., {Bohringer} H., {Edge} A.~C., {Huchra} J.~P., {Briel}
  U.~G., 1996, MNRAS, 281, 799

\bibitem[{{Edge}(1991)}]{edg91}
{Edge} A.~C., 1991, MNRAS, 250, 103

\bibitem[{{Edge} {et~al}\mbox{.}(2002){Edge}, {Wilman}, {Johnstone},
  {Crawford}, {Fabian}, \& {Allen}}]{edg02}
{Edge} A.~C., {Wilman} R.~J., {Johnstone} R.~M., {Crawford} C.~S., {Fabian}
  A.~C., {Allen} S.~W., 2002, MNRAS, 337, 49

\bibitem[{{Edwards} {et~al}\mbox{.}(2016){Edwards}, {Alpert}, {Trierweiler},
  {Abraham}, \& {Beizer}}]{edw16}
{Edwards} L.~O.~V., {Alpert} H.~S., {Trierweiler} I.~L., {Abraham} T., {Beizer}
  V.~G., 2016, MNRAS, 461, 230, \uppercase{P}aper I

\bibitem[{{Edwards} {et~al}\mbox{.}(2007){Edwards}, {Hudson}, {Balogh}, \&
  {Smith}}]{edw07}
{Edwards} L.~O.~V., {Hudson} M.~J., {Balogh} M.~L., {Smith} R.~J., 2007, MNRAS,
  379, 100

\bibitem[{{Edwards} \& {Patton}(2012)}]{edw12}
{Edwards} L.~O.~V., {Patton} D.~R., 2012, MNRAS, 425, 287

\bibitem[{{Edwards} {et~al}\mbox{.}(2009){Edwards}, {Robert}, {Moll{\'a}}, \&
  {McGee}}]{edw09}
{Edwards} L.~O.~V., {Robert} C., {Moll{\'a}} M., {McGee} S.~L., 2009,
  \hbox{MNRAS}, 396, 1953

\bibitem[{{Fabian} {et~al}\mbox{.}(2001){Fabian}, {Sanders}, {Ettori},
  {Taylor}, {Allen}, {Crawford}, {Iwasawa}, \& {Johnstone}}]{fab01}
{Fabian} A.~C., {Sanders} J.~S., {Ettori} S., {Taylor} G.~B., {Allen} S.~W.,
  {Crawford} C.~S., {Iwasawa} K., {Johnstone} R.~M., 2001, MNRAS, 321, L33

\bibitem[{{Fisher}, {Illingworth} \& {Franx}(1995){Fisher}, {Illingworth}, \&
  {Franx}}]{fis95a}
{Fisher} D., {Illingworth} G., {Franx} M., 1995, ApJ, 438, 539

\bibitem[{{Gallazzi} {et~al}\mbox{.}(2005){Gallazzi}, {Charlot}, {Brinchmann},
  {White}, \& {Tremonti}}]{gal05}
{Gallazzi} A., {Charlot} S., {Brinchmann} J., {White} S.~D.~M., {Tremonti}
  C.~A., 2005, MNRAS, 362, 41

\bibitem[{{Gerhard} {et~al}\mbox{.}(2001){Gerhard}, {Kronawitter}, {Saglia}, \&
  {Bender}}]{ger01}
{Gerhard} O., {Kronawitter} A., {Saglia} R.~P., {Bender} R., 2001, AJ, 121,
  1936

\bibitem[{{Gonz{\'a}lez}(1993)}]{gon93}
{Gonz{\'a}lez} J.~J., 1993, PhD thesis, Thesis (PH.D.)--UNIVERSITY OF
  CALIFORNIA, SANTA CRUZ, 1993.Source: Dissertation Abstracts International,
  Volume: 54-05, Section: B, page: 2551.

\bibitem[{{Greene} {et~al}\mbox{.}(2015){Greene}, {Janish}, {Ma}, {McConnell},
  {Blakeslee}, {Thomas}, \& {Murphy}}]{gre15}
{Greene} J.~E., {Janish} R., {Ma} C.-P., {McConnell} N.~J., {Blakeslee} J.~P.,
  {Thomas} J., {Murphy} J.~D., 2015, ApJ, 807, 11

\bibitem[{{Hartke} {et~al}\mbox{.}(2018){Hartke}, {Arnaboldi}, {Gerhard},
  {Agnello}, {Longobardi}, {Coccato}, {Pulsoni}, {Freeman}, \&
  {Merrifield}}]{har18}
{Hartke} J. {et~al.}, 2018, A\&A, 616, A123

\bibitem[{{Hatch}(2007)}]{hat07}
{Hatch} N., 2007, in {Boehringer} H. and {Schuecker} P. and {Pratt} G.W. and
  {Finoguenov} A., eds., ESO Astrophysics Symposia, Heating vs. Cooling in
  Galaxies and Clusters of Galaxies. Springer-Verlag, Garching

\bibitem[{{Hausman} \& {Ostriker}(1978)}]{hau78}
{Hausman} M.~A., {Ostriker} J.~P., 1978, ApJ, 224, 320

\bibitem[{{Hilker} {et~al}\mbox{.}(2018){Hilker}, {Richtler}, {Barbosa},
  {Arnaboldi}, {Coccato}, \& {Mendes de Oliveira}}]{hil18}
{Hilker} M., {Richtler} T., {Barbosa} C.~E., {Arnaboldi} M., {Coccato} L.,
  {Mendes de Oliveira} C., 2018, ArXiv e-prints

\bibitem[{{Hirschmann}, {De Lucia} \& {Fontanot}(2016){Hirschmann}, {De Lucia},
  \& {Fontanot}}]{hir16}
{Hirschmann} M., {De Lucia} G., {Fontanot} F., 2016, MNRAS, 461, 1760

\bibitem[{{Hirschmann} {et~al}\mbox{.}(2015){Hirschmann}, {Naab}, {Ostriker},
  {Forbes}, {Duc}, {Dav{\'e}}, {Oser}, \& {Karabal}}]{hir15}
{Hirschmann} M., {Naab} T., {Ostriker} J.~P., {Forbes} D.~A., {Duc} P.-A.,
  {Dav{\'e}} R., {Oser} L., {Karabal} E., 2015, MNRAS, 449, 528

\bibitem[{{Hopkins} {et~al}\mbox{.}(2009){Hopkins}, {Cox}, {Dutta},
  {Hernquist}, {Kormendy}, \& {Lauer}}]{hop09}
{Hopkins} P.~F., {Cox} T.~J., {Dutta} S.~N., {Hernquist} L., {Kormendy} J.,
  {Lauer} T.~R., 2009, ApJS, 181, 135

\bibitem[{{Hudson} {et~al}\mbox{.}(2010){Hudson}, {Mittal}, {Reiprich},
  {Nulsen}, {Andernach}, \& {Sarazin}}]{hud10}
{Hudson} D.~S., {Mittal} R., {Reiprich} T.~H., {Nulsen} P.~E.~J., {Andernach}
  H., {Sarazin} C.~L., 2010, A\&A, 513, A37

\bibitem[{{Hudson} \& {Ebeling}(1997)}]{hud97}
{Hudson} M.~J., {Ebeling} H., 1997, ApJ, 479, 621

\bibitem[{{Jimmy} {et~al}\mbox{.}(2013){Jimmy}, {Tran}, {Brough}, {Gebhardt},
  {von der Linden}, {Couch}, \& {Sharp}}]{jim13}
{Jimmy}, {Tran} K.-V., {Brough} S., {Gebhardt} K., {von der Linden} A., {Couch}
  W.~J., {Sharp} R.~G., 2013, ApJ, 778, 171

\bibitem[{{Kluge} {et~al}\mbox{.}(2019){Kluge}, {Neureiter}, {Riffeser},
  {Bender}, {Goessl}, {Hopp}, {Schmidt}, {Ries}, \& {Brosch}}]{klu19}
{Kluge} M. {et~al.}, 2019, arXiv e-prints, arXiv:1908.08544

\bibitem[{{Kobayashi}(2004)}]{kob04}
{Kobayashi} C., 2004, MNRAS, 347, 740

\bibitem[{{Krick} \& {Bernstein}(2007)}]{kri07}
{Krick} J.~E., {Bernstein} R.~A., 2007, AJ, 134, 466

\bibitem[{{Krick}, {Bernstein} \& {Pimbblet}(2006){Krick}, {Bernstein}, \&
  {Pimbblet}}]{kri06}
{Krick} J.~E., {Bernstein} R.~A., {Pimbblet} K.~A., 2006, ApJ, 131, 168

\bibitem[{{Krogager} {et~al}\mbox{.}(2014){Krogager}, {Zirm}, {Toft}, {Man}, \&
  {Brammer}}]{kro14}
{Krogager} J.-K., {Zirm} A.~W., {Toft} S., {Man} A., {Brammer} G., 2014, ApJ,
  797, 17

\bibitem[{{La Barbera} {et~al}\mbox{.}(2012){La Barbera}, {Ferreras}, {de
  Carvalho}, {Bruzual}, {Charlot}, {Pasquali}, \& {Merlin}}]{lab12}
{La Barbera} F., {Ferreras} I., {de Carvalho} R.~R., {Bruzual} G., {Charlot}
  S., {Pasquali} A., {Merlin} E., 2012, MNRAS, 426, 2300

\bibitem[{{Laine} {et~al}\mbox{.}(2003){Laine}, {van der Marel}, {Lauer},
  {Postman}, {O'Dea}, \& {Owen}}]{lai03}
{Laine} S., {van der Marel} R.~P., {Lauer} T.~R., {Postman} M., {O'Dea} C.~P.,
  {Owen} F.~N., 2003, AJ, 125, 478

\bibitem[{{Lakhchaura} \& {Singh}(2014)}]{lac14}
{Lakhchaura} K., {Singh} K.~P., 2014, AJ, 147, 156

\bibitem[{{Lauer} {et~al}\mbox{.}(2014){Lauer}, {Postman}, {Strauss}, {Graves},
  \& {Chiasri}}]{lau14}
{Lauer} T.~R., {Postman} M., {Strauss} M.~A., {Graves} G.~J., {Chiasri} N.~E.,
  2014, AJ, 797, 82

\bibitem[{{Longobardi} {et~al}\mbox{.}(2013){Longobardi}, {Arnaboldi},
  {Gerhard}, {Coccato}, {Okamura}, \& {Freeman}}]{lon13}
{Longobardi} A., {Arnaboldi} M., {Gerhard} O., {Coccato} L., {Okamura} S.,
  {Freeman} K.~C., 2013, A\&A, 558, A42

\bibitem[{{Longobardi} {et~al}\mbox{.}(2015){Longobardi}, {Arnaboldi},
  {Gerhard}, \& {Mihos}}]{lon15}
{Longobardi} A., {Arnaboldi} M., {Gerhard} O., {Mihos} J.~C., 2015, A\&A, 579,
  L3

\bibitem[{{Longobardi} {et~al}\mbox{.}(2018){Longobardi}, {Arnaboldi},
  {Gerhard}, {Pulsoni}, \& {S{\"o}ldner-Rembold}}]{lon18}
{Longobardi} A., {Arnaboldi} M., {Gerhard} O., {Pulsoni} C.,
  {S{\"o}ldner-Rembold} I., 2018, A\&A, 620, A111

\bibitem[{{L{\'o}pez-Cruz} {et~al}\mbox{.}(2014){L{\'o}pez-Cruz}, {A{\~n}orve},
  {Birkinshaw}, {Worrall}, {Ibarra-Medel}, {Barkhouse}, {Torres-Papaqui}, \&
  {Motta}}]{lop14}
{L{\'o}pez-Cruz} O., {A{\~n}orve} C., {Birkinshaw} M., {Worrall} D.~M.,
  {Ibarra-Medel} H.~J., {Barkhouse} W.~A., {Torres-Papaqui} J.~P., {Motta} V.,
  2014, ApJL, 795, L31

\bibitem[{{Lotz} {et~al}\mbox{.}(2017){Lotz}, {Koekemoer}, {Coe}, {Grogin},
  {Capak}, {Mack}, {Anderson}, {Avila}, {Barker}, {Borncamp}, {Brammer},
  {Durbin}, {Gunning}, {Hilbert}, {Jenkner}, {Khandrika}, {Levay}, {Lucas},
  {MacKenty}, {Ogaz}, {Porterfield}, {Reid}, {Robberto}, {Royle}, {Smith},
  {Storrie-Lombardi}, {Sunnquist}, {Surace}, {Taylor}, {Williams}, {Bullock},
  {Dickinson}, {Finkelstein}, {Natarajan}, {Richard}, {Robertson}, {Tumlinson},
  {Zitrin}, {Flanagan}, {Sembach}, {Soifer}, \& {Mountain}}]{lot16}
{Lotz} J.~M. {et~al.}, 2017, ApJ, 837, 97

\bibitem[{{Loubser} {et~al}\mbox{.}(2018){Loubser}, {Hoekstra}, {Babul}, \&
  {O'Sullivan}}]{lou18}
{Loubser} S.~I., {Hoekstra} H., {Babul} A., {O'Sullivan} E., 2018, \hbox
  {MNRAS}, 477, 335

\bibitem[{{Loubser} \& {S{\'a}nchez-Bl{\'a}zquez}(2012)}]{lou12}
{Loubser} S.~I., {S{\'a}nchez-Bl{\'a}zquez} P., 2012, MNRAS, 425, 841

\bibitem[{{Loubser} {et~al}\mbox{.}(2009){Loubser}, {S{\'a}nchez-Bl{\'a}zquez},
  {Sansom}, \& {Soechting}}]{lou09}
{Loubser} S.~I., {S{\'a}nchez-Bl{\'a}zquez} P., {Sansom} A.~E., {Soechting}
  I.~K., 2009, MNRAS, 398, 133

\bibitem[{{Loubser} {et~al}\mbox{.}(2008){Loubser}, {Sansom},
  {S{\'a}nchez-Bl{\'a}zquez}, {Soechting}, \& {Bromage}}]{lou08}
{Loubser} S.~I., {Sansom} A.~E., {S{\'a}nchez-Bl{\'a}zquez} P., {Soechting}
  I.~K., {Bromage} G.~E., 2008, MNRAS, 391, 1009

\bibitem[{{Mackie}(1992)}]{mac92}
{Mackie} G., 1992, ApJ, 400, 65

\bibitem[{{Maraston}(2011)}]{mar11}
{Maraston} C., 2011, in Astronomical Society of the Pacific Conference Series,
  Vol. 445, Why Galaxies Care about AGB Stars II: Shining Examples and Common
  Inhabitants, {Kerschbaum} F., {Lebzelter} T., {Wing} R.~F., eds., p. 391

\bibitem[{{Martel}, {Barai} \& {Brito}(2012){Martel}, {Barai}, \&
  {Brito}}]{mar12}
{Martel} H., {Barai} P., {Brito} W., 2012, ApJ, 757, 48

\bibitem[{{Martel}, {Robichaud} \& {Barai}(2014){Martel}, {Robichaud}, \&
  {Barai}}]{mar14}
{Martel} H., {Robichaud} F., {Barai} P., 2014, ApJ, 786, 79

\bibitem[{{McDonald} \& {Veilleux}(2009)}]{mcd09}
{McDonald} M., {Veilleux} S., 2009, ApJL, 703, L172

\bibitem[{{Melnick} {et~al}\mbox{.}(2012){Melnick}, {Giraud}, {Toledo},
  {Selman}, \& {Quintana}}]{mel12}
{Melnick} J., {Giraud} E., {Toledo} I., {Selman} F., {Quintana} H., 2012,
  MNRAS, 427, 850

\bibitem[{{Misgeld} {et~al}\mbox{.}(2011){Misgeld}, {Mieske}, {Hilker},
  {Richtler}, {Georgiev}, \& {Schuberth}}]{mis11}
{Misgeld} I., {Mieske} S., {Hilker} M., {Richtler} T., {Georgiev} I.~Y.,
  {Schuberth} Y., 2011, A\&A, 531, A4

\bibitem[{{Montes} \& {Trujillo}(2014)}]{mon14}
{Montes} M., {Trujillo} I., 2014, ApJ, 794, 137

\bibitem[{{Montes} \& {Trujillo}(2018)}]{mon18}
{Montes} M., {Trujillo} I., 2018, MNRAS, 474, 917

\bibitem[{{Murante} {et~al}\mbox{.}(2004){Murante}, {Arnaboldi}, {Gerhard},
  {Borgani}, {Cheng}, {Diaferio}, {Dolag}, {Moscardini}, {Tormen}, {Tornatore},
  \& {Tozzi}}]{mur04}
{Murante} G. {et~al.}, 2004, ApJL, 607, L83

\bibitem[{{Murante} {et~al}\mbox{.}(2007){Murante}, {Giovalli}, {Gerhard},
  {Arnaboldi}, {Borgani}, \& {Dolag}}]{mur07}
{Murante} G., {Giovalli} M., {Gerhard} O., {Arnaboldi} M., {Borgani} S.,
  {Dolag} K., 2007, MNRAS, 377, 2

\bibitem[{{Navarro}, {Frenk} \& {White}(1996){Navarro}, {Frenk}, \&
  {White}}]{nav96}
{Navarro} J.~F., {Frenk} C.~S., {White} S.~D.~M., 1996, ApJ, 462, 563

\bibitem[{{Newman} {et~al}\mbox{.}(2013){Newman}, {Treu}, {Ellis}, {Sand},
  {Nipoti}, {Richard}, \& {Jullo}}]{new13}
{Newman} A.~B., {Treu} T., {Ellis} R.~S., {Sand} D.~J., {Nipoti} C., {Richard}
  J., {Jullo} E., 2013, ApJ, 765, 24

\bibitem[{{Oliva-Altamirano} {et~al}\mbox{.}(2015){Oliva-Altamirano}, {Brough},
  {Jimmy}, {Couch}, {McDermid}, {Lidman}, {von der Linden}, \& {Sharp}}]{oli15}
{Oliva-Altamirano} P., {Brough} S., {Jimmy}, Kim-Vy T., {Couch} W.~J.,
  {McDermid} R.~M., {Lidman} C., {von der Linden} A., {Sharp} R., 2015, MNRAS,
  449, 3347

\bibitem[{{Oyarz{\'u}n} {et~al}\mbox{.}(2019){Oyarz{\'u}n}, {Bundy},
  {Westfall}, {Belfiore}, {Thomas}, {Maraston}, {Lian}, {Arag{\'o}n-Salamanca},
  {Zheng}, {Gonzalez-Perez}, {Law}, {Drory}, \& {Andrews}}]{oya19}
{Oyarz{\'u}n} G.~A. {et~al.}, 2019, ApJ, 880, 111

\bibitem[{{Padmanabhan} {et~al}\mbox{.}(2004){Padmanabhan}, {Seljak},
  {Strauss}, {Blanton}, {Kauffmann}, {Schlegel}, {Tremonti}, {Bahcall},
  {Bernardi}, {Brinkmann}, {Fukugita}, \& {Ivezi{\'c}}}]{pad04}
{Padmanabhan} N. {et~al.}, 2004, NewAstRev, 9, 329

\bibitem[{{Postman} {et~al}\mbox{.}(2012){Postman}, {Coe}, {Ben{\'{\i}}tez},
  {Bradley}, {Broadhurst}, {Donahue}, {Ford}, {Graur}, {Graves}, {Jouvel},
  {Koekemoer}, {Lemze}, {Medezinski}, {Molino}, {Moustakas}, {Ogaz}, {Riess},
  {Rodney}, {Rosati}, {Umetsu}, {Zheng}, {Zitrin}, {Bartelmann}, {Bouwens},
  {Czakon}, {Golwala}, {Host}, {Infante}, {Jha}, {Jimenez-Teja}, {Kelson},
  {Lahav}, {Lazkoz}, {Maoz}, {McCully}, {Melchior}, {Meneghetti}, {Merten},
  {Moustakas}, {Nonino}, {Patel}, {Reg{\"o}s}, {Sayers}, {Seitz}, \& {Van der
  Wel}}]{pos12}
{Postman} M. {et~al.}, 2012, ApJS, 199, 25

\bibitem[{{Postman} \& {Lauer}(1995)}]{pos95}
{Postman} M., {Lauer} T.~R., 1995, ApJ, 440, 28

\bibitem[{{Presotto} {et~al}\mbox{.}(2014){Presotto}, {Girardi}, {Nonino},
  {Mercurio}, {Grillo}, {Rosati}, {Biviano}, {Annunziatella}, {Balestra},
  {Cui}, {Sartoris}, {Lemze}, {Ascaso}, {Moustakas}, {Ford}, {Fritz}, {Czoske},
  {Ettori}, {Kuchner}, {Lombardi}, {Maier}, {Medezinski}, {Molino},
  {Scodeggio}, {Strazzullo}, {Tozzi}, {Ziegler}, {Bartelmann}, {Benitez},
  {Bradley}, {Brescia}, {Broadhurst}, {Coe}, {Donahue}, {Gobat}, {Graves},
  {Kelson}, {Koekemoer}, {Melchior}, {Meneghetti}, {Merten}, {Moustakas},
  {Munari}, {Postman}, {Reg{\H o}s}, {Seitz}, {Umetsu}, {Zheng}, \&
  {Zitrin}}]{pre14}
{Presotto} V. {et~al.}, 2014, A\&A, 565, A126

\bibitem[{{Puchwein} {et~al}\mbox{.}(2010){Puchwein}, {Springel}, {Sijacki}, \&
  {Dolag}}]{puc10}
{Puchwein} E., {Springel} V., {Sijacki} D., {Dolag} K., 2010, MNRAS, 406, 936

\bibitem[{{Purcell}, {Bullock} \& {Zentner}(2007){Purcell}, {Bullock}, \&
  {Zentner}}]{pur07}
{Purcell} C.~W., {Bullock} J.~S., {Zentner} A.~R., 2007, ApJ, 666, 20

\bibitem[{{Ragone-Figueroa} {et~al}\mbox{.}(2018){Ragone-Figueroa}, {Granato},
  {Ferraro}, {Murante}, {Biffi}, {Borgani}, {Planelles}, \& {Rasia}}]{rag18}
{Ragone-Figueroa} C., {Granato} G.~L., {Ferraro} M.~E., {Murante} G., {Biffi}
  V., {Borgani} S., {Planelles} S., {Rasia} E., 2018, MNRAS, 479, 1125

\bibitem[{{Richtler} {et~al}\mbox{.}(2011){Richtler}, {Salinas}, {Misgeld},
  {Hilker}, {Hau}, {Romanowsky}, {Schuberth}, \& {Spolaor}}]{rit11}
{Richtler} T., {Salinas} R., {Misgeld} I., {Hilker} M., {Hau} G.~K.~T.,
  {Romanowsky} A.~J., {Schuberth} Y., {Spolaor} M., 2011, A\&A, 531, A119

\bibitem[{{Rudick} {et~al}\mbox{.}(2009){Rudick}, {Mihos}, {Frey}, \&
  {McBride}}]{rud09}
{Rudick} C.~S., {Mihos} J.~C., {Frey} L.~H., {McBride} C.~K., 2009, ApJ, 699,
  1518

\bibitem[{{Rudick}, {Mihos} \& {McBride}(2006){Rudick}, {Mihos}, \&
  {McBride}}]{rud06}
{Rudick} C.~S., {Mihos} J.~C., {McBride} C., 2006, ApJ, 648, 936

\bibitem[{{Rudick}, {Mihos} \& {McBride}(2011){Rudick}, {Mihos}, \&
  {McBride}}]{rud11}
{Rudick} C.~S., {Mihos} J.~C., {McBride} C.~K., 2011, ApJ, 732, 48

\bibitem[{{Salom{\'e}} \& {Combes}(2003)}]{sal03}
{Salom{\'e}} P., {Combes} F., 2003, A\&A, 412, 657

\bibitem[{{Schneider} \& {Gunn}(1982)}]{sch82}
{Schneider} D.~P., {Gunn} J.~E., 1982, ApJ, 263, 14

\bibitem[{{Skibba} {et~al}\mbox{.}(2011){Skibba}, {van den Bosch}, {Yang},
  {More}, {Mo}, \& {Fontanot}}]{ski11}
{Skibba} R.~A., {van den Bosch} F.~C., {Yang} X., {More} S., {Mo} H.,
  {Fontanot} F., 2011, MNRAS, 410, 417

\bibitem[{{Smith} {et~al}\mbox{.}(2004){Smith}, {Hudson}, {Nelan}, {Moore},
  {Quinney}, {Wegner}, {Lucey}, {Davies}, {Malecki}, {Schade}, \&
  {Suntzeff}}]{smi04}
{Smith} R.~J. {et~al.}, 2004, AJ, 128, 1558

\bibitem[{{Sommer-Larsen}, {Romeo} \& {Portinari}(2005){Sommer-Larsen},
  {Romeo}, \& {Portinari}}]{som05}
{Sommer-Larsen} J., {Romeo} A.~D., {Portinari} L., 2005, MNRAS, 357, 478

\bibitem[{{Tang} {et~al}\mbox{.}(2018){Tang}, {Lin}, {Cui}, {Kang}, {Wang},
  {Contini}, \& {Yu}}]{tan18}
{Tang} L., {Lin} W., {Cui} W., {Kang} X., {Wang} Y., {Contini} E., {Yu} Y.,
  2018, ApJ, 859, 85

\bibitem[{{Toft} {et~al}\mbox{.}(2014){Toft}, {Smol{\v c}i{\'c}}, {Magnelli},
  {Karim}, {Zirm}, {Michalowski}, {Capak}, {Sheth}, {Schawinski}, {Krogager},
  {Wuyts}, {Sanders}, {Man}, {Lutz}, {Staguhn}, {Berta}, {Mccracken}, {Krpan},
  \& {Riechers}}]{tof14}
{Toft} S. {et~al.}, 2014, ApJ, 782, 68

\bibitem[{{Tonini} {et~al}\mbox{.}(2012){Tonini}, {Bernyk}, {Croton},
  {Maraston}, \& {Thomas}}]{ton12}
{Tonini} C., {Bernyk} M., {Croton} D., {Maraston} C., {Thomas} D., 2012, ApJ,
  759, 43

\bibitem[{{Tonry}(1983)}]{ton83}
{Tonry} J.~L., 1983, ApJ, 266, 58

\bibitem[{{Tremblay} {et~al}\mbox{.}(2015){Tremblay}, {O'Dea}, {Baum},
  {Mittal}, {McDonald}, {Combes}, {Li}, {McNamara}, {Bremer}, {Clarke},
  {Donahue}, {Edge}, {Fabian}, {Hamer}, {Hogan}, {Oonk}, {Quillen}, {Sanders},
  {Salom{\'e}}, \& {Voit}}]{tre15}
{Tremblay} G.~R. {et~al.}, 2015, MNRAS, 451, 3768

\bibitem[{{Tremonti} {et~al}\mbox{.}(2004){Tremonti}, {Heckman}, {Kauffmann},
  {Brinchmann}, {Charlot}, {White}, {Seibert}, {Peng}, {Schlegel}, {Uomoto},
  {Fukugita}, \& {Brinkmann}}]{tre04}
{Tremonti} C.~A. {et~al.}, 2004, ApJ, 613, 898

\bibitem[{{Treu} {et~al}\mbox{.}(2005){Treu}, {Ellis}, {Liao}, \& {van
  Dokkum}}]{tre05}
{Treu} T., {Ellis} R.~S., {Liao} T.~X., {van Dokkum} P.~G., 2005, ApJL, 622, L5

\bibitem[{{Trujillo} {et~al}\mbox{.}(2006){Trujillo}, {Feulner}, {Goranova},
  {Hopp}, {Longhetti}, {Saracco}, {Bender}, {Braito}, {Della Ceca}, {Drory},
  {Mannucci}, \& {Severgnini}}]{tru06}
{Trujillo} I. {et~al.}, 2006, MNRAS, 373, L36

\bibitem[{{Tutukov}, {Dryomov} \& {Dryomova}(2007){Tutukov}, {Dryomov}, \&
  {Dryomova}}]{tut07}
{Tutukov} A.~V., {Dryomov} V.~V., {Dryomova} G.~N., 2007, Astronomy Reports,
  51, 435

\bibitem[{{Tutukov} \& {Fedorova}(2011)}]{tut11}
{Tutukov} A.~V., {Fedorova} A.~V., 2011, Astronomy Reports, 55, 383

\bibitem[{{van de Voort}(2016)}]{van16}
{van de Voort} F., 2016, MNRAS, 462, 778

\bibitem[{{van Dokkum} {et~al}\mbox{.}(2008){van Dokkum}, {Franx}, {Kriek},
  {Holden}, {Illingworth}, {Magee}, {Bouwens}, {Marchesini}, {Quadri},
  {Rudnick}, {Taylor}, \& {Toft}}]{van08}
{van Dokkum} P.~G. {et~al.}, 2008, ApJL, 677, L5

\bibitem[{{Vazdekis} {et~al}\mbox{.}(2015){Vazdekis}, {Coelho}, {Cassisi},
  {Ricciardelli}, {Falc{\'o}n-Barroso}, {S{\'a}nchez-Bl{\'a}zquez}, {La
  Barbera}, {Beasley}, \& {Pietrinferni}}]{vaz15}
{Vazdekis} A. {et~al.}, 2015, MNRAS, 449, 1177

\bibitem[{{Veale} {et~al}\mbox{.}(2018){Veale}, {Ma}, {Greene}, {Thomas},
  {Blakeslee}, {Walsh}, \& {Ito}}]{vea18}
{Veale} M., {Ma} C.-P., {Greene} J.~E., {Thomas} J., {Blakeslee} J.~P., {Walsh}
  J.~L., {Ito} J., 2018, MNRAS, 473, 5446

\bibitem[{{Veale} {et~al}\mbox{.}(2017){Veale}, {Ma}, {Thomas}, {Greene},
  {McConnell}, {Walsh}, {Ito}, {Blakeslee}, \& {Janish}}]{vea17}
{Veale} M. {et~al.}, 2017, MNRAS, 464, 356

\bibitem[{{von der Linden} {et~al}\mbox{.}(2006){von der Linden}, {Best},
  {Kauffmann}, \& {White}}]{von06}
{von der Linden} A., {Best} P.~N., {Kauffmann} G., {White} S.~D.~M., 2006,
  astro-ph/0611196

\bibitem[{{Vulcani} {et~al}\mbox{.}(2014){Vulcani}, {Bundy}, {Lackner},
  {Leauthaud}, {Treu}, {Mei}, {Coccato}, {Kneib}, {Auger}, \& {Nipoti}}]{vul14}
{Vulcani} B. {et~al.}, 2014, ApJ, 797, 62

\bibitem[{{Walcher} {et~al}\mbox{.}(2009){Walcher}, {Coelho}, {Gallazzi}, \&
  {Charlot}}]{wal09}
{Walcher} C.~J., {Coelho} P., {Gallazzi} A., {Charlot} S., 2009, MNRAS, 398,
  L44

\bibitem[{{White}, {Jones} \& {Forman}(1997){White}, {Jones}, \&
  {Forman}}]{whi97}
{White} D.~A., {Jones} C., {Forman} W., 1997, MNRAS, 292, 419

\bibitem[{{Willman} {et~al}\mbox{.}(2004){Willman}, {Governato}, {Wadsley}, \&
  {Quinn}}]{wil04}
{Willman} B., {Governato} F., {Wadsley} J., {Quinn} T., 2004, MNRAS, 355, 159

\bibitem[{{Worthey}(1994)}]{wor94}
{Worthey} G., 1994, ApJS, 95, 107

\end{thebibliography}

\end{document}